\documentclass[
prd,
aps,
reprint,
amsmath,amssymb,
groupedaddress]{revtex4-2}
\usepackage{bm}
\usepackage{color}    
\usepackage{booktabs}
\usepackage{array}
\usepackage{graphicx}  
\usepackage{subfigure}
\usepackage{placeins}
\usepackage{multirow}   
\usepackage{float}
\usepackage[colorlinks,
            linkcolor=red,
            anchorcolor=blue,
            citecolor=cyan,     
            ]{hyperref}
\begin{document}

%%%%%%%%%%%
\title{The Dynamical Instability of Rotating Boson Stars}
\author{Kai-Dong Zhou}
\email{kaidong.zhou@mail.sdu.edu.cn}
\author{Shou-Shan Bao}
\email{ssbao@sdu.edu.cn}
\author{Jian Deng}
\email{jdeng@sdu.edu.cn}
\author{Hong Zhang}
\email{hong.zhang@sdu.edu.cn}
\affiliation{Key Laboratory of Particle Physics and Particle Irradiation (MOE),\\
Institute of Frontier and Interdisciplinary Science, \\
Shandong University, Qingdao, Shandong 266237, China}
%%%%%%%%%%%

\date{\today}

\begin{abstract}
We investigate the dynamical instability of rotating boson stars described by the Gross--Pitaevskii--Poisson equations with contact self-interactions. Through three-dimensional simulations, we confirm that the rotating boson star undergoes a quasiperiodic conversion between ring-like and twin-star-like density configurations in the early nonlinear stage. We develop a systematic linear stability analysis to identify the modes driving this instability and show that repulsive self-interactions could significantly increase the lifetime of the rotating boson stars. We further construct a three-mode Hamiltonian to describe the early nonlinear stage, which explains the quasiperiodic conversion. This analytical framework agrees reasonably well with the simulation results and provides a clear picture to understand the dynamics of rotating boson stars.
\end{abstract}

\maketitle

\section{Introduction}\label{sec:intro}

%%%%%==
Ultralight bosonic dark matter has attracted considerable interest because its wave-like properties may alleviate the small-scale tensions of collisionless cold dark matter \cite{Hu:2000ke,Hui:2016ltb,Chavanis:2025qcg}. Owing to its very large occupation number, bosonic dark matter can be treated as a classical field described by a macroscopic wave function. Numerical simulations show this field can relax into configurations with a core--envelope structure through gravitational cooling \cite{Seidel:1993zk,Schive:2014dra,Levkov:2018kau,Chen:2020cef}. Recent phenomenological studies have explored how such configurations influence stellar-system dynamics \cite{Bar-Or:2018pxz,DuttaChowdhury:2023qxg,Yang:2025bae,Yang:2026wdk}. In these configurations, the central solitonic core is well approximated by a stationary self-gravitating condensate of the bosonic field, commonly known as a gravitational boson condensate.
%%%%%==

%%%%%==
Non-rotating spherical boson configurations have been extensively studied in both relativistic \cite{Kaup:1968zz,Ruffini:1969qy,Gleiser:1988rq,Seidel:1990jh} and nonrelativistic frameworks \cite{Chavanis:2011zi,Harrison2003,Guzman:2004wj,Bernal:2006it,Nambo:2024gvs,Braaten:2015eeu,Zhang:2018slz,Braaten:2019knj}, including their stationary profiles and dynamical stability. Realistic astrophysical systems, however, commonly carry angular momentum. It is natural to ask whether a boson condensate can carry angular momentum. Unlike normal stars undergoing rigid-body rotation, boson fields can carry localized angular momentum only through vortices because of their wave nature. This difference may give rise to distinct dynamical behavior and observational signatures.
%%%%%==

%%%%%==
Rotating boson condensates hosting a central black hole (BH) have attracted widespread interest in the context of BH superradiance \cite{Detweiler:1980uk,Arvanitaki:2010sy,Brito:2015oca,Bao:2022hew,Bao:2023xna}. When the Compton wavelength of an ultralight boson is comparable to the BH radius, quasibound states satisfying the superradiant condition can grow exponentially by extracting rotational energy and angular momentum from the BH, thereby forming a condensate carrying angular momentum. Since this mechanism relies only on gravity, such BH-condensate systems provide sensitive probes of otherwise very weakly coupled ultralight bosons. They can spin down BHs and create characteristic gaps in the BH mass-spin plane \cite{Arvanitaki:2010sy,Cardoso:2018tly,Fernandez:2019qbj,Ng:2019jsx,Ng:2020ruv,Cheng:2022jsw}, emit quasi-monochromatic gravitational waves through boson-pair annihilations \cite{Arvanitaki:2010sy,Arvanitaki:2014wva,Yoshino:2013ofa,Yoshino:2014wwa,Brito:2015oca,Brito:2017wnc, Brito:2017zvb,Guo:2022mpr,Yang:2023vwm,Guo:2024dqd} and transitions between condensate levels \cite{Arvanitaki:2010sy,Arvanitaki:2014wva,Omiya:2024xlz}, and leave characteristic imprints on compact-binary dynamics \cite{Baumann:2018vus,Berti:2019wnn,Baumann:2019ztm,Cardoso:2020hca,Takahashi:2021yhy,Baumann:2022pkl,Tong:2022bbl,Takahashi:2023flk,Xu:2026cky,Xu:2026aic}. Bosonic self-interactions can further give rise to rich nonlinear dynamics, including mode coupling, saturation, and collapse \cite{Yoshino:2012kn,Baryakhtar:2020gao,Omiya:2022mwv,Omiya:2022gwu}.
%%%%%==

%%%%%==
Comparably, the rotating boson condensates under self-gravity, which are referred to as rotating boson stars below, are not very well understood \cite{Silveira:1995dh,Schupp:1995dy,Flores:2024tnu,Liebling:2012fv,Rindler-Daller:2011afd,Schobesberger:2021ghi,Purohit:2023izj}. The astrophysical viability of these stationary vortex configurations depends on their dynamical stability. Relativistic and nonrelativistic real-time simulations have identified a nonaxisymmetric instability, whose effects are weaker with stronger repulsive self-interactions \cite{DiGiovanni:2020ror,Siemonsen:2020hcg,Dmitriev:2021utv}. Rather than relaxing directly into a localized clump, an unstable star first passes through an intermediate stage characterized by a quasiperiodic conversion between ring-like and twin-star-like density configurations \cite{Dmitriev:2021utv}. The perturbation spectrum underlying the instability and the nonlinear mechanism responsible for this conversion remain unclear. Clarifying them is important for determining the lifetime and possible observational signatures of rotating boson stars.
%%%%%==

%%%%%==
The central problem is therefore to identify the perturbation spectrum and determine how the growth of the dominant unstable mode develops into a quasiperiodic conversion between the axisymmetric ring-like and nonaxisymmetric twin-star-like configurations. The present work addresses this problem in two steps. We first develop a systematic linear stability analysis to characterize the perturbation spectrum and eigenfunctions. The stationary background and the dominant unstable eigenmode are then used to construct a three-mode model that explains the appearance of the nonaxisymmetric configuration. Interestingly, the nonaxisymmetric density distribution arises from the interference of axisymmetric modes with different azimuthal numbers. This model further explains the nonlinear saturation of the instability and the reversal of population transfer, thereby providing a qualitative description of the quasiperiodic conversion.
%%%%%==

%%%%%==
The remainder of this paper is organized as follows. Section~\ref{sec:GPP} introduces the nonrelativistic mathematical framework. The stationary rotating solutions are presented in Sec.~\ref{sec:stationary}, and their full three-dimensional evolution is studied in Sec.~\ref{sec:time_evolution}. Section~\ref{sec:linear} develops the linear stability analysis. Section~\ref{sec:nonlinear} derives the three-mode effective dynamics to describe the quasiperiodic conversion and compares its predictions with the full simulations. The final section summarizes the main results.
%%%%%==

\section{GPP equations}\label{sec:GPP}

%%%%%==
Our starting point is the Lagrangian density for a real scalar field minimally coupled to gravity. In natural units ($\hbar=c=1$), it is given by
%%%%%
\begin{equation}\label{eq:full_action}
\mathcal{L} = \sqrt{-g} \left( \frac{R}{16\pi G} + \mathcal{L}_{\rm M} \right),
\end{equation}
%%%%%
where $g\equiv\det(g_{\mu\nu})$ and $R$ is the Ricci scalar. We adopt the metric signature $(-,+,+,+)$. The matter sector is described by
%%%%%
\begin{equation}
\mathcal{L}_{\rm M} = -\frac{1}{2}g^{\mu\nu}\partial_\mu \phi \partial_\nu \phi - \frac{1}{2}\mu^2 \phi^2 - \frac{\lambda_{\rm R}}{4!} \phi^4.
\label{eq:scalar_L}
\end{equation}
%%%%%
We focus on nonrelativistic boson stars in the weak-field regime and adopt the following Newtonian-gauge metric,
%%%%%
\begin{equation}
g_{00}=-(1+2\Phi),\ 
g_{0i}=0,\ 
g_{ij}=(1-2\Phi)\delta_{ij},
\end{equation}
%%%%%
where $\Phi$ is the Newtonian gravitational potential. The real scalar field is decomposed as
\begin{equation}
\phi=\frac{1}{\sqrt{2\mu}}
\left(
\psi e^{-i\mu t}
+
\psi^*e^{i\mu t}
\right),
\end{equation}
%%%%%
where $\psi$ is a slowly varying complex field. Substituting this metric and field decomposition into Eq.~\eqref{eq:full_action}, expanding to leading order and dropping the rapidly oscillating terms yields the nonrelativistic effective Lagrangian density
%%%%%
\begin{equation}\label{eq:lagNR1}
\begin{split}
    \mathcal{L}_{\rm NR}[\psi,\psi^*,\Phi]&= \frac{i}{2} (\psi^* \partial_t \psi - \psi \partial_t \psi^*) - \frac{1}{2\mu} |\nabla \psi|^2  
    \\
    &- \mu \Phi |\psi|^2 - \frac{1}{8\pi G} (\nabla \Phi)^2 - \frac{\lambda_{\rm NR}}{2} |\psi|^4,
\end{split}
\end{equation}
%%%%%
where the nonrelativistic coupling constant is defined as $\lambda_{\rm NR} = \lambda_{\rm R}/8\mu^2$. The same Lagrangian can be obtained more rigorously with a canonical transformation. We refer interested readers to Refs.~\cite{Braaten:2016kzc,Braaten:2018lmj,Namjoo:2017nia} for details.
%%%%%==

%%%%%==
Varying the Lagrangian with respect to $\psi^*$ and $\Phi$ yields the equations of motion
%%%%%
\begin{subequations}\label{eq:dim_GPP}
\begin{align}
    i\partial_t \psi &= -\frac{1}{2\mu} \nabla^2\psi + \mu\Phi\psi + \lambda_{\rm NR} |\psi|^2 \psi, \\
    \nabla^2\Phi &= 4\pi G\mu |\psi|^2. \label{eq:dim_poisson}
\end{align}
\end{subequations}
%%%%%
These are precisely the Gross--Pitaevskii--Poisson (GPP) equations with a quartic self-interaction. Since $\mathcal{L}_{\rm NR}$ contains no time derivative of $\Phi$, the Newtonian potential is not an independent dynamical degree of freedom, but an auxiliary field that can be eliminated from the Lagrangian. Substituting the solution of Eq.~\eqref{eq:dim_poisson} back into Eq.~\eqref{eq:lagNR1} gives
%%%%%
\begin{equation}\label{eq:lagNR}
\begin{split}
    \mathcal{L}_{\rm NR}[\psi, \psi^*] &= \frac{i}{2}\left(\psi^*\partial_t\psi-\psi\partial_t\psi^*\right) - \frac{1}{2\mu} |\nabla \psi|^2 
    \\
    &- \frac{1}{2}\mu\Phi|\psi|^2 - \frac{\lambda_{\rm NR}}{2}|\psi|^4 ,
\end{split}
\end{equation}
%%%%%
where $\Phi$ is now a functional of $\psi$ determined via the Green's function:
\begin{equation}\label{eq:green_func}
    \Phi(\bm x) = 4\pi G \mu \int d^3 \bm x' \mathcal{G}(\bm x - \bm x') |\psi(\bm x')|^2,
\end{equation}
where $\mathcal{G}$ satisfies $\nabla^2 \mathcal{G}(\bm x) = \delta^3(\bm x)$.
%%%%%
A Legendre transformation then yields the Hamiltonian density
\begin{align}\label{eq:hamiltonian}
\mathcal{H}_{\rm NR} = 
-\frac{1}{2\mu}\psi^*\nabla^2\psi
+\frac{1}{2}\mu\Phi|\psi|^2
+\frac{\lambda_{\rm NR}}{2}|\psi|^4 .
\end{align}
%%%%%==

%%%%%==
The Lagrangian in Eq.~\eqref{eq:lagNR} possesses several symmetries that yield conserved quantities. The most relevant ones for the present analysis are the total energy and total particle number,
%%%%%
\begin{align}
E_{tot}&=E_K+E_G+E_I
\\
&=
\int d^3x
\left[
\frac{1}{2\mu}|\nabla\psi|^2
+\frac{1}{2}\mu\Phi|\psi|^2
+\frac{\lambda_{\rm NR}}{2}|\psi|^4
\right], \label{eq:energy_2}
\\
N&=\int d^3x\ |\psi|^2.
\end{align}
%%%%%
In the nonrelativistic limit, the total mass is well approximated by $M\approx \mu N$. The system also exhibits a useful scaling symmetry under the following transformation \cite{Ruffini:1969qy}
%%%%%
\begin{equation}
\begin{split}
&\bm x'=\alpha \bm x,\ t'=\alpha^2 t,\ \psi'(\bm x',t')=\alpha^{-2}\psi(\bm x,t),
\\
&\Phi'(\bm x',t')=\alpha^{-2}\Phi(\bm x,t),\ \lambda_{\rm NR}'=\alpha^2\lambda_{\rm NR},
\end{split}
\end{equation}
%%%%%
where $\alpha$ is an arbitrary positive parameter. This transformation rescales the Lagrangian by only an overall constant factor, leaving the equations of motion invariant.
%%%%%==

%%%%%==
Motivated by this scaling symmetry, we introduce the dimensionless quantities
\begin{equation}\label{eq:dimless_def}
\begin{split}
    &\tilde{\bm x} = \alpha\mu\bm x,\ \tilde{t} = \alpha^2 \mu t,\ \tilde{\psi} = \alpha^{-2} \sqrt{\frac{4\pi G}{\mu}} \psi,
    \\
    &\tilde{\Phi} = \alpha^{-2} \Phi,\ \tilde{\lambda} = \alpha^{2} \frac{1}{4\pi G} \lambda_{\rm NR}.
\end{split}
\end{equation}
In terms of these variables, the GPP equations \eqref{eq:dim_GPP} simplify to
%%%%%
\begin{subequations}\label{eq:dimless_GPP}
\begin{align}
    i\partial_{\tilde{t}} \tilde{\psi} &= -\frac{1}{2} \nabla^2\tilde{\psi} + \tilde{\Phi}\tilde{\psi} + \tilde{\lambda} |\tilde{\psi}|^2 \tilde{\psi}, \\
    \tilde{\nabla}^2\tilde{\Phi} &= |\tilde{\psi}|^2. \label{eq:dimensionless_poisson}
\end{align}
\end{subequations}
%%%%%
Their solutions are characterized by two independent dimensionless quantities, namely the self-interaction strength $\tilde\lambda$ and the total mass $\tilde{M}=\int {\rm d}^3 \tilde{x} |\tilde{\psi}|^2$. Owing to the scaling freedom, one of them can be fixed through an appropriate choice of the scale factor $\alpha$. In this work, we choose
%%%%%
\begin{align}
\alpha= G M \mu. 
\end{align}
The dimensionless total mass is fixed to 
\begin{equation}\label{eq:particle_normalization}
\tilde{M} = \tilde{N}_T = \frac{4\pi G \mu^2}{\alpha} \int d^3 x |\psi|^2 = 4\pi.
\end{equation}
%%%%%
The system is therefore completely characterized by the single dimensionless coupling $\tilde{\lambda}$.
%%%%%==

%%%%%==
In the rest of the paper, all numerical calculations and model constructions are performed using these dimensionless variables. For notational simplicity, we omit the tildes henceforth and use the subscript ``phys" to denote quantities expressed in physical units.
%%%%%==

\section{Stationary solutions}
\label{sec:stationary}

\subsection{Stationary configurations}
%%%%%==
We first construct the stationary solutions of Eqs.~\eqref{eq:dimless_GPP}. Following the standard axisymmetric parametrization for stationary states carrying definite angular momentum along the $z$ axis \cite{Liebling:2012fv}, the wave function is expressed as
%%%%%
\begin{equation}\label{eq:ansatz-1}
\psi(t, \bm x) = \psi_m(r, \theta)  e^{im\varphi} e^{-i \omega_m t}, 
\end{equation}
%%%%%
where $\psi_m$ is the meridional profile depending only on $r$ and $\theta$. The eigenvalue $\omega_m < 0$ represents the binding energy per particle. From the definition of dimensionless time in Eq.~\eqref{eq:dimless_def}, $t=\alpha^2 \mu\, t_{\rm phys}$, the dimensionless eigenenergy is related to its physical counterpart by
%%%%%
\begin{equation}
    \omega_{m,\rm phys} = \alpha^2 \mu\, \omega_m,
\end{equation}
%%%%%
The nonrelativistic limit requires $|\omega_{m,\rm phys}|\ll \mu$, which translates to $\alpha^2|\omega_m| \ll 1$. The azimuthal quantum number $m$ is required to be an integer for the single-valuedness of $\psi$ and is directly related to the conserved angular momentum along the symmetry axis through $J_z=mN$. For $m=0$, the lowest-energy configuration is strictly spherically symmetric. For $m\neq 0$, the regularity of $\psi$ across the spatial domain requires that $\psi_m$ vanishes on the axis of rotation. Hence, configurations with non-zero angular momentum exhibit a toroidal morphology enclosing a vortex line with azimuthal number $m$ \cite{Dmitriev:2021utv}. In this setup, the gravitational potential is static and axisymmetric,
%%%%%
\begin{equation}\label{eq:ansatz-2}
\Phi(t,\bm x) = \Phi_0 (r, \theta),
\end{equation}
%%%%%
where the subscript $0$ indicates no accompaniment of the azimuthal-dependence multiplier. Substituting these expressions into Eq.~\eqref{eq:dimless_GPP} yields 
%%%%%
\begin{subequations}\label{eq:GPP_stationary}
\begin{align}
    \omega_m \psi_m &= -\frac{1}{2} \nabla^2_m \psi_m  + \Phi_0 \psi_m + \lambda \psi^3_m , \label{eq:sch_stationary}
    \\
    \nabla^2_0 \Phi_0 &= \psi_m^2, \label{eq:poisson_stationary}
\end{align}
\end{subequations}
%%%%%
where $\nabla^2_m$ is defined as $\nabla^2 - m^2/(r^2\sin^2\theta)$ .
%%%%%==

\subsection{Energy relations}
\label{subsec:energy-relations}
%%%%%==
We now derive the virial theorem \cite{Wang:2001wq}. Rather than working with an arbitrary functional variation, it is useful to consider a restricted one-parameter family of spatial rescalings, parameterized by a positive real constant $\beta$ through the transformation $\bm{x} \to \beta \bm{x}$. To preserve the total particle number, the wave function must transform as $\psi \to \beta^{-3/2}\psi$. The Poisson equation then dictates that the gravitational potential must scale as $\Phi \to \beta^{-1}\Phi$. Under this transformation, the energy functional in Eq.~\eqref{eq:energy_2} scales as
%%%%%
\begin{equation}\label{eq:variation}
E[\psi_\beta] = \beta^{-2} E_K + \beta^{-1} E_G + \beta^{-3} E_I.
\end{equation}
%%%%%
Evaluating the first and second variations of $E[\psi_\beta]$ with respect to $\beta$ at $\beta=1$ gives
%%%%%
\begin{align}
\frac{d E[\psi_\beta]}{d \beta} \biggl|_{\beta=1} &= -2E_K - E_G - 3 E_I, \label{eq:variation_first}
\\
\frac{d^2 E[\psi_\beta]}{d \beta^2} \biggl|_{\beta=1} &= 6 E_K + 2 E_G + 12 E_I. \label{eq:variation_second}
\end{align}
%%%%%
Since the energy of a stationary solution must be extremal with respect to any allowed variation, including the restricted spatial scaling considered here, setting $dE/d\beta|_{\beta=1}=0$ yields the virial theorem
%%%%%
\begin{equation}
2E_K + E_G + 3 E_I = 0.
\label{eq:virial}
\end{equation}
%%%%%
Using this relation to eliminate $E_G$ in Eq.~\eqref{eq:variation_second} gives
%%%%%
\begin{equation}
  \frac{d^2 E[\psi_\beta]}{d \beta^2} \biggl|_{\beta=1} = -2E_K - 2E_G = 2 E_K + 6 E_I. \label{eq:variation_second2}
\end{equation}
%%%%%
For attractive self-interactions ($E_I<0$), the stationary configurations can be unstable. In contrast, for repulsive self-interactions ($E_I>0$), the second variation is strictly positive, indicating stability against this specific scaling perturbation.
%%%%%==

%%%%%==
Furthermore, multiplying the stationary Schr\"odinger equation \eqref{eq:sch_stationary} by $\psi_m^*$ and integrating over space gives a relation between the eigenvalue and the different energy contributions,
%%%%%
\begin{equation}\label{eq:omega_relation_general}
N \omega_m = E_K + 2 E_G + 2 E_I,
\end{equation}
%%%%%
where $N$ equals $4\pi$ in the present normalization. 
In the non-interacting limit ($\lambda=0$), combining Eqs.~\eqref{eq:virial} and \eqref{eq:omega_relation_general} leads to the relation between the eigenvalue and the total energy
%%%%%
\begin{equation}\label{eq:omega_relation}
    \omega_m = 3E/N,\ \text{for}\ \lambda=0.
\end{equation}
%%%%%==  

\subsection{Boundary conditions and numerical method}
%%%%%==
To solve the coupled stationary GPP system \eqref{eq:GPP_stationary} for rotating boson stars, we first impose physically motivated boundary conditions. Since the equations are invariant under reflection with respect to the equatorial plane ($\theta=\pi/2$), the solutions can be classified according to their parity. It is therefore sufficient to restrict the computational domain to $(r,\theta)\in[0,\infty)\times[0,\pi/2]$, with
%%%%%
\begin{align}\label{eq:BD_parity}
    \partial_\theta \psi_m(r, \pi/2) &= 0\quad \text{for even parity},
    \\
    \psi_m(r, \pi/2) &= 0\quad \text{for odd parity}.
\end{align}
%%%%%
The gravitational potential $\Phi$ is sourced by the mass density $|\psi|^2$, which remains symmetric with respect to the equatorial plane. Hence the potential satisfies the same boundary condition in both parity sectors,
%%%%%
\begin{equation}
\partial_\theta \Phi_0(r, \pi/2) = 0.
\end{equation}
%%%%%
On the symmetry axis, the regularity conditions for the wave function depend on the azimuthal number. For rotating solutions with $m\neq0$, the scalar field must vanish on the $z$ axis,
%%%%%
\begin{equation}
    \psi_m(r, 0) = \psi_m(0, \theta) = 0,
\end{equation}
%%%%%
whereas for the spherically symmetric $m=0$ solution, regularity instead requires $\partial_r\psi_0(0,\theta)=0$ and $\partial_\theta\psi_0(r,0)=0$. The corresponding boundary conditions for the gravitational potential are
%%%%%
\begin{equation}
    \partial_\theta \Phi_0(r,0) = \partial_r \Phi_0(0, \theta) = 0,
\end{equation}
%%%%%
which follow from the symmetry of the source distribution. At infinity, the bound state configuration requires
%%%%%
\begin{equation}\label{eq:BD_infty}
\psi_m(\infty, \theta) = \Phi_0(\infty, \theta) = 0.
\end{equation}
%%%%%
For a given parity sector, the stationary equations together with the boundary conditions in Eqs.~\eqref{eq:BD_parity}--\eqref{eq:BD_infty} determine the two-dimensional profiles of $\psi_m$ and $\Phi_0$.
%%%%%==

%%%%%==
For numerical convenience, we compactify the semi-infinite domain $(r,\theta)\in[0,\infty)\times[0,\pi/2]$ into the unit square $[0,1]\times[0,1]$ through the coordinate transformations $r_c = r/(r + L_r)$ and $\theta_c = 2\theta/\pi$. The scaling parameter $L_r$ is a characteristic length scale optimized to center the wave function within the computational grid. In this work, we choose $L_r = 10$, which provides sufficient resolution for the localized scalar profile without altering the underlying physical properties. The coordinate compactification preserves all boundary and regularity conditions given above.
%%%%%==

%%%%%==
The stationary GPP Eqs.~\eqref{eq:GPP_stationary} are solved using the finite element method (FEM) implemented with the open-source library \texttt{FEniCSx} \cite{BarattaEtal2023,LoggEtal2012}. The basic idea of FEM is to rewrite the differential equations in variational form and then approximate the unknown fields by a finite set of local polynomial basis functions defined on individual mesh cells. In the present calculation, the compactified domain $[0,1]\times[0,1]$ is discretized by a uniform triangular mesh, and the fields are approximated using piecewise quadratic Lagrange ($P_2$) finite elements. The numbers of cells ($N_r$, $N_\theta$) are chosen for both accuracy and computational efficiency. The resulting nonlinear eigenvalue problem is then solved using the self-consistent iterative scheme summarized below.
%%%%%
\begin{enumerate}
%%%%%
\item \textit{Initialization.} An initial scalar field $\psi_m^{(0)}$ is guessed based on the asymptotic behaviors and the parity of the system. Near the rotation axis, regularity requires $\psi_m \sim (r\sin\theta)^m$, whereas a bound-state configuration requires $\psi_m \sim e^{-\sqrt{-2\omega_m}r}$ at infinity. Since $\omega_m$ is not known at this stage, we simply choose $\psi_m \sim e^{-r}$ at large $r$. Then the initial wave functions are
\begin{equation}
    \psi_m^{(0)} = 
    \begin{cases}
        (r\sin\theta)^m e^{-r} & \text{for even parity}, \\
        (r\sin 2\theta)^m e^{-r} & \text{for odd parity}.
    \end{cases}
\end{equation}
Substituting $\psi_m^{(0)}$ into the Poisson equation gives the associated initial gravitational potential $\Phi_0^{(0)}$.
%%%%%
\item \textit{Wave function update.} At the $j$-th iteration ($j\ge 1$), the gravitational and self-interaction contributions obtained from the previous iteration are held fixed, reducing the nonlinear Schrödinger equation to a linear eigenvalue problem, which is solved using the Scalable Library for Eigenvalue Problem Computations (\texttt{SLEPc}) \cite{slepc-users-manual}, with the embedded shift-and-invert spectral transformation to target the desired bound-state eigenvalue.
%%%%%
\item \textit{State selection and normalization.} The obtained eigenstates $\psi^{(j)}_{m,n}\ (n\ge 1)$ are ordered according to the eigenvalues $\omega^{(j)}_{m,n}$, where $n=1$ corresponds to the ground state. The selected eigenfunctions are then normalized according to the global mass constraint in Eq.~\eqref{eq:particle_normalization}.
%%%%%
\item \textit{Potential update.} The normalized wave function $\psi_{m,n}^{(j)}$ is substituted back into the Poisson equation to update the gravitational potential $\Phi_{0,n}^{(j)}$. 
%%%%%
\end{enumerate}
%%%%%
Steps 2--4 are repeated until the relative change in the eigenvalue $\omega_{m,n}$ of the selected state between successive iterations falls below a prescribed tolerance, which is set to $10^{-6}$ in this work.
%%%%%==

\subsection{Numerical convergence tests}
%%%%%
\begin{figure}
  \includegraphics[width=0.45\textwidth]{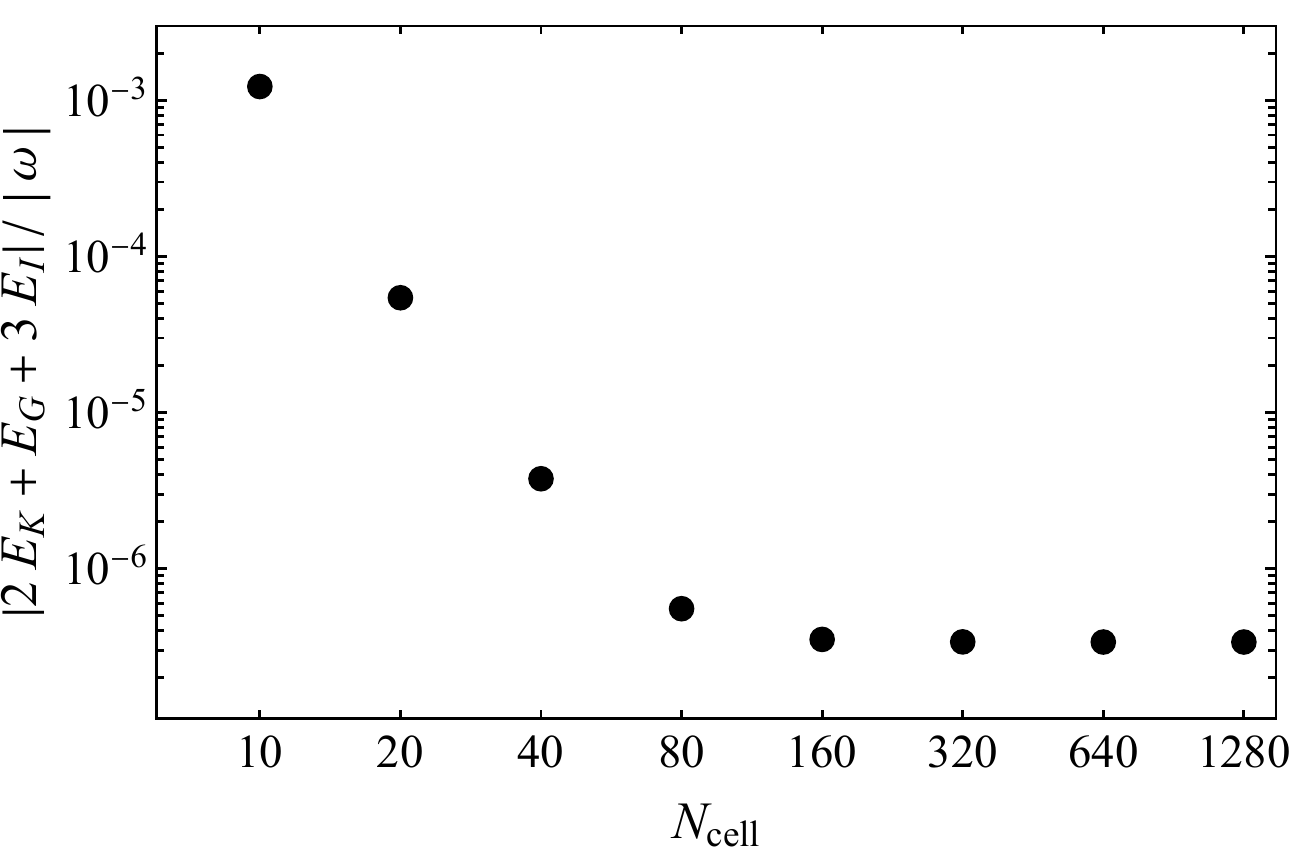}
  \\
  \includegraphics[width=0.45\textwidth]{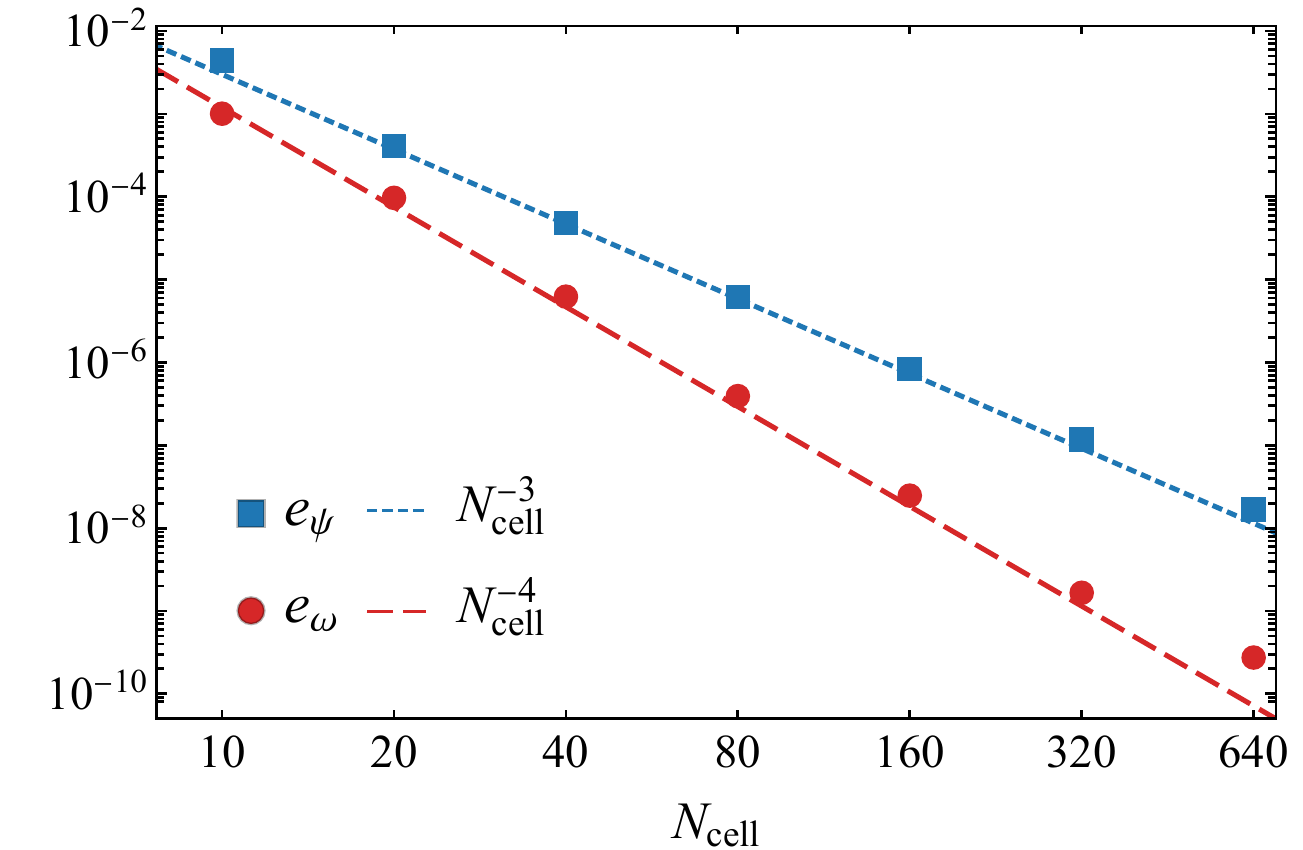}
  \caption{Numerical convergence tests for the even-parity $m=1$ solution with $\lambda=0$. The upper panel shows the normalized violation of the virial identity in Eq.~\eqref{eq:virial} as a function of the number of cells $N_{\rm cell}$. The lower panel shows the relative errors in the eigenenergy and field distribution under mesh refinement, using the finest-mesh solution ($N_{\rm cell}=1280$) as the reference. The dashed lines show the theoretical convergence rates for $P_2$ finite elements: $N_{\rm cell}^{-4}$ for the eigenenergy and $N_{\rm cell}^{-3}$ for the eigenfunction in the $L^2$ norm.}
  \label{fig:converge_static}
\end{figure}
%%%%%

%%%%%==
To verify the numerical scheme, both physical consistency and mesh-refinement convergence tests are performed. Here we use the ground state with $m=1$ and $\lambda=0$ for illustration. 
%%%%%==

%%%%%==
The virial identity \eqref{eq:virial} provides a stringent self-consistency check for the stationary solutions. We consider a sequence of uniformly refined discretizations with equal numbers of cells along the radial and angular directions, $N_r=N_\theta=N_{\rm cell}$, where $N_{\rm cell}$ ranges from $10$ to $1280$. For each discretization, the deviation from the virial relation, normalized by the eigenvalue magnitude $|\omega|$, is used as a dimensionless error measure. This measure is shown in the upper panel of Fig.~\ref{fig:converge_static}. The error decreases as $N_{\rm cell}$ increases, reaching approximately $4\times10^{-7}$ for $N_{\rm cell}\ge 160$. The relation in Eq.~\eqref{eq:omega_relation_general} is tested in the same manner, and its relative error remains below $10^{-12}$ throughout the refinement sequence, further confirming the excellent numerical consistency of the solutions.
%%%%%==

%%%%%==
The convergence of the eigenvalue $\omega$ and the scalar field $\psi$ is also examined through this mesh-refinement procedure. The solution obtained at the finest resolution, i.e. $N_{\rm cell}=1280$, is used as the reference, $\omega_{\rm ref}$ and $\psi_{\rm ref}$. The relative errors for the eigenvalue and the fields are defined by
%%%%%
\begin{equation}
e_\omega= \frac{|\omega-\omega_{\rm ref}|}{|\omega_{\rm ref}|}, 
\ 
e_\psi=\frac{|| \psi - \psi_{\rm ref}||_2}{||\psi_{\rm ref}||_2},
\end{equation}
%%%%%
where $||f||_2 \equiv \left(\int d^3x\; |f|^2 \right)^{1/2}$ is the standard $L^2$ norm of a function $f$. The convergence behavior is shown in the lower panel of Fig.~\ref{fig:converge_static}. As indicated by the dashed reference lines, the eigenvalue error decays as $N{\rm cell}^{-4}$, whereas the eigenfunction error decays as $N_{\rm cell}^{-3}$. These observed rates agree with the standard error estimates for $P_2$ finite element approximations of elliptic eigenvalue problems discussed in Sec.~7.3.1 of Ref.~\cite{hughes2000finite}, thereby providing an additional validation of the numerical implementation.
%%%%%==

%%%%%==
Based on these two tests, the resolution $N_{\rm cell}=320$ is adopted in the subsequent calculations for the balance between accuracy and computational efficiency. The order of typical relative numerical errors is below $10^{-6}$.
%%%%%==

\subsection{Numerical results}
%%%%%
\begin{table}
\begin{centering}
\begin{tabular}{ccc}
\toprule 
$m$ & $\omega_m$(even) & $\omega_m$(odd) \tabularnewline
\midrule
\midrule 
0 & $-0.162769$ & $-$ \tabularnewline
\midrule 
1 & $-0.057084$ & $-0.028970$\tabularnewline
\midrule 
2 & $-0.030626$ & $-0.017097$\tabularnewline
\midrule 
3 & $-0.019610$ & $-0.011611$\tabularnewline
\bottomrule
\end{tabular}
\par\end{centering}
\caption{Ground-state eigenenergies for stationary boson-star solutions with different azimuthal quantum numbers $m$ and parities in the non-interacting case $\lambda=0$.}
\label{tab:eigenvalues}
\end{table}
%%%%%
%%%%%==
For the non-interacting case ($\lambda=0$), Table~\ref{tab:eigenvalues} lists the lowest eigenvalues $\omega_m$ for both even and odd parities with $m\le 3$. As expected, the eigenvalues become less negative with increasing $m$, indicating that the centrifugal barrier associated with angular momentum weakens the binding of the configurations. Our results agree with those reported in Table~I of Ref.~\cite{Flores:2024tnu}, which were obtained using an independent partial-wave expansion. The maximum relative difference is below $2\times10^{-3}$. We also compare our results with those of Ref.~\cite{Dmitriev:2021utv}, where stationary rotating configurations were obtained by combining imaginary-time evolution with projection methods on a three-dimensional Cartesian grid. That work reports total energies rather than the eigenvalues $\omega_m$ used here. In the non-interacting case, the two quantities are related by a factor of $3$ implied by the virial relation~\eqref{eq:omega_relation}. After the conversion, the relative differences of the resulting eigenvalues are below $10^{-2}$.
%%%%%==

\begin{figure}
  \centering
  \begin{tabular}{c@{\hspace{0pt}}c}
    \includegraphics[width=0.24\textwidth]{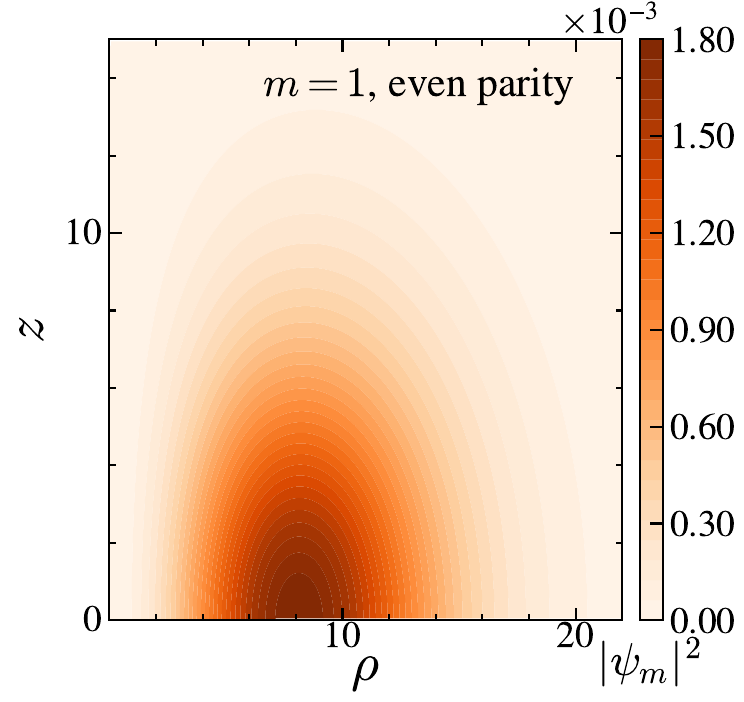} &
    \includegraphics[width=0.24\textwidth]{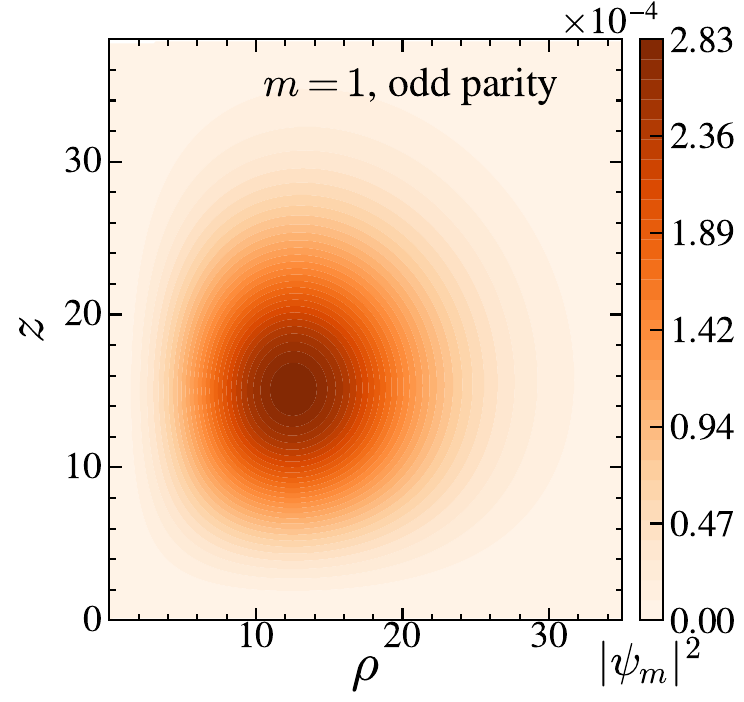} \\[0pt]  % [3pt] 增加行间距
    \includegraphics[width=0.24\textwidth]{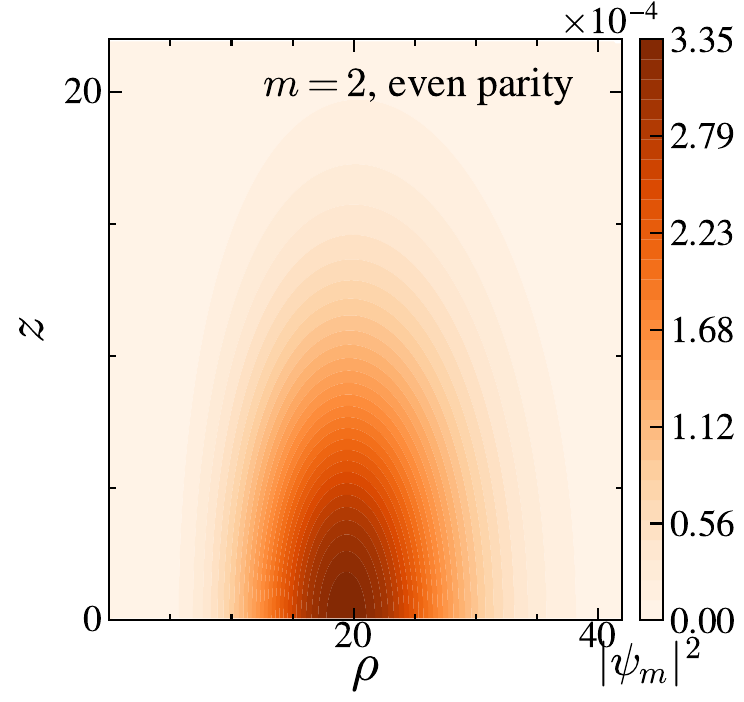} &
    \includegraphics[width=0.24\textwidth]{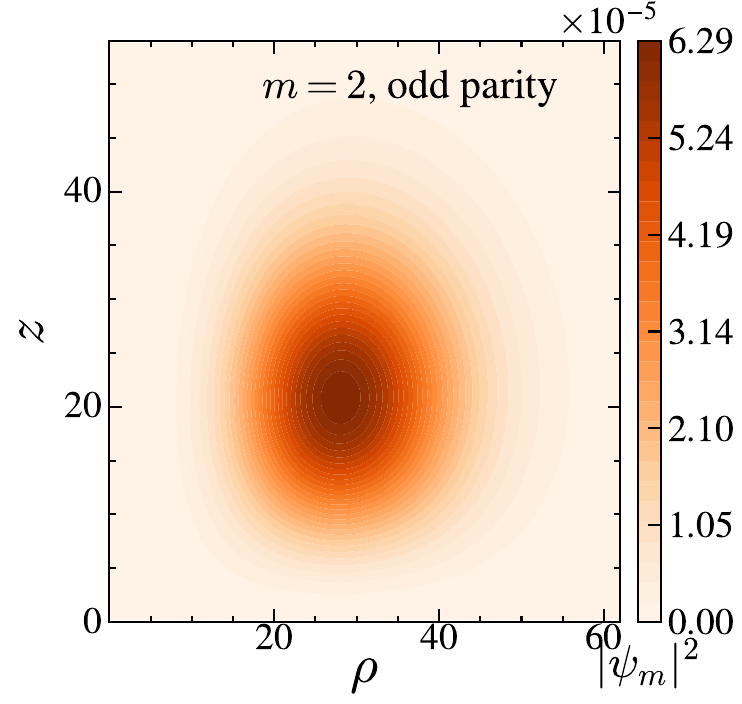} \\[0pt]
    \includegraphics[width=0.24\textwidth]{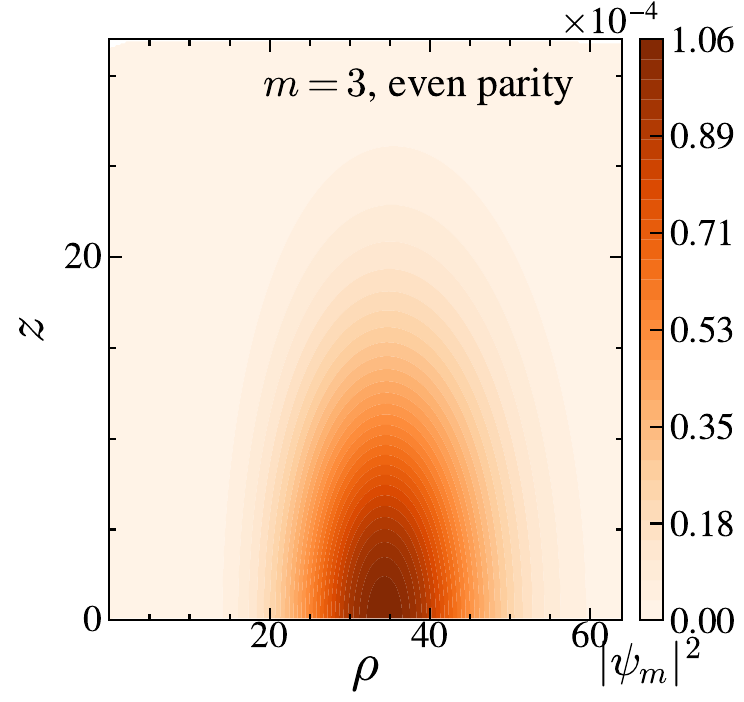} &
    \includegraphics[width=0.24\textwidth]{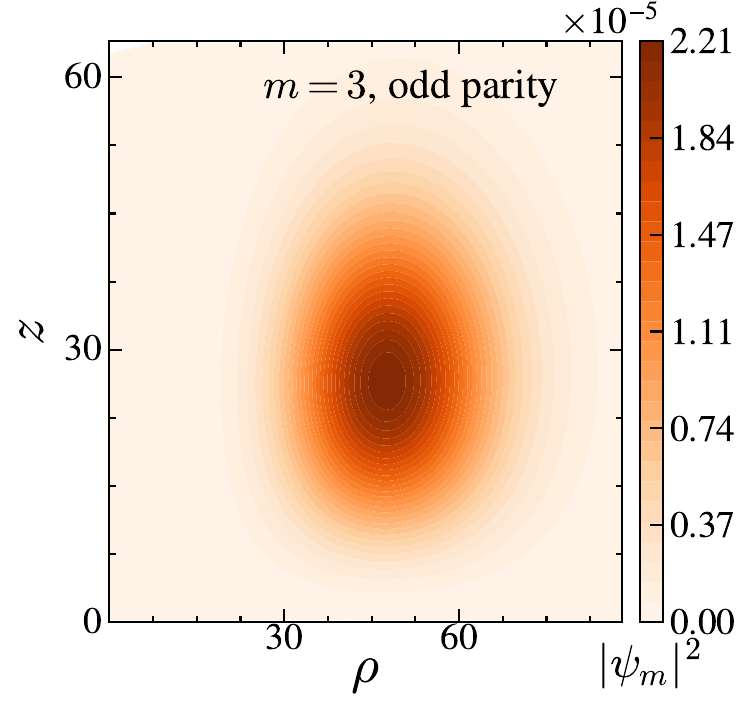}
  \end{tabular}
\caption{Density profiles $|\psi_m|^2$ in the meridional $(\rho,z)$ plane for rotating boson star solutions with $\lambda=0$. From top to bottom, the rows correspond to $m=1,2,3$. The left and right columns show the even- and odd-parity solutions, respectively. The color scale is normalized independently in each panel.}
  \label{fig:profiles_lambda0}
\end{figure}

%%%%%==
Figure~\ref{fig:profiles_lambda0} shows the density profiles $|\psi_m|^2$ of the $m=1,2,3$ solutions in the meridional $(\rho,z)$ plane, where $\rho=r\sin\theta$ and $z=r\cos\theta$. All rotating configurations exhibit toroidal density distributions. For the even-parity solutions, the density reaches its maximum on the equatorial plane at a finite cylindrical radius, corresponding to a single torus in three dimensions. For the odd-parity solutions, the equatorial nodal plane separates the density into two symmetric lobes above and below the plane, corresponding to two tori around the symmetry axis. As $m$ increases, the density maximum moves toward larger radii and the distribution becomes more spatially extended. Since the total mass is fixed by the normalization convention, this broadening of the wave function is accompanied by a decrease in the peak density.
%%%%%==

%%%%%
%%%%%

%%%%%
\begin{figure}
  \centering
  \begin{tabular}{c@{\hspace{0pt}}c}
    \includegraphics[width=0.24\textwidth]{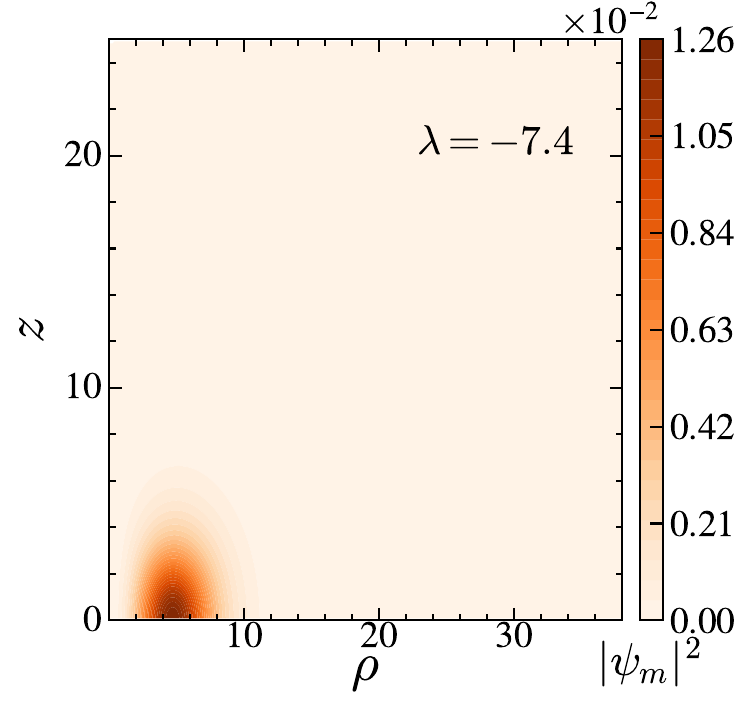} &
    \includegraphics[width=0.24\textwidth]{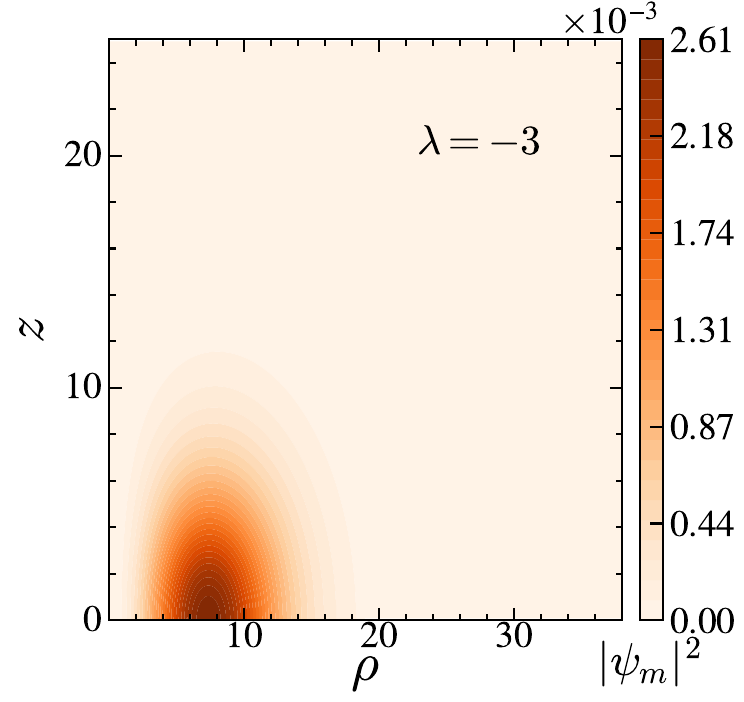} \\[0pt] 
    \includegraphics[width=0.24\textwidth]{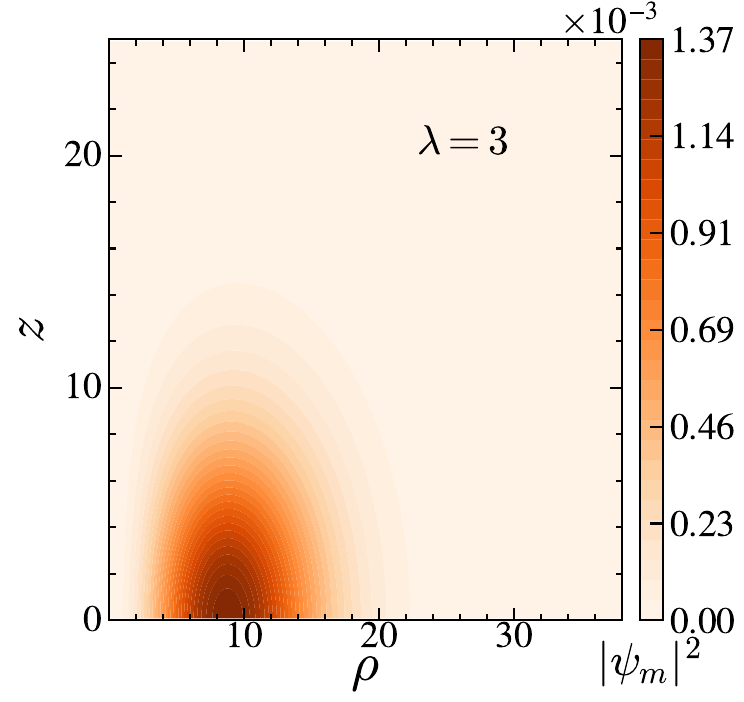} &
    \includegraphics[width=0.24\textwidth]{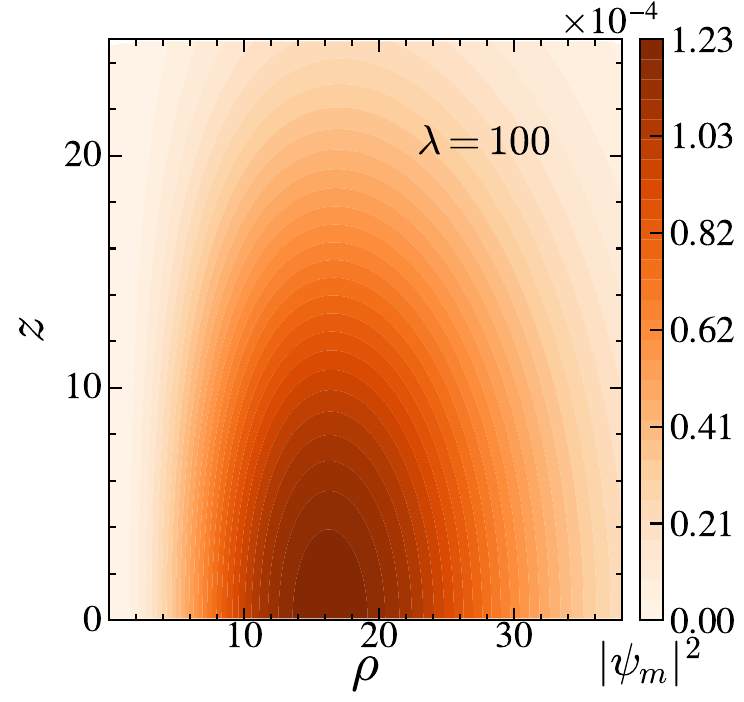} \\[0pt]
  \end{tabular}
\caption{Density profiles $|\psi_1|^2$ of the lowest-energy even-parity $m=1$ configurations in the meridional $(\rho,z)$ plane at different self-interaction strengths. From the upper left to the lower right, the panels correspond to $\lambda=-7.4,-3,3$, and $100$, respectively. The color scale is normalized independently in each panel.}
  \label{fig:profiles_changelambda}
\end{figure}
%%%%%

\begin{figure}
  \includegraphics[width=0.45\textwidth]{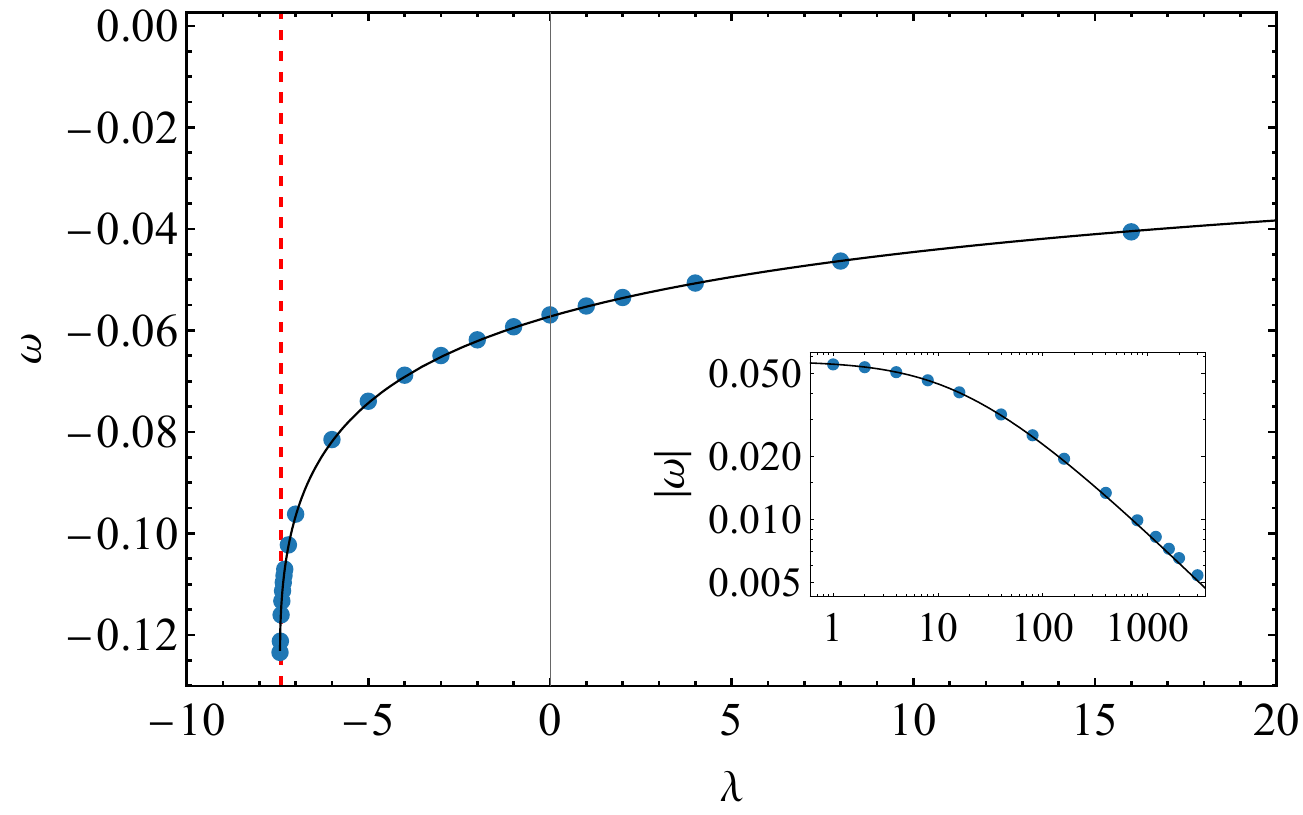}
\caption{Eigenvalue $\omega_m$ of the even-parity $m=1$ state as a function of the self-interaction strength $\lambda$. The dots are the numerical results, while the black solid curve is the empirical relation in Eq.~\eqref{eq:empirical_fit}. The vertical red dashed line marks the critical coupling $\lambda_c=-7.43$. The inset shows the strongly repulsive regime.
}
\label{fig:omega_lambda}
\end{figure}

%%%%%==
To examine the effect of self-interaction, we focus on the lowest-energy even-parity $m=1$ configuration, while the other states exhibit qualitatively similar behavior. The blue dots in Fig.~\ref{fig:omega_lambda} show the numerical dependence of the eigenvalue on $\lambda$. The eigenvalue becomes less negative as $\lambda$ increases, indicating weaker binding. The corresponding density profiles are displayed in Fig.~\ref{fig:profiles_changelambda}. Increasing the strength of the repulsive self-interaction shifts the toroidal density maximum toward larger radii and broadens the configuration, with a corresponding decrease in the peak density.
%%%%%==

\begin{figure}
  \includegraphics[width=0.45\textwidth]{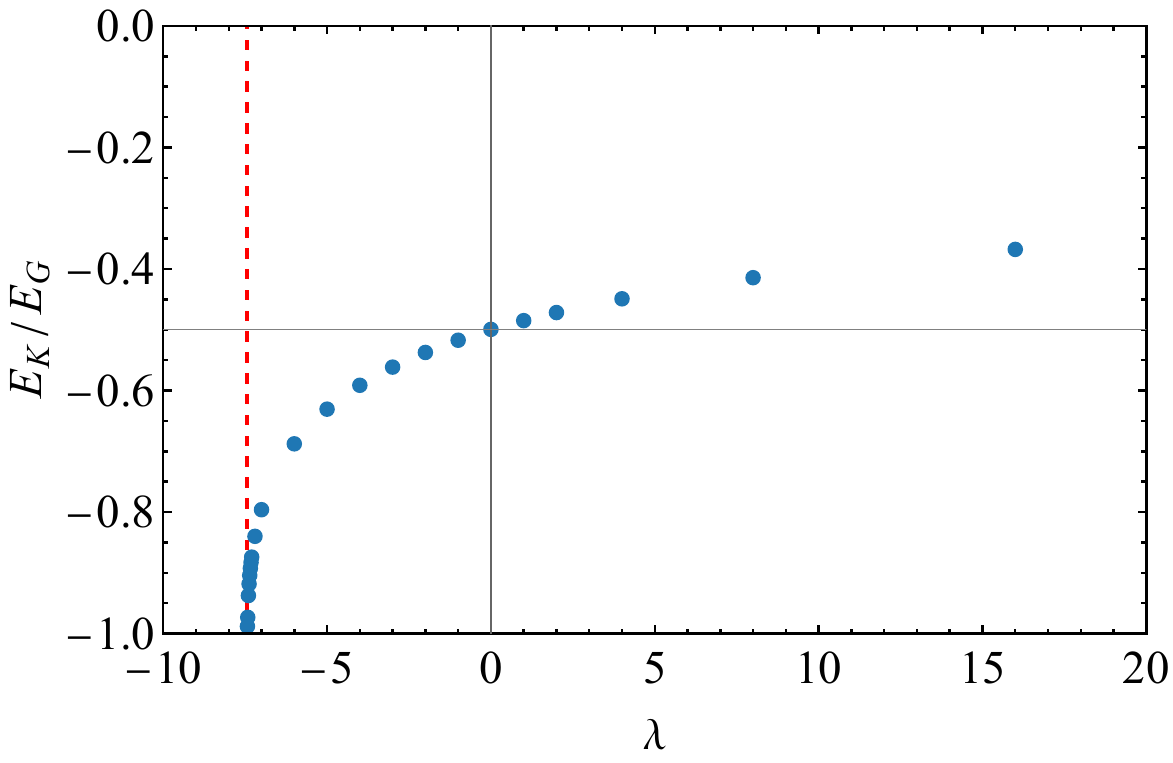}
\caption{Ratio $E_K/E_G$ as a function of the self-interaction strength $\lambda$ for the even-parity $m=1$ state. The two gray solid lines show $E_K/E_G=-1/2$ at $\lambda=0$, while the red dashed line marks the critical coupling $\lambda_c=-7.43$.}
  \label{fig:EkEp_lambda}
\end{figure}

%%%%%==
On the attractive side, the numerical stationary branch approaches a critical coupling $\lambda_c=-7.43$, below which no converged stationary solution is obtained. Physically, this marks the point at which the attractive self-interaction becomes too strong to be balanced by the gradient contribution. This interpretation is quantified by the scaling perturbation. According to Eq.~\eqref{eq:variation_second2}, positivity of the second variation requires $E_K/E_G>-1$, while Fig.~\ref{fig:EkEp_lambda} shows that $E_K/E_G$ approaches $-1$ as $\lambda\to\lambda_c$, indicating marginal stability against this scaling perturbation near the endpoint of the numerical branch. For comparison, after accounting for the factor of $32\pi$ between the normalization conventions, the central value of the critical coupling reported in Ref.~\cite{Dmitriev:2021utv} differs from the present result with relative error $1.1\times10^{-2}$.

%%%%%==
Over the entire stationary branch, the dependence of the eigenvalue on $\lambda$ can be approximately represented by an empirical relation
\begin{equation}\label{eq:empirical_fit}
\omega(\lambda)
=
\frac{\omega_c}
{1+\sqrt{(\lambda-\lambda_c)/\lambda_*}},
\end{equation}
shown by the black solid curve in Fig.~\ref{fig:omega_lambda}. Here $\lambda_c=-7.43$ and $\omega_c=-0.123$ are fixed by the critical endpoint, while a one-parameter fit gives $\lambda_*=5.56$. The maximum relative deviation is about $6\times 10^{-2}$. In the strongly repulsive limit, this expression reduces to $\omega=-0.29 \lambda^{-1/2}$, consistent with the Thomas--Fermi scaling $\omega_m\propto\lambda^{-1/2}$ found in the spherical case \cite{Chavanis:2011zi}.
%%%%%==

%%%%%
%%%%%

%%%%%
%%%%%

\section{Time evolution}
\label{sec:time_evolution}

%%%%%==
The stability analysis based on the energy functional in Sec.~\ref{subsec:energy-relations} is restricted to a special class of scaling perturbations. Rotating boson stars, however, are known to develop nonaxisymmetric dynamical instabilities during nonlinear evolution~\cite{Siemonsen:2020hcg,Dmitriev:2021utv,DiGiovanni:2020ror}. In this section, we investigate their dynamical behavior beyond the variational analysis by focusing on the even-parity $m=1$ solutions, which have the largest binding energy among the configurations with $m\neq0$ and may therefore be of the greatest phenomenological importance. We numerically evolve these solutions in three spatial dimensions for various values of the self-interaction strength $\lambda$.
%%%%%==

\subsection{Numerical method}

%%%%%==
The evolution is carried out on a cubic lattice of side length $L$ with periodic boundary conditions imposed on both $\psi$ and $\Phi$. Periodicity requires the source term in the Poisson equation to have vanishing spatial average. After subtracting the mean density $N/L^3$, the evolution equations become~\cite{Levkov:2018kau}
%%%%%
\begin{subequations}\label{eq:GPP_periodic}
\begin{align}
    i\partial_t \psi &= -\frac{1}{2} \nabla^2\psi + \Phi\psi + \lambda |\psi|^2 \psi,
    \label{eq:sch_periodic}
    \\
    \nabla^2\Phi &= |\psi|^2 - N/L^3.
    \label{eq:poi_periodic}
\end{align}
\end{subequations}
%%%%%
Since the box size is chosen to be much larger than the characteristic size of the condensate, the induced modification of the gravitational potential is approximately uniform within the localized core region. This results in a constant shift of the eigenfrequency $\omega_m$ and does not affect other physical observables. The stationary solutions obtained in Sec.~\ref{sec:stationary} are used as the initial configurations. The profile $\psi_m(r,\theta) e^{im\varphi}$ is interpolated onto the Cartesian grid. This procedure preserves the vortex structure of the rotating stationary state while allowing three-dimensional nonaxisymmetric perturbations to develop during the subsequent evolution.
%%%%%==

\begin{figure}
\centering
\includegraphics[width=0.45\textwidth]{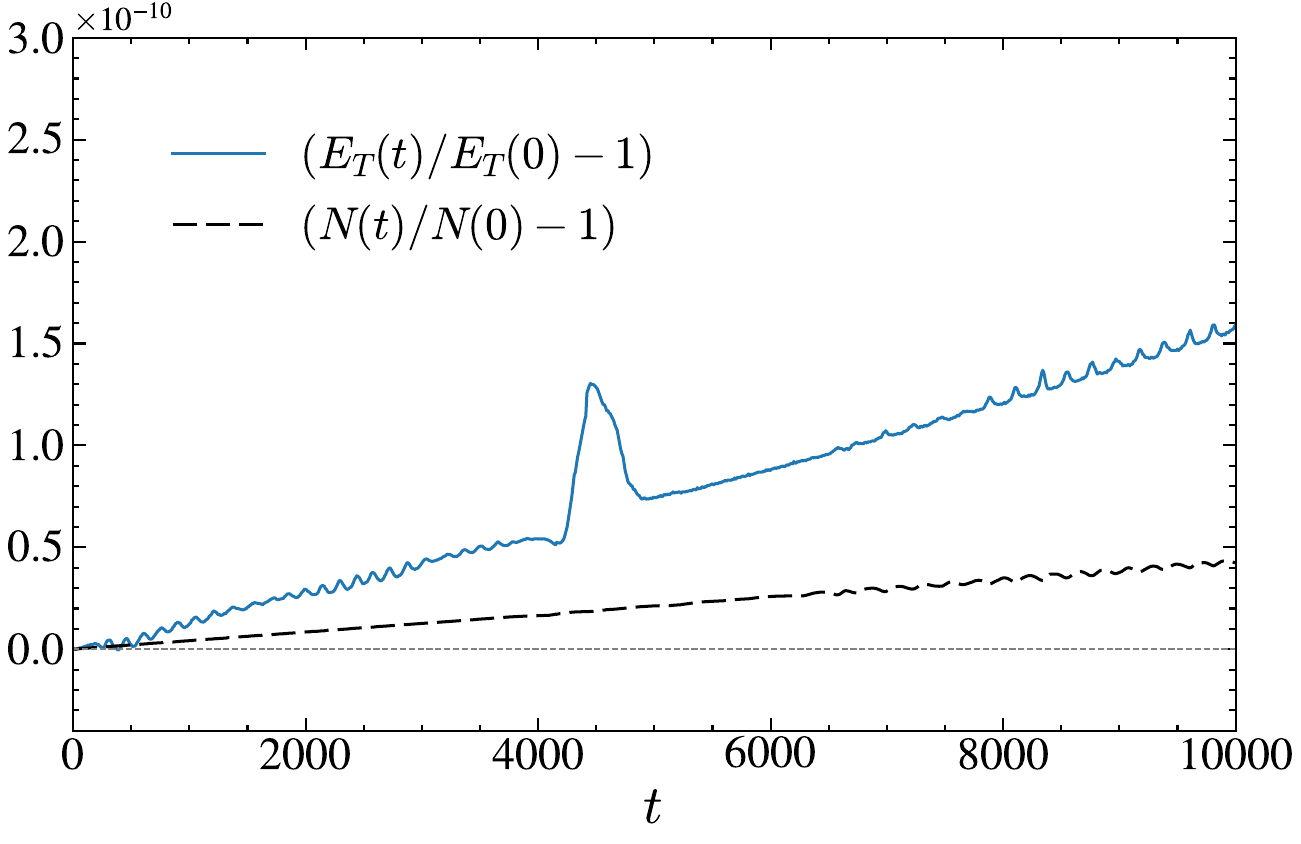}
\caption{Relative deviations from conservation of the total energy and particle number during the numerical time evolution of the $\lambda=0$ rotating solution.}
\label{fig:conservation}
\end{figure}

%%%%%==
The system described by Eqs.~\eqref{eq:GPP_periodic} is evolved using a sixth-order pseudo-spectral operator-splitting scheme~\cite{Yoshida:1990zz,Levkov:2018kau}. The simulations are performed on a uniform Cartesian grid with $256$ points in each spatial direction, a cubic box size $L=255$, and a time step $\Delta t=0.5$. An evolution time of order $10^4$ is sufficient to capture the onset and nonlinear development of the instability. The magnitude of the initial wave function is of order $10^{-16}$ at the numerical boundary, rendering the effects of periodic boundary conditions negligible. For the representative non-interacting case $\lambda=0$, the accumulated relative variations of the total energy and particle number remain below $2\times10^{-10}$ throughout the evolution, as shown in Fig.~\ref{fig:conservation}. The energy error grows approximately linearly, with an abrupt change when the wave-function profile undergoes substantial rearrangement at $t\sim 4000$. This change provides a numerical signature of the onset of the non-axisymmetric instability. By contrast, under the same numerical setup, the spherically symmetric ground state, which is known to be dynamically stable, remains unchanged up to $t=4\times10^4$, and the corresponding conservation-error curves retain nearly constant slopes.
%%%%%==

%%%%%
%%%%%

\subsection{Simulation results}

%%%%%
\begin{figure}
\centering
\subfigure[]{
\includegraphics[width=0.48\textwidth]{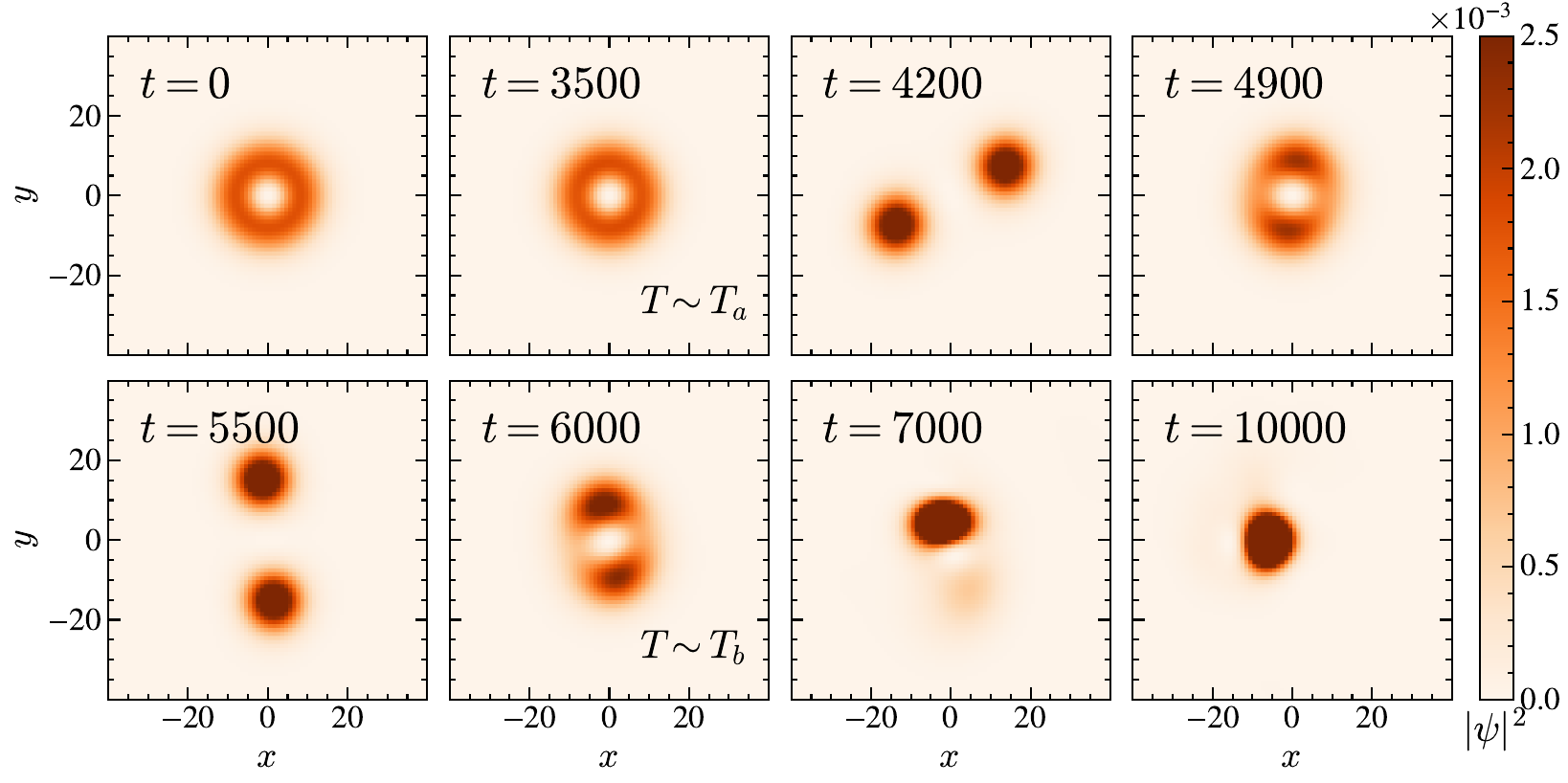}}
\vspace{0pt}
\subfigure[]{
\includegraphics[width=0.48\textwidth]{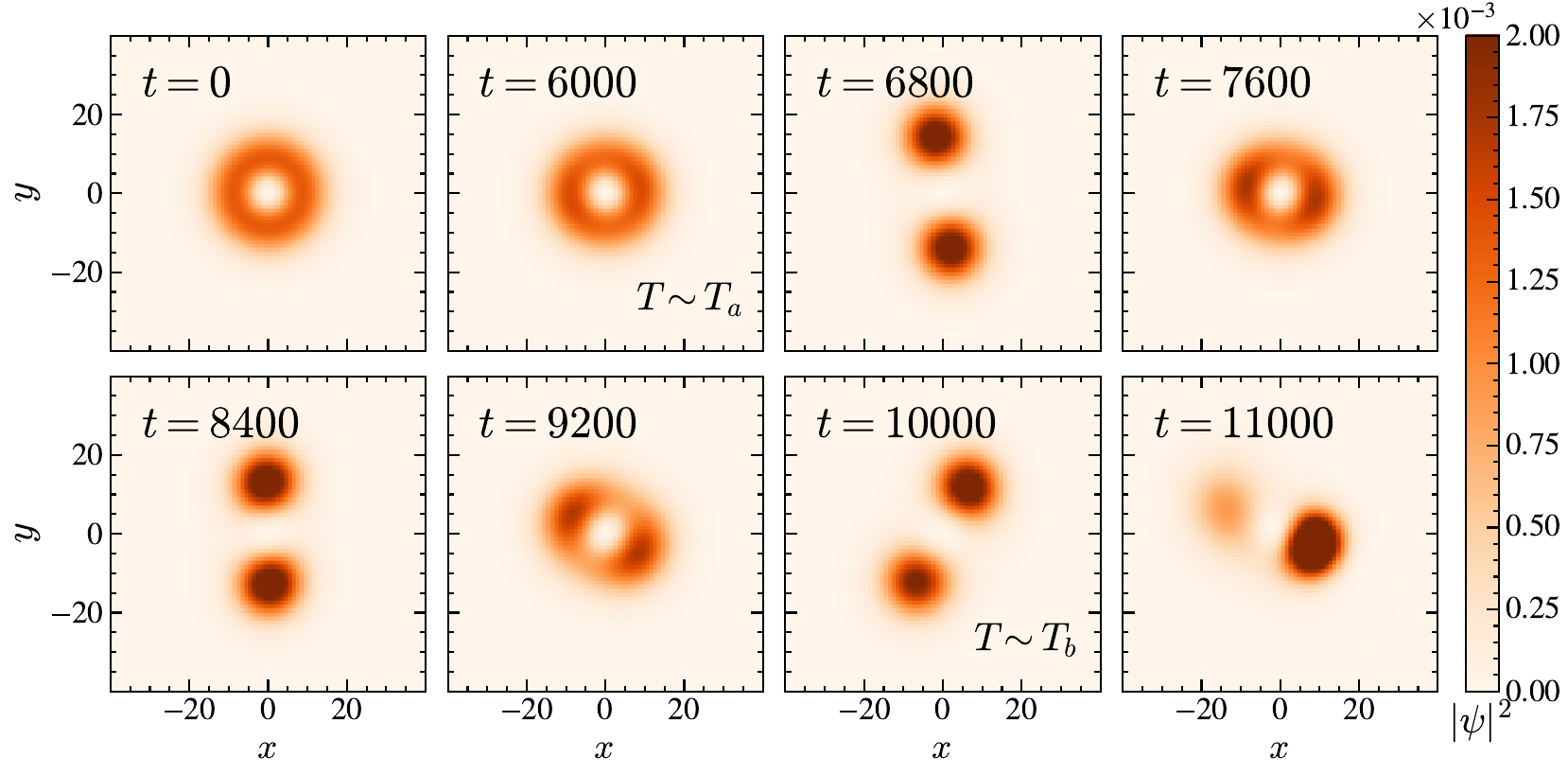}}
\vspace{0pt}
\subfigure[]{
\includegraphics[width=0.48\textwidth]{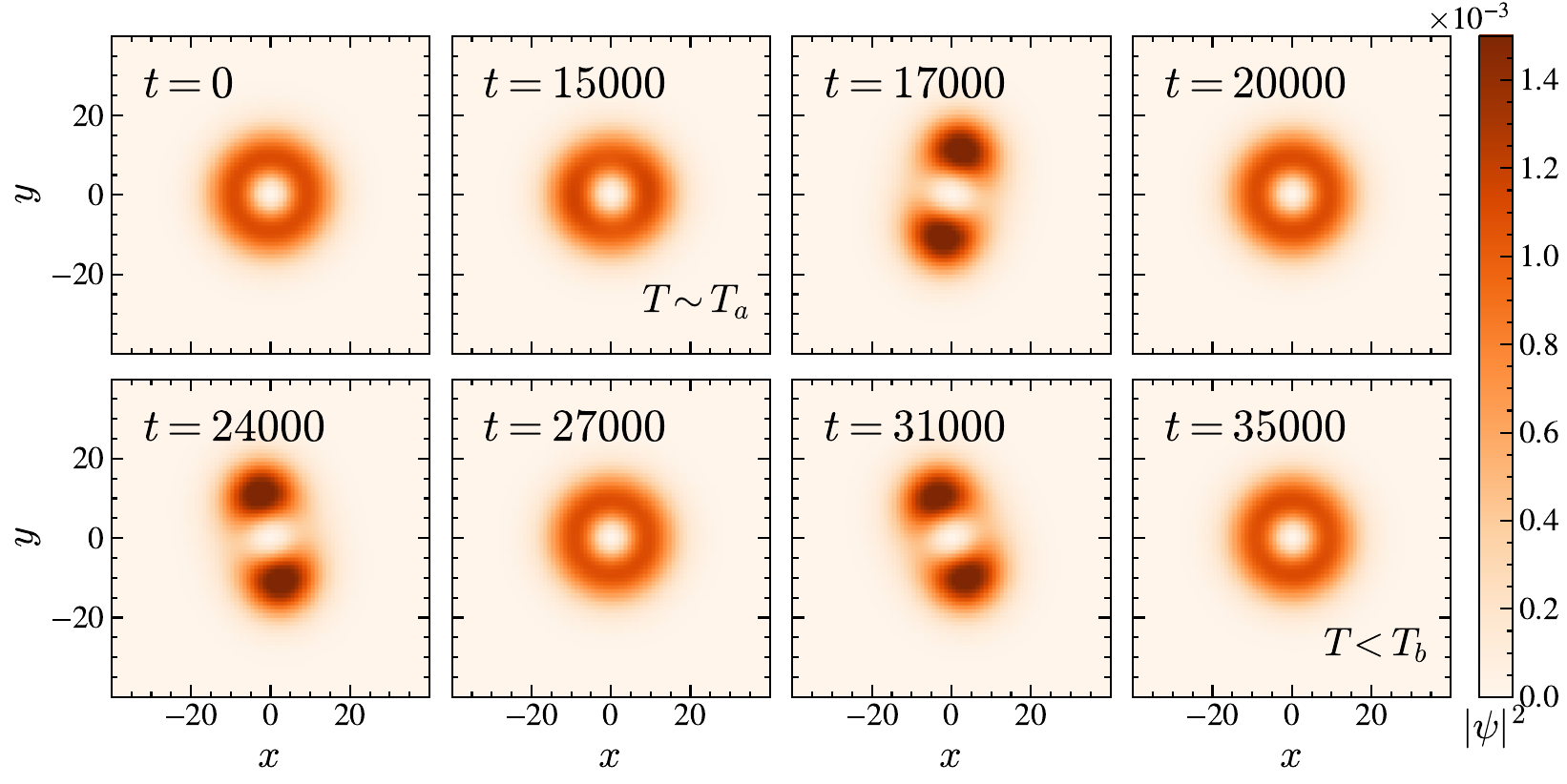}}
\caption{Equatorial density snapshots of $|\psi|^2$ at some representative evolution times for the even-parity $m=1$ solutions with (a) $\lambda=0$, (b) $\lambda=3$, and (c) $\lambda=6$. }
\label{fig:evolution_snapshots}
\end{figure}
%%%%%

%%%%%==
Figure~\ref{fig:evolution_snapshots} illustrates the evolution of $|\psi|^2$ in the equatorial plane for several values of the self-interaction strength $\lambda$. At early times, the rotating toroidal configuration remains nearly stationary for a characteristic timescale $T_a$. Non-axisymmetric deformations then develop, and the density distribution undergoes quasiperiodic transitions between approximately ring-like and twin-star-like configurations. After a longer timescale $T_b$, a significant portion of the condensate is radiated away, and the system eventually relaxes to a single localized clump.
%%%%%==

%%%%%==
Comparing the cases $\lambda=0$, $3$, and $6$ shown in Fig.~\ref{fig:evolution_snapshots}, both $T_a$ and $T_b$ increase substantially with the repulsive self-interaction strength. Further simulations at larger $\lambda$ show that, for $\lambda\ge6.7$, the configurations remain close to their initial axisymmetric states throughout the simulated interval $t\le4\times10^4$, with no visible non-axisymmetric deformation. The repulsive self-interaction therefore significantly suppresses the dynamical instability of the rotating configurations. This behavior has also been observed in previous studies of both relativistic and nonrelativistic rotating boson stars~\cite{Siemonsen:2020hcg,Dmitriev:2021utv}.
%%%%%==

\subsection{Extraction of instability growth rates}

%%%%%==
To quantify the characteristic oscillation and instability timescales, the oscillation frequency and exponential growth rate are extracted directly from the time evolution. In the early stage when the perturbation is small, the time-dependent field can be written schematically as a stationary background plus a small perturbation,
%%%%%
\begin{equation}
\psi(t,\bm{x})
\simeq
e^{-i\omega_m t}
\left[
\psi_m(r,\theta)e^{im\varphi}
+
\delta\psi(\bm{x})
e^{-i\mathrm{Re}(\epsilon)t}
e^{\mathrm{Im}(\epsilon)t}
\right],
\end{equation}
%%%%%
where $\mathrm{Re}(\epsilon)$ gives the oscillation frequency relative to the stationary background, while a positive $\mathrm{Im}(\epsilon)$ signals an unstable mode and determines its exponential growth rate.
%%%%%==

%%%%%
\begin{figure}
\centering
\includegraphics[width=0.45\textwidth]{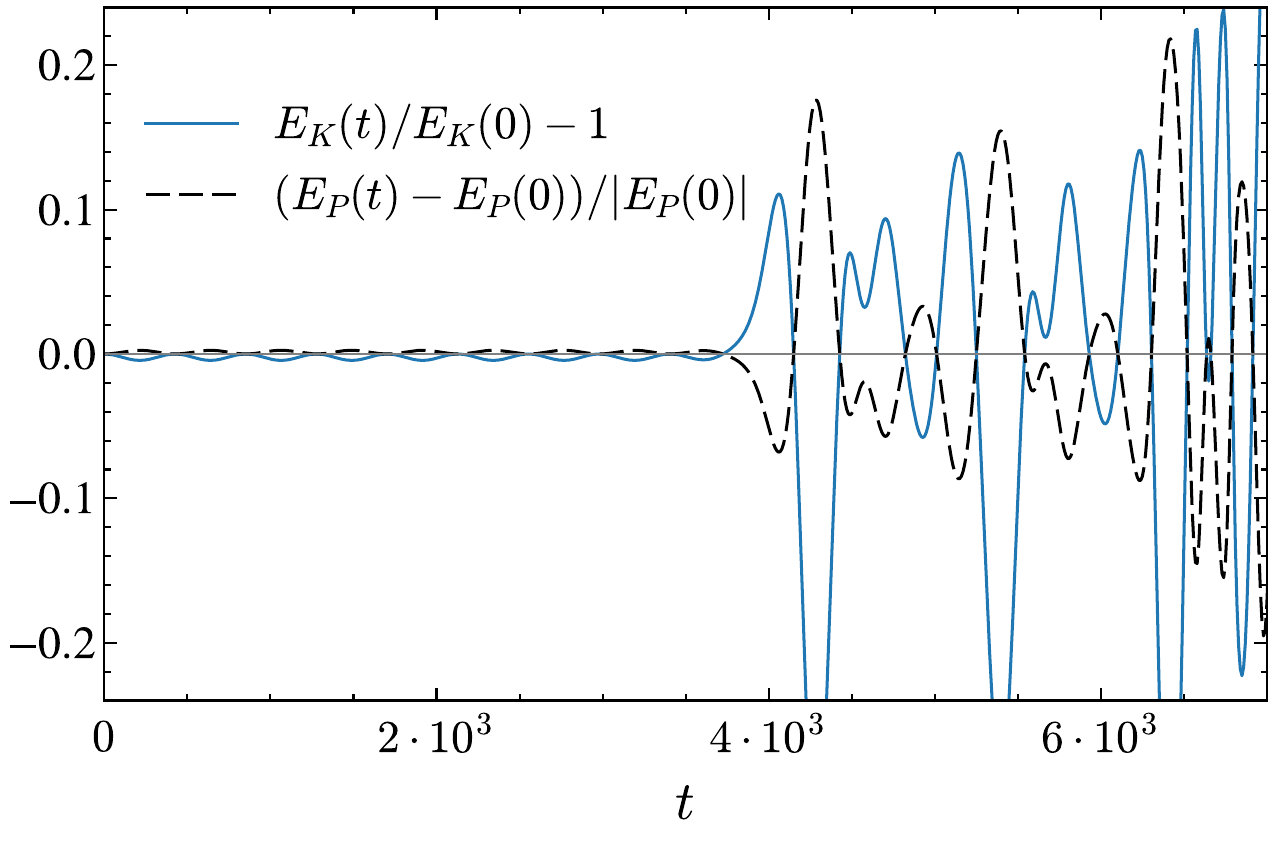}
\caption{Relative variations of the kinetic and potential energies for the $m=1$, $\lambda=0$ solution. The two energy components oscillate in opposite directions while the total energy remains conserved (see Fig.~\ref{fig:conservation}).}
\label{fig:relativeEkEp}
\end{figure}
%%%%%

%%%%%==
Although the total energy remains conserved within numerical error as shown in Fig.~\ref{fig:conservation}, energy is continuously exchanged between its kinetic and potential contributions during the evolution. Figure~\ref{fig:relativeEkEp} shows the relative variations for the $m=1$, $\lambda=0$ solution. In the early stage, the interference between the fixed stationary background and the perturbation produces oscillations of the energy components at the frequency $\mathrm{Re}(\epsilon)$. Up to $t=3500$, the relative exchange between the kinetic and potential energies remains of order $10^{-3}$. A discrete Fourier transform of the signal over this interval yields $\mathrm{Re}(\epsilon)=(1.44\pm0.18)\times10^{-2}$, where the quoted uncertainty is the Fourier frequency resolution set by the finite length of the time interval. At later times, the relative variations increase rapidly to order $10^{-1}$ as the unstable modes become sufficiently large to modify the stationary background and drive the evolution into the nonlinear stage. This abrupt exchange marks the transition from the early linear stage to the subsequent nonlinear evolution, which is also reflected in the sudden change of the relative deviation for the total energy shown in Fig.~\ref{fig:conservation}.
%%%%%==

%%%%%
\begin{figure}
\centering
\includegraphics[width=0.45\textwidth]{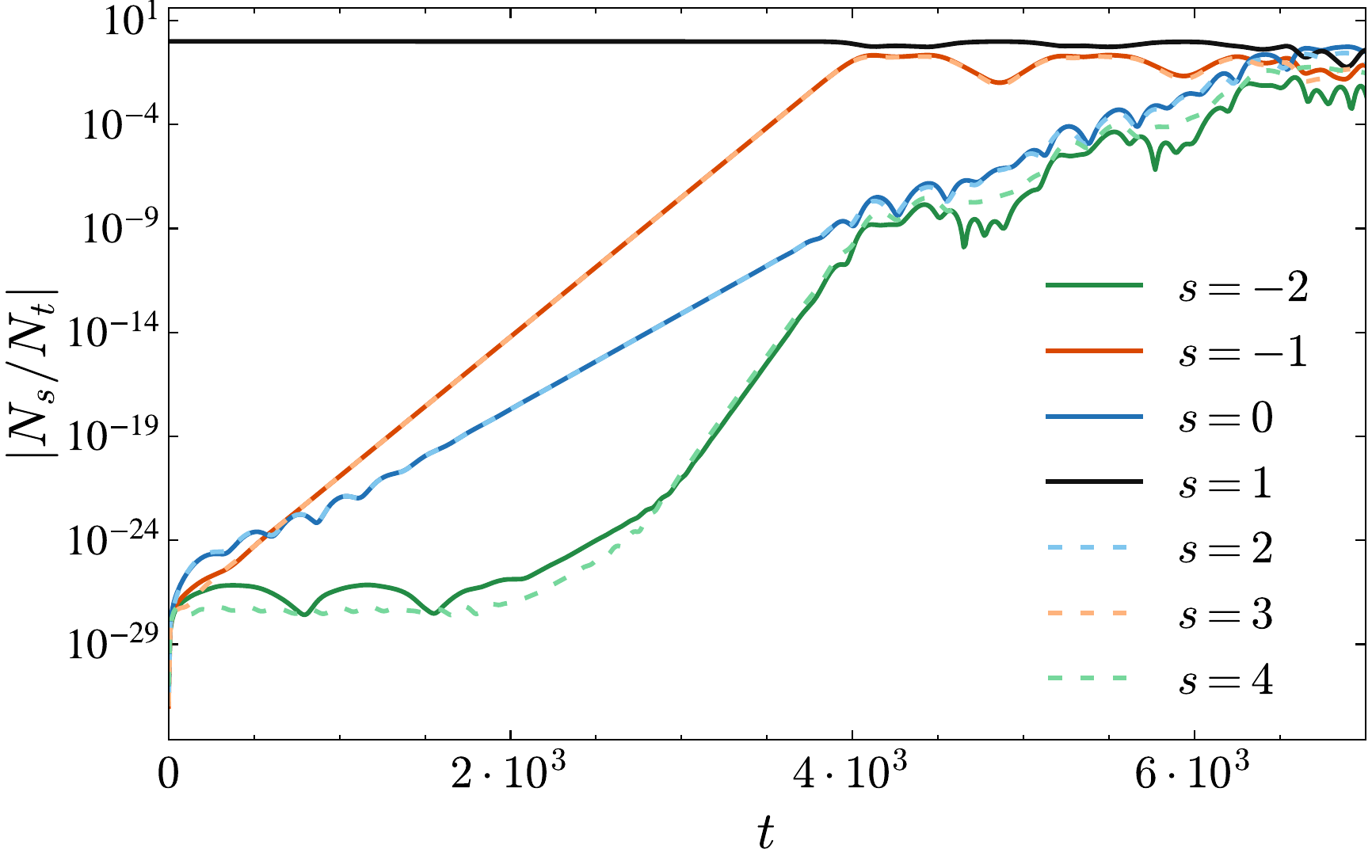}
\caption{Time evolution of the particle number fractions $N_s/N$ in different azimuthal sectors for the $m=1$, $\lambda=0$ solution. }
\label{fig:Ns_t}
\end{figure}
%%%%%

A complementary diagnostic is obtained by decomposing the evolving wave function into azimuthal Fourier components at each time,
%%%%%
\begin{equation}\label{eq:azimuthal_decomposition}
\psi(t,\bm x)=
\sum_s
\psi_s(t,\rho,z)e^{is\varphi}.
\end{equation}
%%%%%
Here, $\psi_s(t,\rho,z)$ denotes the instantaneous azimuthal Fourier component of the evolving field, defined by $\psi_s=(2\pi)^{-1}\int_0^{2\pi} d\varphi\,\psi e^{-is\varphi}$. The particle number contained in each azimuthal sector is defined as $N_s(t)=\int d^3x\,|\psi_s|^2$. The ratio $N_s/N$ measures the fraction of the total particle number carried by the $s$-th azimuthal component. Figure~\ref{fig:Ns_t} shows the evolution of $N_s/N$ for several azimuthal sectors \footnote{The interpolation onto the cubic Cartesian grid preserves only the discrete fourfold rotational symmetry of the grid, leading to numerical leakage from \(s=1\) into \(s=1\pm4j\), with \(j=1,2,\ldots\). At a resolution of $256^3$, these spurious normalized populations remain at approximately $10^{-10}$ during the linear stage and decrease with grid refinement. They are negligible compared with the physically growing sidebands and do not affect the extracted instability rates.}. The small initial populations in the non-background sectors are seeded by discretization errors and subsequently grow exponentially due to the instability. Before late-time radiation becomes appreciable, conservation of the total angular momentum, together with the pairwise excitation of sectors symmetric about the background $s=1$ component, keeps the particle numbers in the pairs $(s=-1,s=3)$ and $(s=0,s=2)$ nearly equal. The nearly overlapping members of each pair are distinguished in Fig.~\ref{fig:Ns_t} by solid curves for $s<1$ and dashed curves for $s>1$. Besides the background $s=1$ component, several additional azimuthal sectors grow exponentially in the early stage. The fastest growth is observed in the $(s=-1,s=3)$ sectors, while weaker growth is also visible in the pair $(s=0,2)$. Since each $N_s$ depends quadratically on the wave-function amplitude, its fitted exponential growth rates are twice the corresponding values of $\mathrm{Im}(\epsilon)$. This gives $\mathrm{Im}(\epsilon)=7.73\times10^{-3}$ for the dominant $s=-1,3$ sectors and $\mathrm{Im}(\epsilon)=5.16\times10^{-3}$ for the subdominant $s=0,2$ sectors.
%%%%%==

%%%%%
\begin{figure}
\centering
\includegraphics[width=0.45\textwidth]{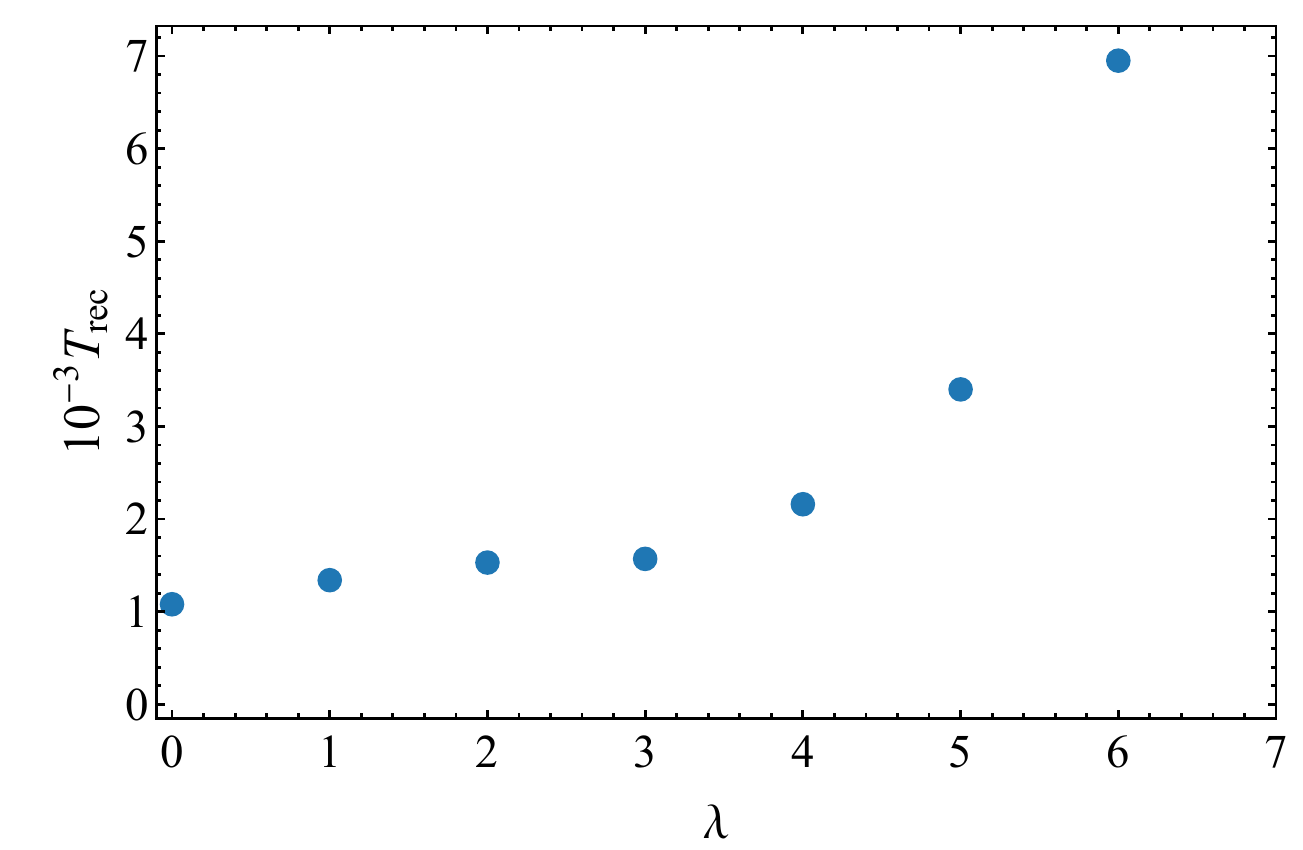}
\caption{
Dependence of the quasiperiodic conversion timescale $T_{\rm conv}$ between ring-like and twin-star-like configurations on the self-interaction strength $\lambda$. The timescale is defined as the interval between consecutive minima in the normalized particle number of the $s=-1$ azimuthal sector.
}
\label{fig:Trec}
\end{figure}
%%%%%

%%%%%==
The rapid growth of the $s=-1$ and $s=3$ sectors provides an intuitive explanation for the early non-axisymmetric deformation seen in Fig.~\ref{fig:evolution_snapshots}. The coherent superposition of these sectors with the background $s=1$ component produces an azimuthal density modulation with two maxima, driving the initially ring-like configuration toward a twin-star-like structure, as pointed out in Ref.~\cite{Dmitriev:2021utv}. Around $t\sim 4000$, the growth of the dominant unstable sectors saturates, and a quasiperiodic exchange of particle number begins between the background mode and the $s=-1$ and $s=3$ sectors. This exchange explains the quasiperiodic conversion between ring-like and twin-star-like configurations before additional azimuthal sectors become significant at later times. The timescale of this quasiperiodic conversion can be quantified from the exchange between the dominant azimuthal sectors. We define the conversion timescale $T_{\rm conv}$ as the interval between two consecutive minima of the normalized particle number in the $s=-1$ sector during the quasiperiodic stage. Figure~\ref{fig:Trec} shows the dependence of $T_{\rm conv}$ on the self-interaction strength $\lambda$. The conversion timescale increases with increasing repulsive self-interaction, indicating that the particle exchange between the background mode and the dominant azimuthal sectors becomes progressively slower.
%%%%%==

%%%%%==
In the next two sections, we will show that the early stage, during which the unstable modes grow exponentially, can be well described by the linearized equations. We thus refer to this interval as the \textit{linear stage}. After the early linear stage, the feedback of the unstable modes on the background becomes significant, marking the onset of what we refer to as the \textit{nonlinear stage}.
%%%%%==

\section{Linear stability analysis}
\label{sec:linear}

%%%%%==
During the early stage of the evolution, perturbations remain sufficiently small, and the stationary background configuration is effectively unchanged. The dynamics can therefore be described within the framework of linear perturbation theory. We employ the Bogoliubov--de Gennes (BdG) formalism \cite{PitaevskiiStringari2016,Castin2001,darksoliton}. Linearizing the Gross--Pitaevskii equation around a stationary solution yields an eigenvalue problem for the perturbation modes. The presence of an eigenvalue with a positive imaginary part indicates a dynamical instability that grows exponentially with time.
%%%%%==

%%%%%==
For condensates confined by external potentials such as harmonic traps, various analytical and numerical methods have been developed for solving the BdG equations \cite{GAO2020109058,XIE2026114813,SADAKA2024108948}. In the boson-star context, the fixed external potential is replaced by the nonlocal self-gravitating potential determined self-consistently through Eq.~\eqref{eq:green_func}. For spherically symmetric nonrelativistic boson stars, the BdG spectrum of radial perturbation modes was computed using spectral methods in Ref.~\cite{Harrison2003}. This analysis was subsequently extended to include self-interactions and nonspherical perturbations through a spherical-harmonic decomposition in Ref.~\cite{Nambo:2024gvs}. For rotating boson stars, the authors of Ref.~\cite{Dmitriev:2021utv} evolved the time-dependent BdG equations and extracted instability growth rates from the exponential amplification of perturbations.
%%%%%==

%%%%%==
In this section, we derive the BdG equations for general rotating boson stars and analyze the mathematical structure of the eigenvalue spectrum. Instead of evolving the time-dependent BdG equations, we formulate the system as an eigenvalue problem and solve it directly using the same FEM framework. The numerical calculations focus on the even-parity $m=1$ configuration. The resulting growth rates are compared with those extracted from the independent time-evolution simulations in Sec.~\ref{sec:time_evolution}.
%%%%%==

\subsection{BdG equations}

%%%%%==
Infinitesimal time-dependent perturbations $\delta\psi(t,\bm{x})$ and $\delta\Phi(t,\bm{x})$ are introduced around the stationary solution,
%%%%%
\begin{align}
\psi(t,\bm x) &= e^{-i\omega_m t}\left[\psi_m(\bm x)e^{im\varphi} + \delta\psi(t,\bm x)\right], \\
\Phi(t,\bm x) &= \Phi_0(\bm x) + \delta\Phi(t,\bm x),
\end{align}
%%%%%
where $\psi_m(r,\theta)$ and $\Phi_0(r,\theta)$ denote the stationary background fields, and $\omega_m$ is the corresponding real eigenfrequency. Substituting these expressions into the time-dependent GPP equations and retaining only terms linear in the perturbations yields
%%%%%
\begin{align}
i\partial_t \delta\psi &= -\frac{1}{2}\nabla^2 \delta\psi + \left(\Phi_0 - \omega_m + 2\lambda \psi_m^2\right)\delta\psi \notag
\\
&+ \lambda \psi_m^2 e^{2im\varphi}\delta\psi^*
+ \psi_m e^{im\varphi}\delta\Phi, \\
\nabla^2 \delta\Phi &= \psi_m \left( e^{im\varphi}\delta\psi^* + e^{-im\varphi}\delta\psi \right).
\end{align}
%%%%%
%%%%%
The definite azimuthal dependence of the background state suggests decomposing the perturbations into sectors with definite angular momentum. Thereby, the perturbations are expanded as
%%%%%
\begin{align}
\delta\psi &= \sum_k \biggl[ A_{m,k}(r,\theta)e^{i(m+k)\varphi}e^{-i\epsilon_k t} \notag
\\
&\qquad + A_{m,-k}^*(r,\theta)e^{i(m-k)\varphi}e^{i\epsilon^*_k t} \biggr], \label{eq:psi_perturbation}
\\
\delta\Phi &= \sum_k \left[ C_k(r,\theta)e^{ik\varphi}e^{-i\epsilon_k t}
+ C_k^*(r,\theta)e^{-ik\varphi}e^{i\epsilon^*_k t} \right],
\end{align}
%%%%%
where $\epsilon_k$ is a generally complex eigenvalue of the BdG problem. The integer $k$ labels the azimuthal perturbation sector, while $A_{m,\pm k}(r,\theta)$ and $C_k(r,\theta)$ are generally complex amplitudes depending only on $r$ and $\theta$. Substitution into the linearized equations shows that, for each fixed perturbation sector $k$, the components with azimuthal numbers $m+k$ and $m-k$ are coupled, leading to the BdG equations
%%%%%
\begin{subequations}\label{eq:BdG_system}
\begin{align}
\epsilon_k A_{m,k} &= L_{m+k}A_{m,k} + \lambda \psi_m^2 A_{m,-k} + \psi_m C_k, \label{eq:psi_m+k}
\\
\epsilon_k A_{m,-k} &= -L_{m-k}A_{m,-k} - \lambda \psi_m^2 A_{m,k} - \psi_m C_k, \label{eq:psi_m-k}
\\
\nabla_k^2 C_k &= \psi_m(A_{m,k} + A_{m,-k}), \label{eq:C_k}
\end{align}
\end{subequations}
%%%%%
where
%%%%%
\begin{equation}
  L_s \equiv -\frac{1}{2}\nabla_s^2 + \Phi_0 + 2\lambda \psi_m^2 - \omega_m,
\end{equation}
and $\nabla_s^2$ denotes $\nabla^2-s^2/(r^2\sin^2\theta)$, as before.
%%%%%==

%%%%%==
Equations.~\eqref{eq:BdG_system} describe a linear eigenvalue problem once the stationary background $(\psi_m,\Phi_0)$ is fixed. The BdG equations contain $\Phi_0$ and $\omega_m$ only through the difference $\Phi_0-\omega_m$. Since the periodic boundary conditions used in numerical simulations shift both quantities by the same amount, the difference stays the same, such that the BdG perturbation frequencies can be compared directly with those extracted from the time-evolution simulations.
%%%%%==

%%%%%==
The spectrum of this eigenvalue problem exhibits a characteristic symmetry structure. Suppose that an eigenmode is labeled by $(k,\epsilon_k,A_{m,k},A_{m,-k})$. The perturbation expansion in Eq.~\eqref{eq:psi_perturbation} then implies that $(-k,-\epsilon^*_k,A_{m,-k}^*,A_{m,k}^*)$ is also an eigenmode. Using the fact that the background field $\psi_m$ is real, complex conjugation of the BdG equations further implies that $(k,\epsilon^*_k,A_{m,k}^*,A_{m,-k}^*)$ and $(-k,-\epsilon_k,A_{m,-k},A_{m,k})$ are also eigenmodes. Interestingly, this symmetry structure persists for a general complex background as a consequence of the block structure of the BdG operator~\cite{Castin2001}. Generally, for fixed $k\neq0$, the eigenvalues appear in pairs $\{ \epsilon_k, \epsilon^*_k \}$. Then any decaying mode with $\mathrm{Im}(\epsilon_k)<0$ is accompanied by a growing mode with $\mathrm{Im}(\epsilon_k)>0$. Given $\epsilon_k$, the eigenvalues for the opposite sector $-k$ can be simply obtained as $\{ -\epsilon_k, -\epsilon^*_k \}$. For special cases with $m=0$ or $k=0$, additional symmetries exist and the eigenvalues appear in a quadruplet $\{ \epsilon_k, \epsilon^*_k, -\epsilon_k, -\epsilon^*_k \}$. Therefore, the system is dynamically stable only when all eigenvalues are real. In addition, the spectrum for $k=0$ also has a trivial zero-energy mode $(0,0,a\psi_m,-a\psi_m)$ with $a\in \mathbb{C}$, corresponding to a gauge transformation that does not produce any physical excitation.
%%%%%==

%%%%%==
Another useful constraint can be obtained from the combination of the BdG equations. We multiply Eq.~\eqref{eq:psi_m+k} by $A_{m,k}^*$ and Eq.~\eqref{eq:psi_m-k} by $A_{m,-k}^*$, subtract the resulting equations, and integrate over space to obtain
%%%%%
\begin{align}
&\epsilon_k \int d^3x \left(|A_{m,k}|^2-|A_{m,-k}|^2\right) \notag
\\
&=
\int d^3x\, A_{m,k}^* L_{m+k}A_{m,k}
+
\int d^3x\, A_{m,-k}^* L_{m-k}A_{m,-k}
\notag\\
&\quad
+2\lambda\int d^3x\,\psi_m^2
\,\mathrm{Re}\left(A_{m,k}^*A_{m,-k}\right)
+\int d^3x\, C_k \nabla_k^2 C_k^* ,
\label{eq:bdg_constraint}
\end{align}
%%%%%
where the last term has been simplified using Eq.~\eqref{eq:C_k}. The self-adjointness of $L_s$ and $\nabla_s^2$ implies that the right-hand side of Eq.~\eqref{eq:bdg_constraint} is real. It follows that
%%%%%
\begin{equation}
(\epsilon_k-\epsilon^*_{k})
\int d^3x
\left(|A_{m,k}|^2-|A_{m,-k}|^2\right)
=0.
\label{eq:bdg_krein}
\end{equation}
%%%%%
As a consequence, any dynamically unstable mode with $\mathrm{Im}(\epsilon_k)\neq0$ must satisfy
%%%%%
\begin{equation}
\int d^3x
\left(|A_{m,k}|^2-|A_{m,-k}|^2\right)=0.
\label{eq:bdg_unstable_constraint}
\end{equation}
We further define
\begin{equation}
  Q = \frac{|\int d^3 x \left(|A_{m,k}|^2-|A_{m,-k}|^2\right)|}{\int d^3 x \left(|A_{m,k}|^2+|A_{m,-k}|^2\right)},
\end{equation}
%%%%%
which provides a useful diagnostic for identifying unstable eigenmodes. It is also invariant under the spectral symmetries, so all members of a spectral quadruplet have the same value. 
%%%%%==

\subsection{Numerical method}

%%%%%==
Boundary conditions on the symmetry axis are fixed by the azimuthal index $(m\pm k)$ carried by each perturbation field. A Dirichlet condition is imposed on the $z$ axis when this index is nonzero, whereas a Neumann condition is imposed when it vanishes. The remaining boundary conditions are the same as those used in the stationary eigenmode calculation. For the even-parity background considered here, the perturbation amplitudes can be further decomposed into even- and odd-parity sectors under equatorial reflection.
%%%%%==

%%%%%==
The computational domain and mesh are identical to those used for the stationary background in Sec.~\ref{sec:stationary} ($N_r = N_\theta = 320$).  The three fields $(A_{m,k},A_{m,-k},C_k)$ are discretized in the product space $V_h = V_1 \times V_1 \times V_1$, where $V_1$ is the $P_2$ Lagrange finite-element space. Discretization of the BdG system in $V_h$ yields the generalized eigenvalue problem
%%%%%
\begin{equation}\label{eq:matrix_eigenvalue_problem}
\epsilon M u = K u,
\end{equation}
%%%%%
where $u$ is the combined coefficient vector obtained by expanding the three fields in the $P_2$ basis of $V_1$. The matrices $K$ and $M$ inherit a $3\times 3$ block structure from the BdG Eq.~\eqref{eq:BdG_system}. In particular, $M$ is singular because the Poisson equation~\eqref{eq:C_k} does not involve the eigenvalue parameter $\epsilon$. The eigenvalue problem is then solved with SLEPc using a shift-and-invert spectral transformation with a shift of $10^{-4}$ to target eigenvalues near the origin. The problem type is set to \texttt{EPS\_PGNHEP}, which uses the $M$-induced inner product together with eigenvector purification to preserve the symmetry structure of the eigenvalue problem~\cite{slepc-users-manual}. In practice, this yields smaller residuals of \eqref{eq:matrix_eigenvalue_problem} than those obtained with the default solver.
%%%%%==

\subsection{Numerical Results}

%%%%%
\begin{table}
\centering
\renewcommand{\arraystretch}{1.3}
\begin{tabular}{c|c|c}
\noalign{\hrule height 0.75pt}
% \hline
$\quad k\quad$ & $\epsilon_k$ & $Q$ 
\tabularnewline
\hline
\hline 
\multirow{3}{*}{0} & $\pm 0.068251$  & $0.399$ \\
                    & $\pm 0.10572$ & $0.998$ \\
                    & $\pm 0.12060$ & $0.962$ \\
                    % & $\pm 3.853\times 10^{-2}$ & $0.997$ \\
\hline
\multirow{3}{*}{1}  & $\pm 0.11110$  & $0.991$ \\
                    & $\pm 0.13143$ & $0.997$ \\
                    & $\pm 0.13352$ & $0.999$ \\
\hline 
\multirow{3}{*}{2}  & $\pm 0.10572$  & $0.998$ \\
                    & $\pm 0.13129$ & $0.999$ \\
                    & $\pm 0.14272$ & $0.999$ \\
\hline
\multirow{3}{*}{3} & $\pm 0.13143$  & $0.999$ \\
                    & $\pm 0.14272$ & $0.999$ \\
                    & $\pm 0.14885$ & $0.999$ \\
\noalign{\hrule height 0.75pt}
\end{tabular}
\caption{BdG spectrum for even-parity perturbations of the non-interacting ($\lambda=0$) spherically symmetric $m=0$ ground state. The index $k$ labels the azimuthal perturbation sector, and $\epsilon_k$ is the corresponding BdG eigenvalue obtained from Eqs.~\eqref{eq:BdG_system}. The quantity $Q$ measures the norm imbalance between the two BdG components and must vanish for dynamically unstable modes.} 
\label{tab:bdg_m=0}
\end{table}
%%%%%

%%%%%
\begin{table}
\centering
\renewcommand{\arraystretch}{1.3}
\begin{tabular}{c|c|c}
\noalign{\hrule height 0.75pt}
% \hline
$\quad k\quad$ & $\epsilon_k$ & $Q$ 
\tabularnewline
\hline
\hline 
\multirow{3}{*}{0} & $\pm 1.482\times 10^{-2}$  & $0.277$ \\
                    & $\pm 2.791\times 10^{-2}$ & $0.983$ \\
                    & $\pm 3.402\times 10^{-2}$ & $0.977$ \\
                    % & $\pm 3.853\times 10^{-2}$ & $0.997$ \\
\hline
\multirow{3}{*}{1}  & $- 1.722\times 10^{-2}$  & $0.966$ \\
                    & $(2.452\pm 0.529 i)\times 10^{-2}$  & $7.046\times 10^{-9}$ \\
                    & $- 2.913\times 10^{-2}$ & $0.984$ \\
\hline 
\multirow{3}{*}{2}  & $(1.620\pm 0.773i)\times 10^{-2}$  & $3.337\times 10^{-14}$ \\
                    & $- 2.154\times 10^{-2}$ & $0.998$ \\
                    & $- 2.875\times 10^{-2}$ & $0.998$ \\
\hline
\multirow{3}{*}{3} & $- 8.144 \times 10^{-3}$  & $0.980$ \\
                    & $- 3.029\times 10^{-2}$ & $0.999$ \\
                    & $\                                                                            3.487\times 10^{-2}$ & $0.991$ \\
\noalign{\hrule height 0.75pt}
\end{tabular}
\caption{BdG spectrum for even-parity perturbations of the noninteracting ($\lambda=0$) rotating $m=1$ background. The notation is the same as in Table~\ref{tab:bdg_m=0}.}
\label{tab:bdg_m=1}
\end{table}
%%%%%

%%%%%==
The calculated BdG spectra for the spherical $m=0$ and rotating $m=1$ backgrounds are summarized in Tables~\ref{tab:bdg_m=0} and~\ref{tab:bdg_m=1}, respectively. The numerical results confirm the spectral symmetry structure discussed below Eq.~\eqref{eq:BdG_system}. Given that the typical numerical error of the background solution is of order $10^{-6}$, symmetry-related eigenvalues with relative differences below $10^{-5}$ are grouped into the same spectral multiplet and listed in a single row. The zero mode associated with a gauge transformation is omitted since the magnitude of its eigenvalue is below $10^{-5}$, and only the three nonzero spectral multiplets with the smallest $|\epsilon_k|$ are listed for each $k$ sector.
%%%%%==

%%%%%==
The lowest-energy $m=0$ configuration is first studied as a validation test, since the nonrelativistic spherically symmetric ground state without self-interaction is well known to be linearly stable \cite{Guzman:2004wj,Bernal:2006it,Harrison2003,Nambo:2024gvs}. For the even-parity perturbation sectors considered here, the resulting eigenvalues are all real within numerical accuracy, as listed in Table~\ref{tab:bdg_m=0}. The near degeneracy of some eigenvalues across different $k$ sectors originates from the spherical symmetry of the background. In the exact spherical problem, $l$ is a good quantum number, and modes with the same $l$ but different azimuthal indices $k$ are degenerate. 
%%%%%==

%%%%%==
Because the numerical formulation adopted here is axisymmetric rather than strictly spherical, even the $k=0$ sector contains both spherically symmetric and nonspherical axisymmetric modes.
Only the nearly spherically symmetric eigenfunctions, corresponding to the first and third modes in the $k=0$ sector in Table~\ref{tab:bdg_m=0}, can be directly compared with one-dimensional spherical calculations in the literature. Their eigenvalues agree with those of Ref.~\cite{Harrison2003} to the reported significant digits after accounting for the factor of 2 arising from the different normalization. The lowest nearly spherically symmetric perturbation mode, corresponding to the first mode in the $k=0$ sector of Table~\ref{tab:bdg_m=0}, was also computed in Ref.~\cite{Guzman:2004wj}. After conversion to the present normalization, their reported quantities yield the corresponding dimensionless frequency $\epsilon=\sigma/M^2 = 0.068569$, which differs from the present result by a relative amount of $5\times10^{-3}$. 
The odd-parity perturbation sectors are also examined and found to contain only real eigenvalues. Taken together, these results show that the spherically symmetric ground state is linearly stable against both spherical and nonspherical perturbations. This conclusion also agrees with the spherical-harmonic analysis of Ref.~\cite{Nambo:2024gvs}. These comparisons for the spherical background confirm the reliability of our numerical method and its implementation for computing the BdG spectra of boson stars.
%%%%%==

%%%%%==
Next, we consider the even-parity rotating background with $m=1$ in the absence of self-interaction. The even-parity perturbation spectra in the azimuthal sectors $k=0,1,2,3$ are presented in Table~\ref{tab:bdg_m=1}. The odd-parity perturbation sectors are also examined in this background and contain only real eigenvalues. The negative-$k$ sectors are related to the listed ones by the symmetry structure of the spectrum and are omitted.
%%%%%==

%%%%%==
Most eigenvalues in Table~\ref{tab:bdg_m=1} are real and are associated with stable oscillatory perturbation modes. The lowest-frequency excitation in the $k=0$ sector has $\mathrm{Re}(\epsilon)=1.482\times10^{-2}$, which agrees with the oscillation frequency extracted from the energy-exchange dynamics during the linear stage shown in Fig.~\ref{fig:relativeEkEp} within the Fourier frequency resolution.  
%%%%%==

%%%%%
\begin{figure}
  \centering
  \begin{tabular}{c@{\hspace{0pt}}c}
    \includegraphics[width=0.24\textwidth]{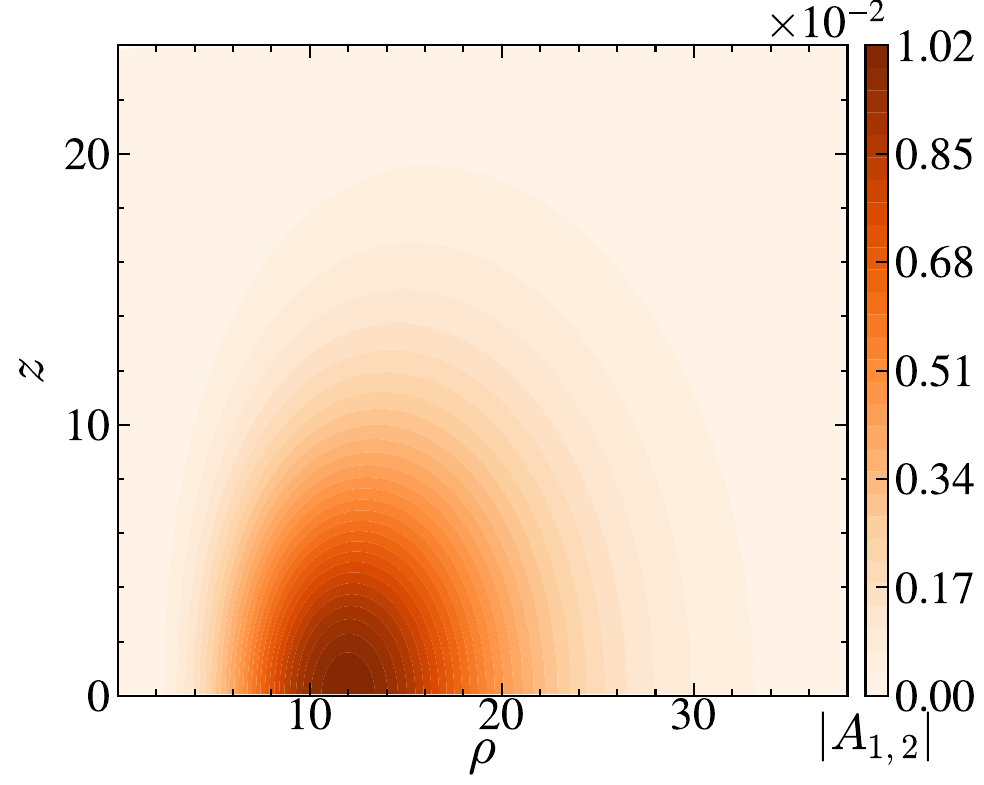} &
    \includegraphics[width=0.24\textwidth]{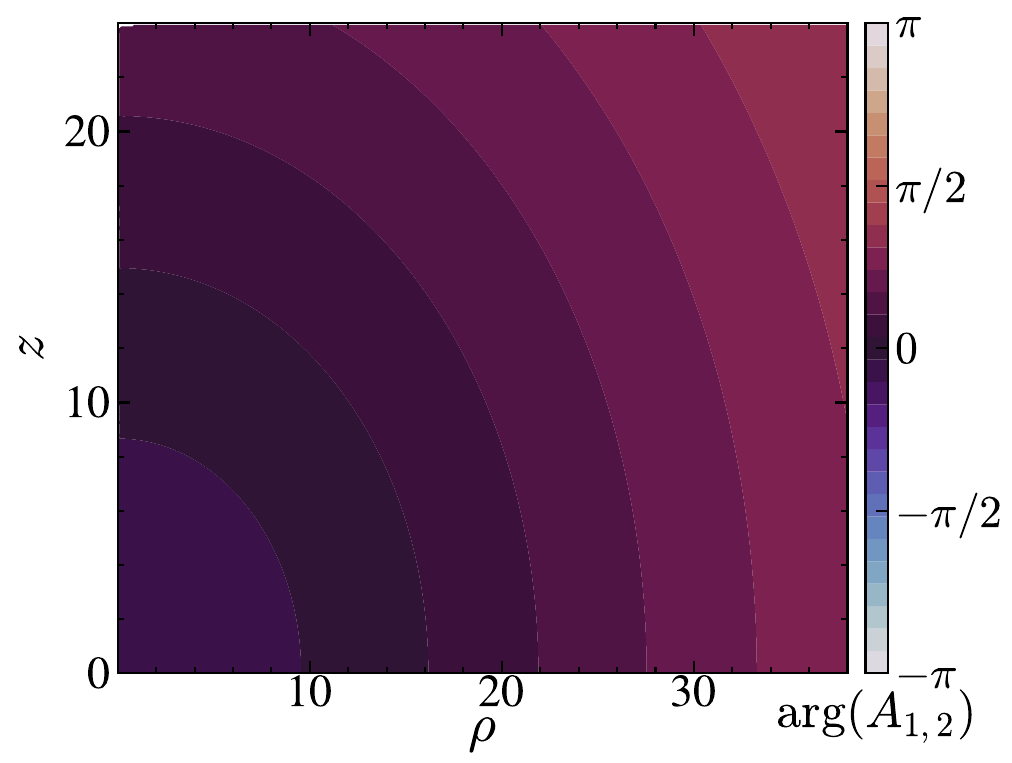} \\[0pt]  
    \includegraphics[width=0.24\textwidth]{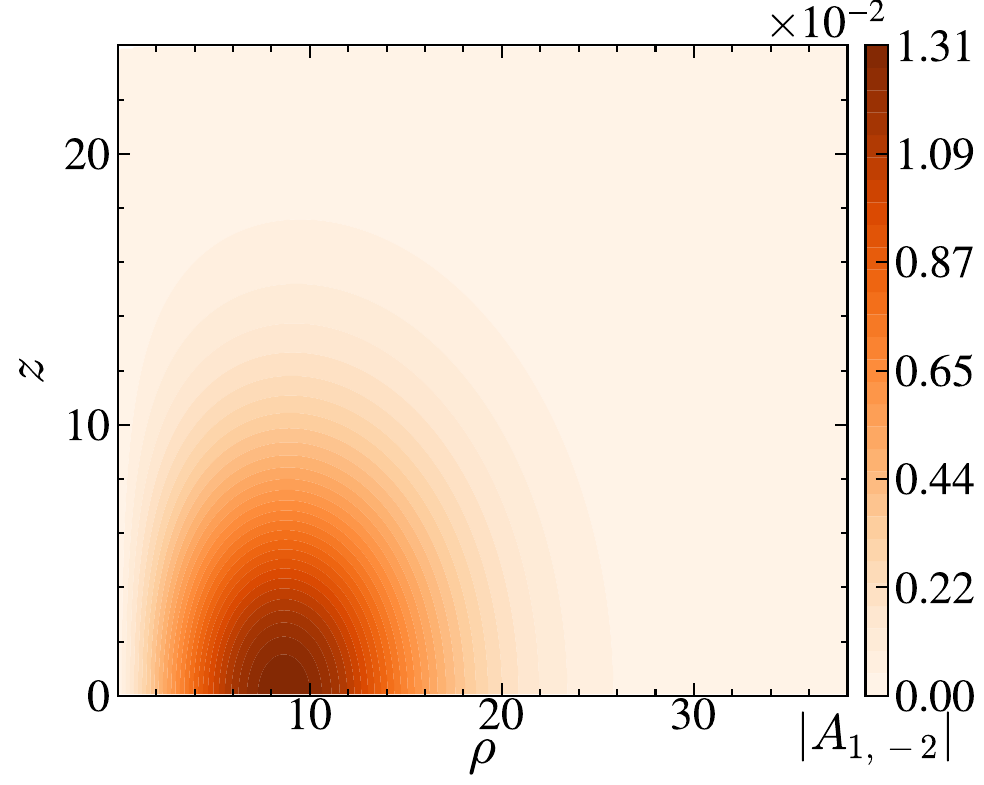} &
    \includegraphics[width=0.24\textwidth]{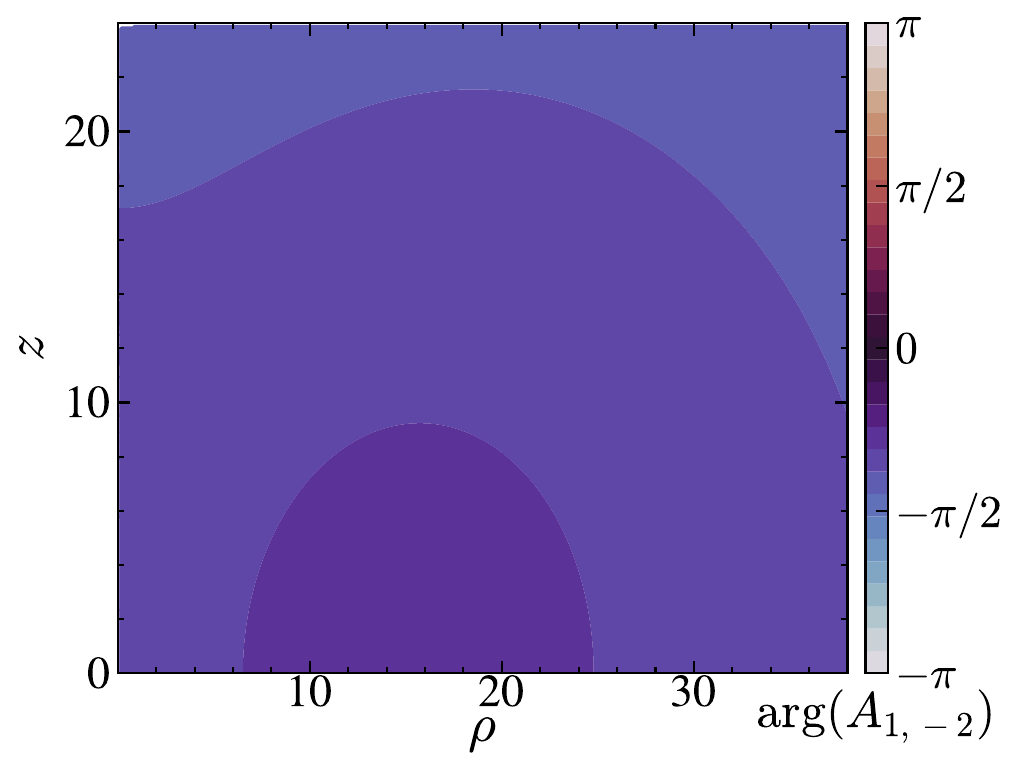}
  \end{tabular}
\caption{The dominant unstable $k=2$ BdG mode is shown for the even-parity $m=1$ background at $\lambda=0$. The top and bottom rows show the components $A_{m,k}$ and $A_{m,-k}$, respectively. The left and right columns show their magnitudes and phases.}
\label{fig:bdg_mode}
\end{figure}
%%%%%

%%%%%==
Complex eigenvalues are found in the $k=1$ and $k=2$ sectors. Their imaginary parts agree with the growth rates extracted from the particle numbers of the azimuthal modes in Fig.~\ref{fig:Ns_t}, with relative errors below $3\times10^{-2}$. The $k=2$ mode has the largest imaginary part and dominates the leading linear instability. Although the $k=3$ eigenvalues given in Table~\ref{tab:bdg_m=1} are real for the stationary $m=1$ background, their populations increase in the late-time simulation, as shown in Fig.~\ref{fig:Ns_t}. This occurs because the evolving configuration is no longer dominated solely by the $m=1$ mode, allowing nonlinear mode coupling to populate other azimuthal sectors. The eigenfunctions of the leading $k=2$ unstable mode, normalized with $\int d^3x |A_{m,\pm k}|^2=1$, are shown in Fig.~\ref{fig:bdg_mode}. This simultaneous normalization is possible because the two components have equal norms for complex-eigenvalue modes, as required by Eq.~\eqref{eq:bdg_unstable_constraint}.
%%%%%==

%%%%%
\begin{figure}[]
\centering
\includegraphics[width=0.45\textwidth]{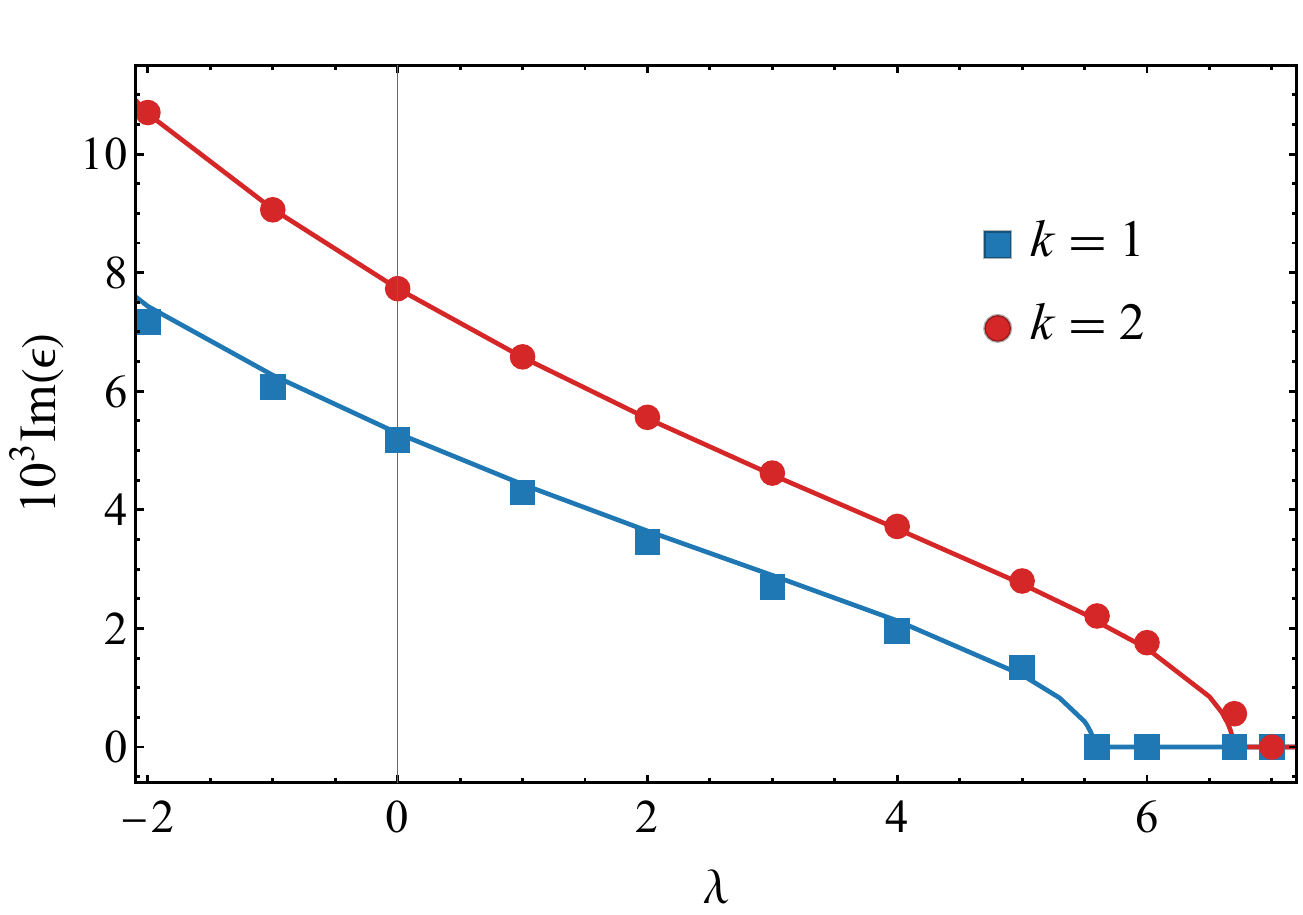}
\caption{Growth rates of the $k=1$ and $k=2$ instabilities as functions of the self-interaction strength $\lambda$. The blue and red curves are the positive imaginary parts of the corresponding BdG eigenvalues, respectively. The symbols denote growth rates obtained from exponential fits to the corresponding azimuthal-sector populations during the linear stage of the full time-evolution simulations.}
\label{fig:growth_rate}
\end{figure}
%%%%%

%%%%%==
Finally, we assess the dependence of these instabilities on the self-interaction strength. Figure~\ref{fig:growth_rate} shows the magnitude of the imaginary part of the two unstable modes in Table~\ref{tab:bdg_m=1} as a function of $\lambda$. For $k=1$, the imaginary part decreases with increasing $\lambda$ and becomes numerically indistinguishable from zero ($<10^{-5}$) at $\lambda \ge 5.58$, indicating that no unstable mode remains in this sector.  For $k=2$, the unstable mode persists up to larger values of $\lambda$ and disappears at $\lambda \ge 6.69$. This threshold agrees with that inferred from the fastest-growing $k=2$ mode reported in Ref.~\cite{Dmitriev:2021utv}, after accounting for a normalization factor of $32\pi$. The symbols in Fig.~\ref{fig:growth_rate} denote growth rates extracted from exponential fits to the azimuthal mode decomposition, following the procedure illustrated in Fig.~\ref{fig:Ns_t}. The BdG predictions agree well with these values, confirming that the early-time amplification of perturbations is governed by the unstable BdG modes.
%%%%%==

%%%%%==
The difference between the two growth rates also explains the characteristic times $T_a$ and $T_b$ introduced in Sec.~\ref{sec:time_evolution}. Because $\mathrm{Im}(\epsilon_2)>\mathrm{Im}(\epsilon_1)$, the $k=2$ mode grows faster, and its amplitude becomes comparable to that of the stationary background first. This gives the estimate $T_a\sim[\mathrm{Im}(\epsilon_2)]^{-1}$ for the end of the linear stage. After $T_a$, the $k=2$ mode drives the quasiperiodic conversion. The more slowly growing $k=1$ mode becomes significant near $T_b$. Assuming comparable initial perturbation amplitudes, this gives $T_b-T_a\sim[\mathrm{Im}(\epsilon_1)]^{-1}-[\mathrm{Im}(\epsilon_2)]^{-1}$.
%%%%%==

\section{Nonlinear Evolution and Multi-Mode Framework}
\label{sec:nonlinear}

%%%%%==
The linearized framework described in the previous section is valid up to the characteristic timescale $T_a$, during which the unstable perturbations are sufficiently small, and the rotating toroidal configuration remains nearly axisymmetric. After this time, the continued growth of these unstable modes causes nonlinear effects to become significant. As illustrated in Fig.~\ref{fig:evolution_snapshots}, this nonlinear stage ($T_a< t < T_b$) is characterized by a quasiperiodic conversion between approximately ring-like and twin-star-like density distributions. Figure~\ref{fig:Ns_t} also shows that particle number is exchanged quasiperiodically among different azimuthal sectors, rather than being transferred predominantly from the background sector to the unstable sidebands as in the linear stage. In this section, we develop an analytical framework beyond the BdG formalism to describe this behavior.
%%%%%==  

%%%%%==
We first reduce the infinite-dimensional field dynamics to a set of coupled ordinary differential equations using a variational method. Such a reduction has been widely applied to Bose--Einstein condensates confined in double-well traps \cite{Smerzi1997,Ostrovskaya2000}. In this approach, the time-dependent field is expanded in a prescribed set of stationary modes, which may be either localized states associated with the individual wells or global modes of the composite trap. With these spatial profiles held fixed, the dynamics is described by the time-dependent mode populations and relative phases, providing an effective description of tunneling and population exchange among the retained modes.
%%%%%==

%%%%%==
In the present problem, however, only the background configuration is an exact stationary solution of the GPP equations, while the other relevant modes are unstable BdG excitations built on this background. We therefore construct the reduced basis using the exact stationary background together with the sideband profiles of the dominant unstable mode. For the even-parity $m=1$ background considered here, the dominant instability lies in the $k=2$ sector, leading to the following three-mode variational ansatz 
%%%%%
\begin{equation}\label{eq:ansatz_specific}
\psi(t,\bm x) = \sum_{s \in \{1, 3, -1\}} \sqrt{N_s(t)} f_s(r,\theta) e^{i\Omega_s(t,\bm x)} e^{is\varphi}.
\end{equation}
%%%%%
Here, $f_1(r,\theta)$ is the normalized spatial profile of the stationary background satisfying Eq.~\eqref{eq:GPP_stationary}. The profiles $f_3(r,\theta)\equiv|A_{1,2}(r,\theta)|$ and $f_{-1}(r,\theta)\equiv|A_{1,-2}(r,\theta)|$ are the moduli of the two sideband components of the dominant unstable $k=2$ BdG eigenmode obtained from Eqs.~\eqref{eq:BdG_system}. The time-dependent mode populations and phase fields are represented by $N_s(t)$ and $\Omega_s(t,\bm x)$, respectively. The interference between the $s=1$ background and the $s=-1,3$ sidebands generates the leading $k=2$ azimuthal deformation underlying the quasiperiodic conversion between ring-like and twin-star-like configurations. Beyond $T_b$, additional azimuthal sectors acquire appreciable populations and matter emission becomes significant, so the three-mode truncation in Eq.~\eqref{eq:ansatz_specific} is no longer sufficient.
%%%%%==

%%%%%==
The remainder of this section is organized as follows. First, we introduce an approximation by neglecting the spatial dependence of the phases $\Omega_s(t,\bm{x})$. Next, we substitute the ansatz in Eq.~\eqref{eq:ansatz_specific} into the system's Lagrangian density to derive an effective Hamiltonian in terms of the mode populations $N_s(t)$ and phases $\Omega_s(t)$. Finally, we analyze the resulting dynamical equations to determine the evolutionary trajectories of the population dynamics, and compare the analytical predictions with the full three-dimensional simulation.
%%%%%==

\subsection{Spatial phase coherence}

%%%%%
\begin{figure}[t]
\centering
\includegraphics[width=0.45\textwidth]{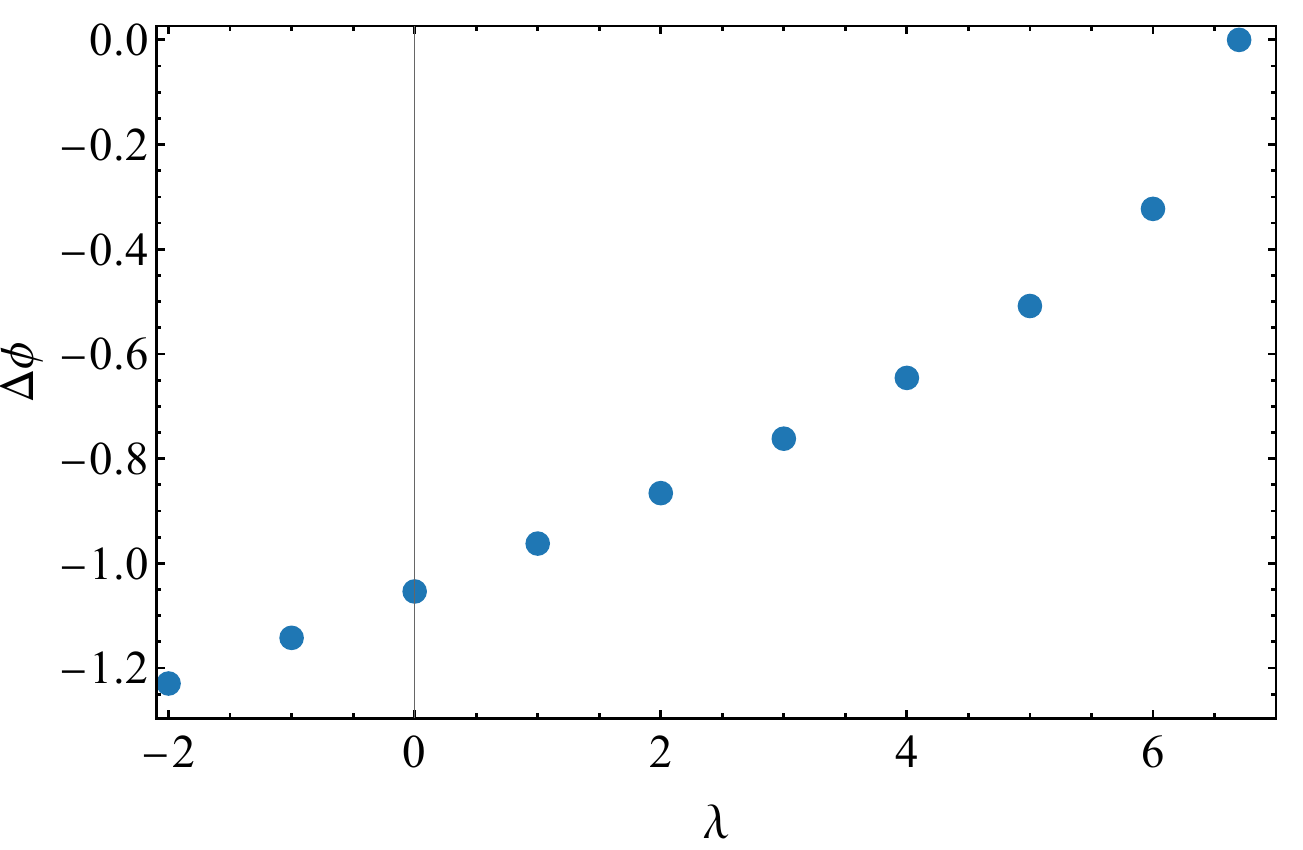}
\caption{Phase difference $\Delta\phi=\arg(A_{1,-2})-\arg(A_{1,2})$ between the two sideband components of the $k=2$ BdG mode for the even-parity rotating $m=1$ background as a function of the self-interaction strength $\lambda$. }
\label{fig:deltaphi}
\end{figure}
%%%%%%

%%%%%==
The three-mode truncation alone does not yet produce a finite-dimensional dynamical system because the phases remain position-dependent. Few-mode reductions commonly adopt a coherent-phase approximation \cite{Smerzi1997}, in which the spatial phase profile of each retained mode is replaced by a single time-dependent variable, thereby reducing the field dynamics to a finite set of collective coordinates. In our case, the phases of the two sideband components of the dominant unstable $k=2$ BdG eigenmode are observed to be nearly uniform over the high-density region of the condensate, as illustrated for $\lambda=0$ in Fig.~\ref{fig:bdg_mode}. Quantitatively, the phase-gradient contribution to the total energy, $\int f_s^2 (\nabla \Omega_s)^2 d^3x$, is approximately two orders of magnitude smaller than the corresponding amplitude-gradient contribution, $\int (\nabla f_s)^2 d^3x$. Hence, we neglect the space dependence of $\Omega_s$ and set $\Omega_s(t, \bm{x}) \to \bar\Omega_s(t)$. Since the phase is defined modulo $2\pi$, the global phase of each mode is evaluated using the density-weighted circular mean
%%%%%
\begin{equation}\label{eq:phase_average}
\bar\Omega_s(t)
=
\arg\left(
\int d^3x\,f_s^2(\bm{x})e^{i\Omega_s(t,\bm{x})}
\right).
\end{equation}
%%%%%
The two sideband components $A_{1,2}(r,\theta)$ and $A_{1,-2}(r,\theta)$ of the unstable $k=2$ BdG mode are generally complex-valued. Under the spatial phase-coherence approximation, the arbitrary overall phase of the BdG eigenvector can be chosen to remove the phase of $A_{1,2}$. With this choice, the phase of $A_{1,-2}$ is equal to the phase difference $\Delta\phi\equiv\arg(A_{1,-2})-\arg(A_{1,2})$. It varies with $\lambda$ since the corresponding BdG system in Eqs.~\eqref{eq:BdG_system} depends on $\lambda$, as shown in Fig.~\ref{fig:deltaphi}. As $\lambda$ increases, $\Delta\phi$ increases from negative values and reaches zero at $\lambda=6.69$. For $\lambda\geq6.69$, the eigenvalue of this mode is purely real, and its two sideband components can both be chosen to be real and have the same sign.
%%%%%

\begin{widetext}
%%%%%==
Next, we project the continuous BdG equations~\eqref{eq:BdG_system} onto this uniform-phase basis. Multiplying Eqs.\eqref{eq:psi_m+k} and \eqref{eq:psi_m-k} by $f_3(\bm{x})$ and $f_{-1}(\bm{x})$, respectively, and integrating the resulting equations over space yields a pair of coupled integral equations,
%%%%%
\begin{subequations}\label{eq:bdg_projected_complex}
\begin{align}
\epsilon =& \int f_3(\bm x) \left( -\frac{1}{2}\nabla^2_3 + N \mathcal{G}_0 \left[ f^2_1(\bm x) \right] + 2N \lambda f^2_1(\bm x) - \omega_1 \right) f_3(\bm x) d^3{\bm x} + \lambda N e^{i\Delta\phi} \int f^2_1(\bm x) f_3(\bm x) f_{-1}(\bm x) d^3{\bm x} \notag
\\
&+ N \int f_1(\bm x) f_3(\bm x) \mathcal{G}_{2} \left[ f_1(\bm x) f_3(\bm x) \right] d^3{\bm x} + N  e^{i\Delta\phi} \int f_1(\bm x) f_3(\bm x) \mathcal{G}_{2} \left[ f_1(\bm x) f_{-1}(\bm x) \right] d^3{\bm x},
\\
\epsilon =& -\int f_{-1}(\bm x) \left( -\frac{1}{2}\nabla^2_{-1} + N \mathcal{G}_0 \left[ f^2_1(\bm x) \right] + 2N \lambda f^2_1(\bm x) - \omega_1 \right) f_{-1}(\bm x) d^3{\bm x} - \lambda N e^{-i\Delta\phi} \int f^2_1(\bm x) f_{3} (\bm x) f_{-1}(\bm x) d^3{\bm x} \notag
\\
&- N \int f_1(\bm x) f_{-1}(\bm x) \mathcal{G}_{2} \left[ f_1(\bm x) f_{-1}(\bm x) \right] d^3{\bm x} - N  e^{-i\Delta\phi} \int f_1(\bm x) f_{-1}(\bm x) \mathcal{G}_{2} \left[ f_1(\bm x) f_{3}(\bm x) \right] d^3{\bm x}.
\end{align}  
\end{subequations}
%%%%%
Here, we introduce the Green's-function operator $\mathcal{G}_s$, defined by the relation $\nabla_s^2\mathcal{G}_s[f]=f$ for an arbitrary smooth source function $f(\bm{x})$. In terms of this operator, the gravitational potentials can be written as $\Phi_0=N\mathcal{G}_0[f_1^2]$ and $C_k=\sqrt{N}\mathcal{G}_k[f_1(f_3+f_{-1}e^{i\Delta\phi})]$, allowing the interaction terms to be expressed entirely in terms of the mode amplitudes. In practice, the action of $\mathcal{G}_s$ is evaluated numerically by solving the corresponding Poisson-type equations. Separating the complex eigenvalue $\epsilon$ into its real and imaginary parts yields the following equations,
%%%%%
\begin{subequations}\label{eq:bdg_real_imag}
\begin{align}
\omega_1 + \mathrm{Re}(\epsilon) &= \int d^3x\ f_3 \left( -\frac{1}{2}\nabla^2_3 + N \mathcal{G}_0[f_1^2] + 2N \lambda f_1^2 \right) f_3 + \lambda N \cos(\Delta\phi) \int d^3x\ \left(f_1^2 f_3 f_{-1}\right) \notag 
\\
&\quad + N \int d^3x\ f_1 f_3 \mathcal{G}_2[f_1 f_3] + N \cos(\Delta\phi) \int d^3x\ f_1 f_3 \mathcal{G}_2[f_1 f_{-1}] , 
\\
\omega_1 - \mathrm{Re}(\epsilon) &= \int d^3x\ f_{-1} \left( -\frac{1}{2}\nabla^2_{-1} + N \mathcal{G}_0[f_1^2] + 2N \lambda f_1^2 \right) f_{-1} + \lambda N \cos(\Delta\phi) \int d^3x\ f_1^2 f_{-1} f_3 \notag 
\\
&\quad + N \int d^3x\ f_1 f_{-1} \mathcal{G}_2[f_1 f_{-1}] + N \cos(\Delta\phi) \int d^3x\ f_1 f_{-1} \mathcal{G}_2[f_1 f_3] , 
\\
\mathrm{Im}(\epsilon) &= \sin(\Delta\phi) \left( \lambda N \int d^3x\ f_1^2 f_3 f_{-1}  + N \int d^3x\ f_1 f_3 \mathcal{G}_2[f_1 f_{-1}] \right), \label{eq:bdg_imag}
\end{align}
\end{subequations}
%%%%%
where the spatial dependence of the mode profiles $f_s$ is suppressed for brevity. These relations explicitly map the BdG eigenvalue $\epsilon$ to the overlap integrals of the three interacting modes. Notably, taking the imaginary part of either equation in Eqs.~\eqref{eq:bdg_projected_complex} yields the same expression for $\mathrm{Im}(\epsilon)$ in Eq.~\eqref{eq:bdg_imag}, due to the self-adjoint symmetry of the Green's function operator $\int d^3x\, u \mathcal{G}_s[v] = \int d^3x\, v \mathcal{G}_s[u]$.
%%%%%==
\end{widetext}

\subsection{Lagrangian Reduction and Effective Hamiltonian}

%%%%%==
Conservation of total particle number and angular momentum requires the two sidebands to have equal populations, $N_{-1}(t)=N_3(t)\equiv n(t)$. The background population is then given by $N_1(t)=N-2n(t)$, leaving $n(t)$ as the only independent population variable. Substituting the ansatz in Eq.~\eqref{eq:ansatz_specific} into the time-derivative part of the Lagrangian density in Eq.~\eqref{eq:lagNR} and integrating over space, one obtains
%%%%%
\begin{align}\label{eq:lag_reduced}
\int d^3x \frac{i}{2}(\psi^*\partial_t\psi-\psi\partial_t\psi^*) = -N \frac{d\bar{\Omega}_1}{dt} + n(t) \frac{d\Delta\bar{\Omega}}{dt},
\end{align}
%%%%%
where the relative phase is defined as $\Delta\bar{\Omega}(t) = 2\bar{\Omega}_1(t) - \bar{\Omega}_3(t) - \bar{\Omega}_{-1}(t)$. The first term on the right-hand side of Eq.~\eqref{eq:lag_reduced} is a total time derivative and does not affect the equations of motion. The second term explicitly identifies $n$ and $\Delta\bar\Omega$ as a canonically conjugate pair. This establishes the canonical structure of the reduced dynamical system, allowing us to construct the effective Hamiltonian by integrating the Hamiltonian density in Eq.~\eqref{eq:hamiltonian} under the three-mode ansatz.
%%%%%==

\begin{widetext}
%%%%%==
To derive the explicit form of this effective Hamiltonian, we first evaluate the individual energy contributions in Eq.~\eqref{eq:hamiltonian} sequentially. The spatial kinetic energy term evaluates to
%%%%%
\begin{equation}\label{eq:kinetic_term}
  \int d^3x \left( -\frac{1}{2} \psi^* \nabla^2 \psi \right) = \int d^3x \left( -\frac{1}{2}(N-2n)  f_1 \nabla^2_1 f_1 -\frac{1}{2} n f_3 \nabla^2_3 f_3 -\frac{1}{2} n f_{-1} \nabla^2_{-1} f_{-1} \right) 
\end{equation}
%%%%%
The three-mode ansatz in Eq.~\eqref{eq:ansatz_specific} retains the azimuthal sectors $s=1,-1,3$. The pairwise products entering the density $|\psi|^2$ generate only the harmonics $s=0,\pm2,$ and $\pm4$. Each density harmonic sources the gravitational potential in the same azimuthal sector. With the three-mode truncation, the self-consistent gravitational potential can thus be expanded as
%%%%%
\begin{equation}
  \Phi(t,\bm x) = \Phi_0(r,\theta) + \sum_{s \in \{2, 4\}} \left( \Phi_s(r,\theta) e^{is\varphi} + \Phi^*_s(r,\theta) e^{-is\varphi} \right),
\end{equation}
%%%%%
where each harmonic $\Phi_s$ is determined by the corresponding source term,
%%%%%
\begin{subequations} \label{eq:poisson_modes}
  \begin{align}
  \nabla^2_0 \Phi_0 &= N_1 f^2_1 + N_3 f^2_3 + N_{-1} f^2_{-1}, \\
  \nabla^2_2 \Phi_2 &= \sqrt{N_1 N_3} e^{-i\bar\Omega_1 + i\bar\Omega_3} f_1 f_3 + \sqrt{N_1 N_{-1}} e^{i\bar\Omega_1 - i\bar\Omega_{-1}} f_1 f_{-1}, \\
  \nabla^2_4 \Phi_4 &= \sqrt{N_3 N_{-1}} e^{i\bar\Omega_3 - i\bar\Omega_{-1}} f_3 f_{-1}.
\end{align}
\end{subequations}
%%%%%
The gravitational potential energy term is then given by
%%%%%
\begin{align}
\begin{split}
  \int d^3x \left( \frac{1}{2} \Phi \psi^* \psi \right) =& \int d^3x \frac{1}{2} \Phi_0 \left[ N_1 f^2_1 + n f^2_3 + n f^2_{-1} \right]
  \\
  &+ \int d^3x \frac{1}{2} \left[ \sqrt{N_1 N_3} \Phi_2 e^{i\bar\Omega_1 - i\bar\Omega_3} f_1 f_3 + \sqrt{N_1 N_{-1}} \Phi_2 e^{-i\bar\Omega_1 + i\bar\Omega_{-1}} f_1 f_{-1} + \mathrm{c.c.} \right]
  \\
  &+ \int d^3x \frac{1}{2} \left[ \sqrt{N_3 N_{-1}} \Phi_4 e^{i\bar\Omega_3 - i\bar\Omega_{-1}} f_3 f_{-1} + \mathrm{c.c.} \right].
\end{split}
\end{align}
%%%%%
After substituting the solutions of the Poisson equations~\eqref{eq:poisson_modes} and using the symmetry of the Green’s-function operators $\mathcal{G}s$, we can express this energy contribution entirely in terms of the mode amplitudes $f_j$. Setting $N_1=N-2n$ and $N_3=N_{-1}=n$, we obtain
%%%%%
\begin{align}
\begin{split}\label{eq:gravitational_term}
    \int d^3x \left( \frac{1}{2} \Phi \psi^* \psi \right) =&  \int d^3x\ \frac{1}{2} N^2 f^2_1 \mathcal{G}_0 [f^2_1] + \int d^3x\ n N (-2 f^2_1 + f^2_3 +f^2_{-1}) \mathcal{G}_0 [f^2_1]
  \\
  &+ \int d^3x \frac{1}{2} n^2 \left(2f^2_1 - f^2_3 - f^2_{-1}\right) \mathcal{G}_0 \left[2f^2_1 - f^2_3 - f^2_{-1}\right] 
  \\
  &+ \int d^3x\ (N-2n) n \left( f_1 f_3 \mathcal{G}_2 \left[ f_1 f_3 \right] + f_1 f_{-1} \mathcal{G}_2 \left[ f_1 f_{-1} \right] + 2\cos\Delta\bar\Omega f_1 f_3 \mathcal{G}_2 \left[ f_1 f_{-1} \right] \right)
  \\
  &+ \int d^3x\ n^2 f_3 f_{-1} \mathcal{G}_4 \left[ f_3 f_{-1} \right].
\end{split}
\end{align}
%%%%%
The contact self-interaction energy is straightforwardly evaluated as
%%%%%
\begin{align}
\begin{split}\label{eq:self_interaction_term}
  \int d^3x \left( \frac{1}{2}\lambda |\psi|^4 \right) =& \int d^3x \frac{1}{2} \lambda \biggl[ (N-2n)^2 f^4_1 + n^2 f^4_3 + n^2 f^4_{-1} + 4 (N-2n) n f^2_1 (f^2_3 + f^2_{-1}) + 4n^2 f^2_3 f^2_{-1} 
\\
&+4\cos\Delta\bar\Omega (N-2n) n f^2_1 f_3 f_{-1} \biggr]
\end{split}
\end{align}
%%%%%==

%%%%%==
Combining the contributions from Eqs.~\eqref{eq:kinetic_term}, \eqref{eq:gravitational_term}, and \eqref{eq:self_interaction_term}, and arranging the result by powers of the population $n$ gives the total effective Hamiltonian in polynomial form
%%%%%
\begin{equation}\label{eq:hamiltonian_polynomial}
  H = H^{(0)} + H^{(1)} n + H^{(2)} n^2.
\end{equation}
%%%%%
The integral coefficients in this expansion represent the effective coupling strengths of various elastic two-body scattering processes among the three basis modes, mediated by self-gravity and the contact interaction. The zeroth-order term $H^{(0)}$ represents the energy of the pure background state
%%%%%
\begin{equation}
  H^{(0)} = \int d^3x \left[ -\frac{1}{2} N f_1 \nabla^2_1 f_1 + \frac{1}{2} N^2 f^2_1 \mathcal{G}_0 f^2_1 + \frac{1}{2} \lambda N^2 f^4_1 \right]
\end{equation}
%%%%%
The linear term $H^{(1)}$ can be significantly simplified using the linear BdG relations~\eqref{eq:bdg_real_imag} to eliminate redundant integrals, giving
%%%%%
\begin{align}
  H^{(1)} =& 2N \left(\cos\Delta\bar\Omega - \cos\Delta\phi \right) K
\end{align}
%%%%%
where $K$ is the coupling constant for the population-altering channel, $1+1 \leftrightarrow 3+(-1)$. In this channel, two background particles are converted into a pair of sideband particles, or vice versa, directly driving the dynamic growth or depletion of $n$. This process is mediated by both the self-gravity and the contact interaction, yielding
%%%%%
\begin{equation}\label{eq:K}
  K = \int d^3x \left( f_1 f_3 \mathcal{G}_2 \left[ f_1 f_{-1} \right] 
      + \lambda f^2_1 f_3 f_{-1} \right).
\end{equation}
%%%%%
The second-order term $H^{(2)}$ captures the remaining combined effects and is given by
%%%%%
\begin{equation}
  H^{(2)} = U - 4K\cos\Delta\bar\Omega,
\end{equation}
%%%%%
where $U$ collects the contributions from all population-preserving scattering 
processes (e.g., $1+3 \to 1+3$, $1+(-1) \to 1+(-1)$, and $3+(-1) \to 3+(-1)$),
%%%%%
\begin{align}\label{eq:U}
\begin{split}
    U = &\ \frac{1}{2} \int d^3x \left( 2f^2_1 - f^2_3 - f^2_{-1} \right) \mathcal{G}_0 \left[ 2f^2_1 - f^2_3 - f^2_{-1} \right] \\
    &+ \int d^3x \biggl( -2 f_1 f_3 \mathcal{G}_2 \left[ f_1 f_3 \right] - 2 f_1 f_{-1} \mathcal{G}_2 \left[ f_1 f_{-1} \right] + f_3 f_{-1} \mathcal{G}_4 \left[ f_3 f_{-1} \right] \biggr) \\
    &+ \frac{\lambda}{2} \int d^3x \left[ \left( 2f^2_1 - f^2_3 - f^2_{-1} \right)^2 + 2 f^2_3 f^2_{-1} - 4 f^2_1 f^2_3 - 4 f^2_1 f^2_{-1} \right].
\end{split}
\end{align}
%%%%%==

%%%%%==
Physically, the polynomial dependence of $H$ on $n$ shown in Eq.~\eqref{eq:hamiltonian_polynomial} originates from the $2 \to 2$ mode coupling between the background and sidebands. The interaction energy associated with this process stems from the field combination $\int d^3x\, \psi_1^{*2} \psi_{-1} \psi_3  = (N - 2n)n C$, where $C$ denotes the effective overlap integral of these modes. This polynomial structure accounts for both the growth of $n$ in the linear stage and its saturation in the nonlinear stage. For the self-gravitating configurations considered here, the effective coupling is negative $C<0$. At the beginning ($n \to 0$), the linear term makes the energy decrease toward larger $n$, corresponding to the initial instability. As the sideband population $n$ accumulates, the quadratic term bends the energy curve upward and produces a stable minimum at a finite population $n_*$. The total energy remains conserved, and trajectories surrounding this minimum describe periodic population exchange in the reduced model. In the next subsection, we will examine this behavior numerically and compare it with the quasiperiodic exchange observed in the full simulation.
%%%%%==
\end{widetext}

%%%%%==
Altogether, the effective Hamiltonian in Eq.~\eqref{eq:hamiltonian_polynomial} can be rewritten as
%%%%%
\begin{equation}\label{eq:hamiltonian_effective}
\begin{split}
    H_{\rm eff} &=  2n N K (\cos\Delta\bar\Omega - \cos\Delta\phi) 
    \\
    &\quad + n^2 \left[ U - 4 \cos\Delta\bar\Omega K \right],
\end{split}
\end{equation}
%%%%%
where the constant term $H^{(0)}$ has been omitted because it does not affect the dynamics. The corresponding equations of motion for the canonical pair $(n, \Delta\bar{\Omega})$ are given in the Hamiltonian form by
%%%%%
\begin{equation}
    \frac{dn}{dt} = -\frac{\partial H_{\rm eff}}{\partial \Delta\bar\Omega},\ 
    \frac{d\Delta\bar\Omega}{dt} = \frac{\partial H_{\rm eff}}{\partial n},
\end{equation}
%%%%%
which yield
%%%%%
\begin{subequations}\label{eq:canonical_eom}
\begin{align}
    \frac{dz}{dt} & = 2Nz(1-2z) K \sin\Delta\bar\Omega , \label{eq:dzdt}
    \\
    \frac{d\Delta\bar\Omega}{dt} & = 2 N K (\cos\Delta\bar\Omega - \cos\Delta\phi) \notag
    \\
    &\quad + 2 N z (U - 4 K \cos\Delta\bar\Omega). \label{eq:dDeltadt}
\end{align}
\end{subequations}
%%%%%
where $z = n/N \in [0,1/2]$ is the normalized population of each sideband. 
%%%%%==

%%%%%==
To solve the reduced dynamics, one must specify the initial values of $z$ and $\Delta\bar\Omega$. The evolution considered here starts from the $m=1$ background, so the initial sideband population $z$ is infinitesimal. Determining the initial value of $\Delta\bar\Omega$ requires a more careful analysis. A generic initial perturbation in the $k=2$ sector contains a superposition of stable and unstable BdG eigenmodes. The unstable eigenmode rapidly becomes dominant because it grows exponentially, causing the phase difference between the two BdG sideband components to approach the value $\Delta\phi$. Once this unstable mode dominates, the phases of the background and sideband components can be obtained by comparing the three-mode ansatz in Eq.~\eqref{eq:ansatz_specific} with the linear perturbation expansion in Eq.~\eqref{eq:psi_perturbation}. This comparison gives
%%%%%
\begin{subequations}\label{eq:linear_phase_ev}
  \begin{align}
  \bar{\Omega}_1(t) &= -\omega_1 t, \\
  \bar{\Omega}_3(t) &= -(\omega_1 + \mathrm{Re}(\epsilon)) t, \\
  \bar{\Omega}_{-1}(t) &= -(\omega_1 - \mathrm{Re}(\epsilon)) t - \Delta\phi,
\end{align}
\end{subequations}
%%%%%
The minus sign of the phase difference in $\bar\Omega_{-1}$ follows from the complex conjugation of the $m-k$ component in Eq.~\eqref{eq:psi_perturbation}. Substituting these three phase expressions into the definition of $\Delta\bar\Omega$ cancels all time-dependent terms and gives $\Delta\bar\Omega(t<T_a)=\Delta\phi$. Thus the sideband phase difference $\Delta\phi$ selected by the unstable BdG eigenmode provides the initial value of the dynamical relative phase $\Delta\bar\Omega$.
%%%%%==

%%%%%==
The qualitative behavior of the reduced dynamics can be understood directly from Eqs.~\eqref{eq:canonical_eom}. During the linear stage, $z\ll1$ and $\Delta\bar\Omega=\Delta\phi$. Equation~\eqref{eq:dzdt} reduces to an exponential growth equation for $z$, with a growth rate of $2NK\sin\Delta\phi$. According to Eq.~\eqref{eq:bdg_imag}, this rate is equal to $2\operatorname{Im}(\epsilon)$, so the reduced equations recover the early-time exponential growth of the two dominant sidebands observed in Fig.~\ref{fig:Ns_t}. As $z$ reaches a finite value, the population-dependent term in Eq.~\eqref{eq:dDeltadt} becomes significant and shifts $\Delta\bar\Omega$ away from $\Delta\phi$. This phase evolution eventually changes the sign of $\sin\Delta\bar\Omega$, and hence that of $\dot z$, causing the sideband population to decrease after reaching its maximum. The reduced system then exhibits periodic population exchange, resulting in the quasiperiodic oscillations observed in the full field simulation.
%%%%%==

\subsection{Comparison of Reduced and Full Dynamics}

\begin{figure}
\centering
\includegraphics[width=0.45\textwidth]{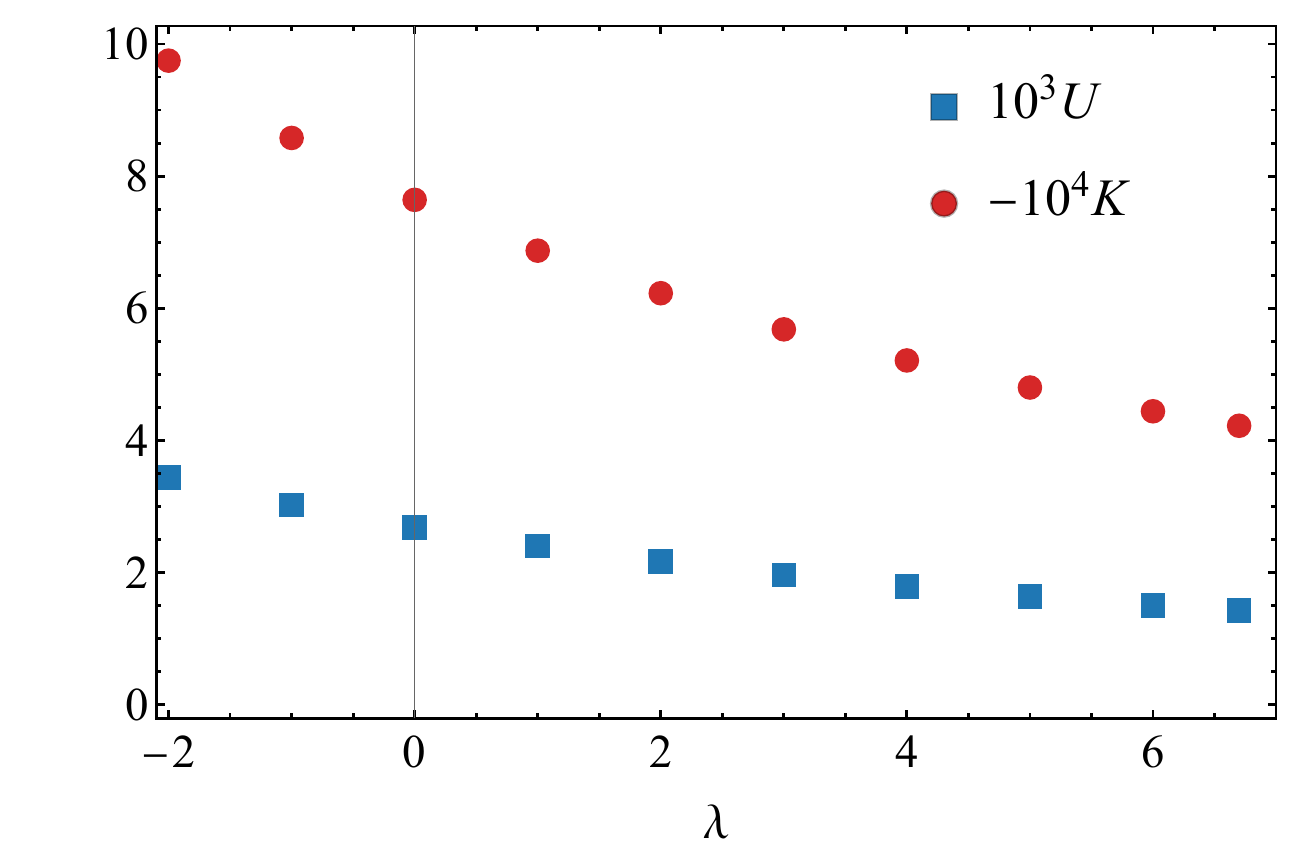}
\caption{Effective coupling coefficients $K$ and $U$ in the three-mode Hamiltonian $H_{\rm eff}$ as functions of the self-interaction strength $\lambda$. The coefficients are defined in Eqs.~\eqref{eq:K} and \eqref{eq:U}.}
\label{fig:U&K}
\end{figure}

%%%%%==
This subsection examines the phase-space structure of the effective Hamiltonian and compares the reduced trajectories with those obtained from the full three-dimensional simulations. Constructing the phase portraits of $H_{\rm eff}$ requires the three parameters $\Delta\phi$, $K$, and $U$ shown in Eq.~\eqref{eq:hamiltonian_effective}. The relative phase $\Delta\phi$ is calculated from the two sideband components of the unstable $k=2$ BdG eigenmode, while $K$ and $U$ are obtained from overlap integrals involving the stationary background and the sideband profiles. These three quantities are not independent. For the even-parity $m=1$ family considered here, the dimensionless formulation introduced in Sec.~\ref{sec:GPP} leaves $\lambda$ as the only free parameter. At each value of $\lambda$, solving the stationary equations~\eqref{eq:GPP_stationary} and BdG equations~\eqref{eq:BdG_system} determines the three basis profiles and hence the values of $\Delta\phi$, $K$, and $U$. The dependence of $\Delta\phi$ on $\lambda$ is shown in Fig.~\ref{fig:deltaphi}, while the corresponding results for $K$ and $U$ are presented in Fig.~\ref{fig:U&K}. The magnitudes of both coefficients decrease with increasing $\lambda$, while their ratio remains nearly constant at $K/U\approx -0.29$, with a relative variation below $5\%$ over the range considered.
%%%%%==

%%%%%
%%%%%

%%%%%
\begin{figure*}
\centering
\includegraphics[width=0.9\textwidth]{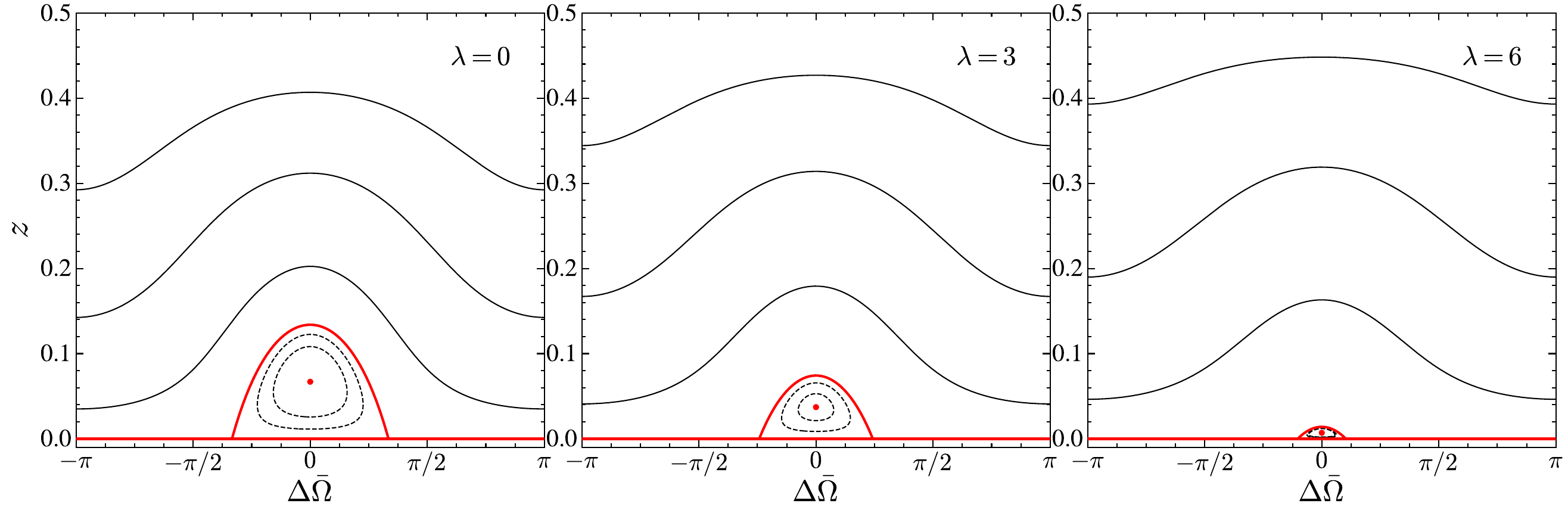}
\caption{Phase portraits of the reduced Hamiltonian in the $(\Delta\bar\Omega,z)$ plane for $\lambda=0,3,6$. The red solid curves denote $H_{\rm eff}=0$. The black solid and dashed curves denote contours with $H_{\rm eff}>0$ and $H_{\rm eff}<0$, respectively. The red dots mark the stable fixed points.}
\label{fig:reduced_phase_portraits}
\end{figure*}
%%%%%

%%%%%
\begin{figure*}
\centering
\includegraphics[width=0.9\textwidth]{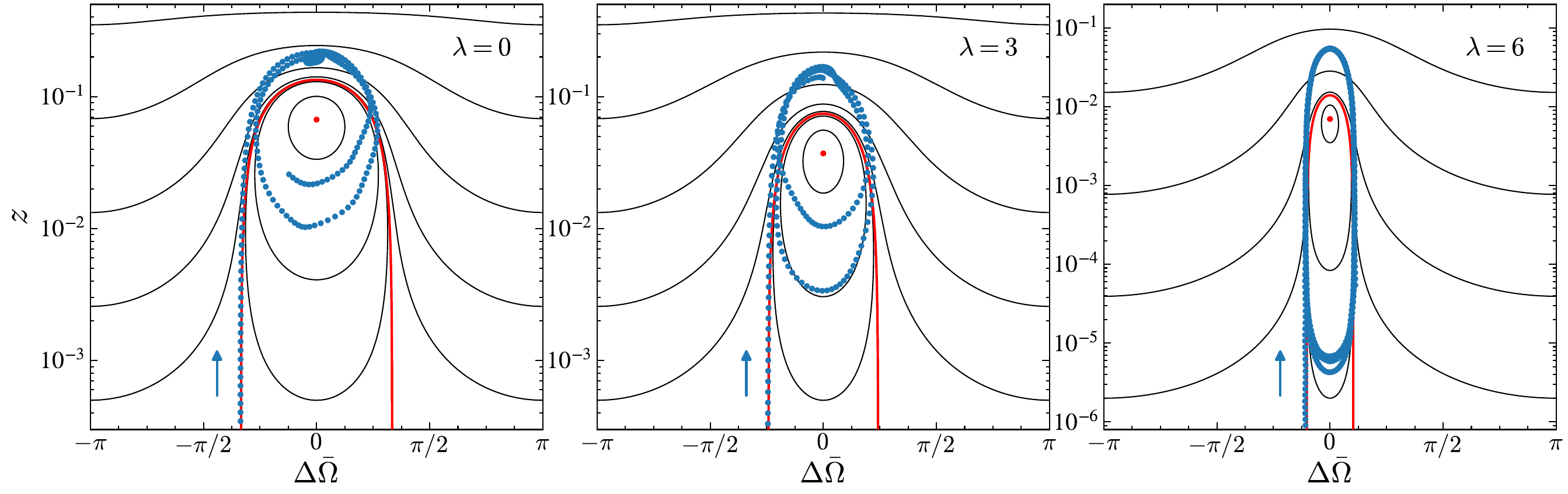}
\caption{Phase-space comparison between the reduced model and the full three-dimensional simulations for $\lambda=0,3,6$. The blue dots show the $(\Delta\bar\Omega,z)$ trajectories extracted from the simulations, and the blue arrows indicate the direction of increasing time. The black curves represent contours of constant $H_{\rm eff}$, while the red curves denote $H_{\rm eff}=0$. }
\label{fig:phase_space_comparison}
\end{figure*} 
%%%%%

%%%%%==
Since the initial value of $z$ is infinitesimal, the effective Hamiltonian approaches zero according to Eq.~\eqref{eq:hamiltonian_effective}, and an evolution governed exactly by $H_{\rm eff}$ would follow the corresponding zero-energy contour. Fig.~\ref{fig:reduced_phase_portraits} shows the phase portraits in the canonical variables $\Delta\bar\Omega\in(-\pi,\pi]$ and $z\in[0,1/2]$ for $\lambda=0, 3, 6$. 
The red contour $H_{\rm eff}=0$ consists of a horizontal line at $z=0$ and a nontrivial arched branch. The horizontal line represents the idealized background with no sideband perturbation. For an infinitesimal but nonzero initial value of $z$, the dominant unstable mode rapidly drives $\Delta\bar\Omega$ toward $\Delta\phi$. The reduced evolution then starts from $(\Delta\bar\Omega,z)=(\Delta\phi,0^+)$ and follows the arched zero-energy branch. The black solid contours above this branch have $H_{\rm eff}>0$ and correspond to running-phase trajectories, along which the unwrapped relative phase increases monotonically. The black dashed contours enclosed by the red curves have $H_{\rm eff}<0$ and form closed trajectories around the stable fixed point marked by the red point. Along these closed trajectories, the relative phase remains bounded and the sideband population oscillates. 
The interior fixed point can be obtained by setting the right-hand sides of Eqs.~\eqref{eq:canonical_eom} to zero,
%%%%%
\begin{equation}
    \Delta\bar\Omega_* = 0,\ z_* = \frac{K(\cos\Delta\phi-1)}{U-4K}.
\end{equation}
%%%%%
As $\lambda$ increases, the stable fixed point shifts toward smaller $z$. Since the ratio $K/U$ varies only weakly over the range considered, this shift is mainly controlled by $\Delta\phi$. For $K/U=-0.29$, the fixed-point population is approximately $z_* = -0.13(\cos\Delta\phi-1)$. As $\lambda$ approaches $6.69$, $\Delta\phi$ tends to vanish, driving $z_*$ toward zero. 
%%%%%==

%%%%%==
The phase portraits discussed above are obtained under the three-mode and fixed-profile approximations. The full GPP system, however, contains additional azimuthal sectors and spatial degrees of freedom. Consequently, the trajectories extracted from the full simulations are not expected to follow the red zero-energy contour exactly. Figure~\ref{fig:phase_space_comparison} compares the trajectories predicted by $H_{\rm eff}$ with those extracted from the full simulations using the azimuthal decomposition in Eq.~\eqref{eq:azimuthal_decomposition}. The black curves denote contours of constant $H_{\rm eff}$, while the blue dots trace the phase-space trajectories extracted from the simulations. For $\lambda=0$, data are included up to $T_b=6\times10^3$. Beyond $T_b$, additional azimuthal sectors acquire appreciable populations, as shown in Fig.~\ref{fig:Ns_t}, and the quasiperiodic conversion between the ring-like and twin-star-like configurations disappears, as seen in Fig.~\ref{fig:evolution_snapshots}(a). For $\lambda=3$, the comparison is restricted to $t\leq T_b=10^4$ for the same reason. For $\lambda=6$, no corresponding breakdown occurs before the simulation ends at $t=4\times10^4$, so the complete evolution range is included. The quasiperiodic conversion remains visible throughout this interval, as shown in Fig.~\ref{fig:evolution_snapshots}(c).
%%%%%==

%%%%%==
During the linear stage, the trajectories extracted from the full simulations follow the red contour $H_{\rm eff}=0$. The relative phase stays near $\Delta\phi$, while the sideband population grows. After the system enters the nonlinear stage, the trajectories first move above the zero-energy contour, reaching values of $z$ larger than those predicted by the effective model. They then depart from the red contour and form bounded loops around the corresponding stable fixed point. These loops approximately follow nearby contours with slightly negative $H_{\rm eff}$, consistent with energy transfer from the three retained modes to additional azimuthal modes. This bounded motion produces recurrent population exchange around a finite sideband population. As $\lambda$ approaches $6.69$, the stable fixed point moves toward smaller $z_*$ and eventually merges with $z=0$. Consequently, the population exchange occurs at smaller sideband populations, reducing the amplitude of the $k=2$ density deformation. Hence, the two density peaks of the twin-star-like configuration become less clearly separated, as seen in Fig.~\ref{fig:evolution_snapshots}(c). At the critical value of $\lambda$, the $k=2$ instability is suppressed, and the associated quasiperiodic conversion disappears.
%%%%%==

%%%%%==
Although the effective model qualitatively explains the recurrent population exchange underlying the quasiperiodic behavior during the nonlinear stage, the numerical trajectories exhibit two distinct deviations from its predictions. Energy transfer to additional modes can account for the first deviation, namely the formation of bounded loops inside the zero-energy contour. But such energy transfer cannot explain the second deviation, in which the trajectories cross above the red contour and reach larger values of $z$. This second deviation reveals a limitation of the fixed-profile approximation, in which the spatial profiles $f_s$ are held fixed. This approximation overestimates the energy cost of transferring particles from the background into the sideband sectors, resulting in the energy minimum at a smaller sideband population than that observed in the full evolution. This discrepancy can be further attributed to the adjustment of the density spatial distribution. As shown in Fig.~\ref{fig:bdg_mode}, the $s=3$ sideband profile extends to larger radii than the stationary $m=1$ background. As the sideband populations increase, the density distribution broadens and the self-gravitational potential becomes shallower, allowing the mode profiles to expand accordingly. This self-adjustment lowers the energy cost of population transfer relative to the fixed-basis estimate. Similar dilatational dynamics have been studied as breathing modes in systems with time-dependent external potentials \cite{Castin:1996zz,Asakawa:2023tkm}. Extending the present model to include this breathing mode may improve its quantitative agreement with the full simulations. Such an extension will be explored in future work.
%%%%%==

\section{Summary}

%%%%%==
In this work, we have investigated the stationary configurations and dynamical evolution of nonrelativistic rotating boson stars with contact self-interaction. The even-parity $m=1$ configuration, which is the lowest-energy rotating state, serves as the principal example. Previous studies showed that this configuration develops a nonaxisymmetric instability and indicated that perturbations in the $k=2$ sector have the largest growth rate~\cite{Siemonsen:2020hcg,Dmitriev:2021utv}. The present work extends these results in two directions. First, we formulate a systematic BdG framework for the linear stability analysis of rotating boson stars. This framework determines the eigenvalues and eigenfunctions of both stable and unstable perturbation modes. Second, we examine the previously unexplored early nonlinear stage. During this stage, the density distribution undergoes a quasiperiodic conversion between ring-like and twin-star-like configurations, accompanied by population exchange between the $m=1$ background and the $m=1\pm2$ sidebands. This behavior is described by an effective Hamiltonian constructed through a three-mode variational reduction based on the stationary background and the dominant unstable BdG eigenmode.
%%%%%==

%%%%%==
Following the mathematical framework introduced in Sec.~\ref{sec:GPP}, stationary solutions are obtained in Sec.~\ref{sec:stationary} for different azimuthal numbers $m$, parity sectors, and interaction strengths $\lambda$. Both even- and odd-parity rotating solutions exhibit toroidal density distributions. The eigenvalues of the noninteracting solutions are summarized in Table~\ref{tab:eigenvalues}. Repulsive interactions broaden the torus and make the eigenvalues less negative, as shown in Fig.~\ref{fig:omega_lambda}. In the strongly repulsive regime, the solutions approach Thomas--Fermi scaling. For the even-parity $m=1$ configuration, the attractive branch exists only for $\lambda\geq-7.43$, where $\lambda\equiv GM^2\mu^2\lambda_{\rm NR}/(4\pi)$ and $\lambda_{\rm NR}$ is the dimensionless contact interaction strength in the nonrelativistic Lagrangian density given in Eq.~\eqref{eq:lagNR}.
%%%%%==

%%%%%==
In Sec.~\ref{sec:time_evolution}, the full three-dimensional simulations reveal three characteristic stages, as shown in Fig.~\ref{fig:evolution_snapshots}. During the linear stage $t<T_a$, the sideband populations grow exponentially but remain much smaller than the background population, leaving the toroidal density nearly invariant. During the quasiperiodic nonlinear stage $T_a<t<T_b$, the density alternates between ring-like and twin-star-like configurations. The azimuthal decomposition in Fig.~\ref{fig:Ns_t} shows that this conversion is accompanied by population exchange between the $m=1$ background and the $m=-1,3$ sidebands. After $T_b$, additional azimuthal sectors acquire appreciable populations and the quasiperiodic conversion terminates. Increasing the repulsive interaction delays both $T_a$ and $T_b$ and suppresses the development of the nonaxisymmetric instability.
%%%%%==

%%%%%==
In Sec.~\ref{sec:linear}, the linear stage before $T_a$ is analyzed using the BdG framework. The spectra in Table~\ref{tab:bdg_m=0} show that the spherical ground state is stable in all azimuthal perturbation sectors considered. For the even-parity $m=1$ background, complex eigenvalues are observed in the $k=1$ and $k=2$ sectors, as shown in Table~\ref{tab:bdg_m=1}. The eigenvalues of the $k=2$ mode have the largest imaginary part, and its two sideband components correspond to the rapidly growing $m=-1$ and $m=3$ sectors observed in the full evolution. The calculated imaginary parts agree with the growth rates extracted from the simulations. As $\lambda$ increases, the imaginary parts of the $k=1$ eigenvalues vanish for $\lambda\geq5.58$, whereas those of the $k=2$ eigenvalues vanish for $\lambda\geq6.69$. The BdG analysis thus determines the mode selection and growth rates during the linear stage.
%%%%%==

%%%%%==
In Sec.~\ref{sec:nonlinear}, the quasiperiodic nonlinear stage between $T_a$ and $T_b$ is analyzed using the three-mode variational reduction. The wave function is expanded in the stationary $m=1$ background and the two sideband profiles of its dominant unstable $k=2$ BdG eigenmode. The spatial phase-coherence approximation replaces each phase field by a global phase. Particle number and angular momentum conservation then reduce the dynamical variables to the canonical pair $(z,\Delta\bar\Omega)$, where $z$ is the normalized sideband population and $\Delta\bar\Omega$ is the global relative phase. In the linear limit, the resulting effective Hamiltonian in Eq.~\eqref{eq:hamiltonian_effective} reproduces the initial exponential growth. At finite sideband population, it describes the nonlinear saturation of this growth and the subsequent population exchange between the background and sideband sectors. Comparison with the full simulations in Fig.~\ref{fig:phase_space_comparison} shows that the reduced model qualitatively captures the phase-space evolution in the quasiperiodic stage.
%%%%%==

%%%%%==
Although this work focuses on the even-parity $m=1$ background, both the BdG analysis of the linear stage and the few-mode description of the early nonlinear stage can be applied to other rotating configurations. We also examined the noninteracting odd-parity $m=1$ background and found unstable modes in perturbation sectors of both equatorial parities, with the largest growth rate occurring in the odd-parity $k=2$ sector. For a higher-$m$ background, the dominant instability may occur in a different $k$ sector~\cite{Dmitriev:2021utv}. Once this sector is identified, the corresponding reduced basis can be constructed from the background $m$ mode and its $m\pm k$ sidebands.
%%%%%==

%%%%%==
Our results have two main implications. First, strong repulsive self-interactions can significantly increase the lifetime of a nonrelativistic rotating boson star. Specifically, for the even-parity $m=1$ configuration, the characteristic time $T_a$ provides an estimate of its dimensionless lifetime. The corresponding physical lifetime is
%%%%%
\begin{equation}
\begin{split}
T_{a,\rm phys}
&=10^{8}\,\mathrm{yr}
\left(\frac{10^{-22}\,\mathrm{eV}}{\mu}\right)^3
\left(\frac{10^9\,M_\odot}{M}\right)^2
\frac{10^{-3}}{\mathrm{Im}(\epsilon)}
\\
&=10^{-2}\,\mathrm{yr}
\left(\frac{10^{-6}\,\mathrm{eV}}{\mu}\right)^3
\left(\frac{10^{-10}\,M_\odot}{M}\right)^2
\frac{10^{-3}}{\mathrm{Im}(\epsilon)},
\end{split}
\end{equation}
%%%%%
where \(\mathrm{Im}(\epsilon)\) is the dimensionless growth rate of the \(k=2\) sector shown in Fig.~\ref{fig:growth_rate}. For ultralight bosonic dark matter, \(T_{a,\rm phys}\) can be comparable to the age of the Universe and could have important implications for galaxy formation. 
%%%%%==

%%%%%==
Second, an interesting nonlinear phenomenon occurs even when \(T_{a,\rm phys}\) is not sufficiently long. Before the rotating boson star relaxes into a single localized clump, its density can undergo a quasiperiodic conversion between distinct configurations, as shown in Fig.~\ref{fig:evolution_snapshots}. For the specific example considered here, the duration of this stage depends on the competition between the \(k=1\) and \(k=2\) instabilities. In the interval \(5.58<\lambda<6.69\), the \(k=2\) mode is the only unstable channel, allowing the system to undergo more conversion cycles. The corresponding physical conversion period is
%%%%%
\begin{equation}
\begin{split}
T_{{\rm conv},\rm phys}
&=10^8\,\mathrm{yr}
\left(\frac{10^{-22}\,\mathrm{eV}}{\mu}\right)^3
\left(\frac{10^9\,M_\odot}{M}\right)^2
\frac{T_{\rm conv}}{10^3}
\\
&=10^{-2}\,\mathrm{yr}
\left(\frac{10^{-6}\,\mathrm{eV}}{\mu}\right)^3
\left(\frac{10^{-10}\,M_\odot}{M}\right)^2
\frac{T_{\rm conv}}{10^3},
\end{split}
\end{equation}
%%%%%
where \(T_{\rm conv}\) is the dimensionless conversion period given in Fig.~\ref{fig:Trec}. These long-lived oscillations of the nonaxisymmetric density distribution may produce characteristic observational signatures of rotating boson stars.
%%%%%==

\acknowledgments
%%%%%==
We thank Yin-Da Guo, Guo-Dong Zhang for helpful discussions. This work is supported by the National Natural Science Foundation of China (Grants Nos. 12447105 and 12575148)
%%%%%==

\bibliography{RotatingBosonStars.bib}

%apsrev4-2.bst 2019-01-14 (MD) hand-edited version of apsrev4-1.bst
%Control: key (0)
%Control: author (8) initials jnrlst
%Control: editor formatted (1) identically to author
%Control: production of article title (0) allowed
%Control: page (0) single
%Control: year (1) truncated
%Control: production of eprint (0) enabled
\begin{thebibliography}{86}%
\makeatletter
\providecommand \@ifxundefined [1]{%
 \@ifx{#1\undefined}
}%
\providecommand \@ifnum [1]{%
 \ifnum #1\expandafter \@firstoftwo
 \else \expandafter \@secondoftwo
 \fi
}%
\providecommand \@ifx [1]{%
 \ifx #1\expandafter \@firstoftwo
 \else \expandafter \@secondoftwo
 \fi
}%
\providecommand \natexlab [1]{#1}%
\providecommand \enquote  [1]{``#1''}%
\providecommand \bibnamefont  [1]{#1}%
\providecommand \bibfnamefont [1]{#1}%
\providecommand \citenamefont [1]{#1}%
\providecommand \href@noop [0]{\@secondoftwo}%
\providecommand \href [0]{\begingroup \@sanitize@url \@href}%
\providecommand \@href[1]{\@@startlink{#1}\@@href}%
\providecommand \@@href[1]{\endgroup#1\@@endlink}%
\providecommand \@sanitize@url [0]{\catcode `\\12\catcode `\$12\catcode
  `\&12\catcode `\#12\catcode `\^12\catcode `\_12\catcode `\%12\relax}%
\providecommand \@@startlink[1]{}%
\providecommand \@@endlink[0]{}%
\providecommand \url  [0]{\begingroup\@sanitize@url \@url }%
\providecommand \@url [1]{\endgroup\@href {#1}{\urlprefix }}%
\providecommand \urlprefix  [0]{URL }%
\providecommand \Eprint [0]{\href }%
\providecommand \doibase [0]{https://doi.org/}%
\providecommand \selectlanguage [0]{\@gobble}%
\providecommand \bibinfo  [0]{\@secondoftwo}%
\providecommand \bibfield  [0]{\@secondoftwo}%
\providecommand \translation [1]{[#1]}%
\providecommand \BibitemOpen [0]{}%
\providecommand \bibitemStop [0]{}%
\providecommand \bibitemNoStop [0]{.\EOS\space}%
\providecommand \EOS [0]{\spacefactor3000\relax}%
\providecommand \BibitemShut  [1]{\csname bibitem#1\endcsname}%
\let\auto@bib@innerbib\@empty
%</preamble>
\bibitem [{\citenamefont {Hu}\ \emph {et~al.}(2000)\citenamefont {Hu},
  \citenamefont {Barkana},\ and\ \citenamefont {Gruzinov}}]{Hu:2000ke}%
  \BibitemOpen
  \bibfield  {author} {\bibinfo {author} {\bibfnamefont {W.}~\bibnamefont
  {Hu}}, \bibinfo {author} {\bibfnamefont {R.}~\bibnamefont {Barkana}},\ and\
  \bibinfo {author} {\bibfnamefont {A.}~\bibnamefont {Gruzinov}},\ }\bibfield
  {title} {\bibinfo {title} {{Cold and fuzzy dark matter}},\ }\href
  {https://doi.org/10.1103/PhysRevLett.85.1158} {\bibfield  {journal} {\bibinfo
   {journal} {Phys. Rev. Lett.}\ }\textbf {\bibinfo {volume} {85}},\ \bibinfo
  {pages} {1158} (\bibinfo {year} {2000})},\ \Eprint
  {https://arxiv.org/abs/astro-ph/0003365} {arXiv:astro-ph/0003365}
  \BibitemShut {NoStop}%
\bibitem [{\citenamefont {Hui}\ \emph {et~al.}(2017)\citenamefont {Hui},
  \citenamefont {Ostriker}, \citenamefont {Tremaine},\ and\ \citenamefont
  {Witten}}]{Hui:2016ltb}%
  \BibitemOpen
  \bibfield  {author} {\bibinfo {author} {\bibfnamefont {L.}~\bibnamefont
  {Hui}}, \bibinfo {author} {\bibfnamefont {J.~P.}\ \bibnamefont {Ostriker}},
  \bibinfo {author} {\bibfnamefont {S.}~\bibnamefont {Tremaine}},\ and\
  \bibinfo {author} {\bibfnamefont {E.}~\bibnamefont {Witten}},\ }\bibfield
  {title} {\bibinfo {title} {{Ultralight scalars as cosmological dark
  matter}},\ }\href {https://doi.org/10.1103/PhysRevD.95.043541} {\bibfield
  {journal} {\bibinfo  {journal} {Phys. Rev. D}\ }\textbf {\bibinfo {volume}
  {95}},\ \bibinfo {pages} {043541} (\bibinfo {year} {2017})},\ \Eprint
  {https://arxiv.org/abs/1610.08297} {arXiv:1610.08297 [astro-ph.CO]}
  \BibitemShut {NoStop}%
\bibitem [{\citenamefont {Chavanis}(2025)}]{Chavanis:2025qcg}%
  \BibitemOpen
  \bibfield  {author} {\bibinfo {author} {\bibfnamefont {P.-H.}\ \bibnamefont
  {Chavanis}},\ }\bibfield  {title} {\bibinfo {title} {{A review of basic
  results on the Bose{\textendash}Einstein condensate dark matter model}},\
  }\href {https://doi.org/10.3389/fspas.2025.1538434} {\bibfield  {journal}
  {\bibinfo  {journal} {Front. Astron. Space Sci.}\ }\textbf {\bibinfo {volume}
  {12}},\ \bibinfo {pages} {1538434} (\bibinfo {year} {2025})}\BibitemShut
  {NoStop}%
\bibitem [{\citenamefont {Seidel}\ and\ \citenamefont
  {Suen}(1994)}]{Seidel:1993zk}%
  \BibitemOpen
  \bibfield  {author} {\bibinfo {author} {\bibfnamefont {E.}~\bibnamefont
  {Seidel}}\ and\ \bibinfo {author} {\bibfnamefont {W.-M.}\ \bibnamefont
  {Suen}},\ }\bibfield  {title} {\bibinfo {title} {{Formation of solitonic
  stars through gravitational cooling}},\ }\href
  {https://doi.org/10.1103/PhysRevLett.72.2516} {\bibfield  {journal} {\bibinfo
   {journal} {Phys. Rev. Lett.}\ }\textbf {\bibinfo {volume} {72}},\ \bibinfo
  {pages} {2516} (\bibinfo {year} {1994})},\ \Eprint
  {https://arxiv.org/abs/gr-qc/9309015} {arXiv:gr-qc/9309015} \BibitemShut
  {NoStop}%
\bibitem [{\citenamefont {Schive}\ \emph {et~al.}(2014)\citenamefont {Schive},
  \citenamefont {Chiueh},\ and\ \citenamefont {Broadhurst}}]{Schive:2014dra}%
  \BibitemOpen
  \bibfield  {author} {\bibinfo {author} {\bibfnamefont {H.-Y.}\ \bibnamefont
  {Schive}}, \bibinfo {author} {\bibfnamefont {T.}~\bibnamefont {Chiueh}},\
  and\ \bibinfo {author} {\bibfnamefont {T.}~\bibnamefont {Broadhurst}},\
  }\bibfield  {title} {\bibinfo {title} {{Cosmic Structure as the Quantum
  Interference of a Coherent Dark Wave}},\ }\href
  {https://doi.org/10.1038/nphys2996} {\bibfield  {journal} {\bibinfo
  {journal} {Nature Phys.}\ }\textbf {\bibinfo {volume} {10}},\ \bibinfo
  {pages} {496} (\bibinfo {year} {2014})},\ \Eprint
  {https://arxiv.org/abs/1406.6586} {arXiv:1406.6586 [astro-ph.GA]}
  \BibitemShut {NoStop}%
\bibitem [{\citenamefont {Levkov}\ \emph {et~al.}(2018)\citenamefont {Levkov},
  \citenamefont {Panin},\ and\ \citenamefont {Tkachev}}]{Levkov:2018kau}%
  \BibitemOpen
  \bibfield  {author} {\bibinfo {author} {\bibfnamefont {D.~G.}\ \bibnamefont
  {Levkov}}, \bibinfo {author} {\bibfnamefont {A.~G.}\ \bibnamefont {Panin}},\
  and\ \bibinfo {author} {\bibfnamefont {I.~I.}\ \bibnamefont {Tkachev}},\
  }\bibfield  {title} {\bibinfo {title} {{Gravitational Bose-Einstein
  condensation in the kinetic regime}},\ }\href
  {https://doi.org/10.1103/PhysRevLett.121.151301} {\bibfield  {journal}
  {\bibinfo  {journal} {Phys. Rev. Lett.}\ }\textbf {\bibinfo {volume} {121}},\
  \bibinfo {pages} {151301} (\bibinfo {year} {2018})},\ \Eprint
  {https://arxiv.org/abs/1804.05857} {arXiv:1804.05857 [astro-ph.CO]}
  \BibitemShut {NoStop}%
\bibitem [{\citenamefont {Chen}\ \emph {et~al.}(2021)\citenamefont {Chen},
  \citenamefont {Du}, \citenamefont {Lentz}, \citenamefont {Marsh},\ and\
  \citenamefont {Niemeyer}}]{Chen:2020cef}%
  \BibitemOpen
  \bibfield  {author} {\bibinfo {author} {\bibfnamefont {J.}~\bibnamefont
  {Chen}}, \bibinfo {author} {\bibfnamefont {X.}~\bibnamefont {Du}}, \bibinfo
  {author} {\bibfnamefont {E.~W.}\ \bibnamefont {Lentz}}, \bibinfo {author}
  {\bibfnamefont {D.~J.~E.}\ \bibnamefont {Marsh}},\ and\ \bibinfo {author}
  {\bibfnamefont {J.~C.}\ \bibnamefont {Niemeyer}},\ }\bibfield  {title}
  {\bibinfo {title} {{New insights into the formation and growth of boson stars
  in dark matter halos}},\ }\href {https://doi.org/10.1103/PhysRevD.104.083022}
  {\bibfield  {journal} {\bibinfo  {journal} {Phys. Rev. D}\ }\textbf {\bibinfo
  {volume} {104}},\ \bibinfo {pages} {083022} (\bibinfo {year} {2021})},\
  \Eprint {https://arxiv.org/abs/2011.01333} {arXiv:2011.01333 [astro-ph.CO]}
  \BibitemShut {NoStop}%
\bibitem [{\citenamefont {Bar-Or}\ \emph {et~al.}(2019)\citenamefont {Bar-Or},
  \citenamefont {Fouvry},\ and\ \citenamefont {Tremaine}}]{Bar-Or:2018pxz}%
  \BibitemOpen
  \bibfield  {author} {\bibinfo {author} {\bibfnamefont {B.}~\bibnamefont
  {Bar-Or}}, \bibinfo {author} {\bibfnamefont {J.-B.}\ \bibnamefont {Fouvry}},\
  and\ \bibinfo {author} {\bibfnamefont {S.}~\bibnamefont {Tremaine}},\
  }\bibfield  {title} {\bibinfo {title} {{Relaxation in a Fuzzy Dark Matter
  Halo}},\ }\href {https://doi.org/10.3847/1538-4357/aaf28c} {\bibfield
  {journal} {\bibinfo  {journal} {Astrophys. J.}\ }\textbf {\bibinfo {volume}
  {871}},\ \bibinfo {pages} {28} (\bibinfo {year} {2019})},\ \Eprint
  {https://arxiv.org/abs/1809.07673} {arXiv:1809.07673 [astro-ph.GA]}
  \BibitemShut {NoStop}%
\bibitem [{\citenamefont {Dutta~Chowdhury}\ \emph {et~al.}(2023)\citenamefont
  {Dutta~Chowdhury}, \citenamefont {van~den Bosch}, \citenamefont {van Dokkum},
  \citenamefont {Robles}, \citenamefont {Schive},\ and\ \citenamefont
  {Chiueh}}]{DuttaChowdhury:2023qxg}%
  \BibitemOpen
  \bibfield  {author} {\bibinfo {author} {\bibfnamefont {D.}~\bibnamefont
  {Dutta~Chowdhury}}, \bibinfo {author} {\bibfnamefont {F.~C.}\ \bibnamefont
  {van~den Bosch}}, \bibinfo {author} {\bibfnamefont {P.}~\bibnamefont {van
  Dokkum}}, \bibinfo {author} {\bibfnamefont {V.~H.}\ \bibnamefont {Robles}},
  \bibinfo {author} {\bibfnamefont {H.-Y.}\ \bibnamefont {Schive}},\ and\
  \bibinfo {author} {\bibfnamefont {T.}~\bibnamefont {Chiueh}},\ }\bibfield
  {title} {\bibinfo {title} {{On the Dynamical Heating of Dwarf Galaxies in a
  Fuzzy Dark Matter Halo}},\ }\href {https://doi.org/10.3847/1538-4357/acc73d}
  {\bibfield  {journal} {\bibinfo  {journal} {Astrophys. J.}\ }\textbf
  {\bibinfo {volume} {949}},\ \bibinfo {pages} {68} (\bibinfo {year} {2023})},\
  \Eprint {https://arxiv.org/abs/2303.08846} {arXiv:2303.08846 [astro-ph.GA]}
  \BibitemShut {NoStop}%
\bibitem [{\citenamefont {Yang}\ \emph {et~al.}(2025)\citenamefont {Yang},
  \citenamefont {Zhang}, \citenamefont {Bi},\ and\ \citenamefont
  {Yin}}]{Yang:2025bae}%
  \BibitemOpen
  \bibfield  {author} {\bibinfo {author} {\bibfnamefont {Y.-M.}\ \bibnamefont
  {Yang}}, \bibinfo {author} {\bibfnamefont {Z.-C.}\ \bibnamefont {Zhang}},
  \bibinfo {author} {\bibfnamefont {X.-J.}\ \bibnamefont {Bi}},\ and\ \bibinfo
  {author} {\bibfnamefont {P.-F.}\ \bibnamefont {Yin}},\ }\bibfield  {title}
  {\bibinfo {title} {{Tidal Suppression of Fuzzy Dark Matter Heating in Milky
  Way Satellite Galaxies}},\ }\href {https://doi.org/10.3847/2041-8213/adff81}
  {\bibfield  {journal} {\bibinfo  {journal} {Astrophys. J. Lett.}\ }\textbf
  {\bibinfo {volume} {990}},\ \bibinfo {pages} {L67} (\bibinfo {year}
  {2025})},\ \Eprint {https://arxiv.org/abs/2507.01686} {arXiv:2507.01686
  [astro-ph.CO]} \BibitemShut {NoStop}%
\bibitem [{\citenamefont {Yang}\ \emph {et~al.}(2026)\citenamefont {Yang},
  \citenamefont {Bi}, \citenamefont {Wang},\ and\ \citenamefont
  {Yin}}]{Yang:2026wdk}%
  \BibitemOpen
  \bibfield  {author} {\bibinfo {author} {\bibfnamefont {Y.-M.}\ \bibnamefont
  {Yang}}, \bibinfo {author} {\bibfnamefont {X.-J.}\ \bibnamefont {Bi}},
  \bibinfo {author} {\bibfnamefont {L.}~\bibnamefont {Wang}},\ and\ \bibinfo
  {author} {\bibfnamefont {P.-F.}\ \bibnamefont {Yin}},\ }\bibfield  {title}
  {\bibinfo {title} {{Collapse versus Disruption: The Fate of Compact Stellar
  Systems in Ultralight Dark Matter Halos}},\ }\href@noop {} {\  (\bibinfo
  {year} {2026})},\ \Eprint {https://arxiv.org/abs/2601.09403}
  {arXiv:2601.09403 [astro-ph.CO]} \BibitemShut {NoStop}%
\bibitem [{\citenamefont {Kaup}(1968)}]{Kaup:1968zz}%
  \BibitemOpen
  \bibfield  {author} {\bibinfo {author} {\bibfnamefont {D.~J.}\ \bibnamefont
  {Kaup}},\ }\bibfield  {title} {\bibinfo {title} {{Klein-Gordon Geon}},\
  }\href {https://doi.org/10.1103/PhysRev.172.1331} {\bibfield  {journal}
  {\bibinfo  {journal} {Phys. Rev.}\ }\textbf {\bibinfo {volume} {172}},\
  \bibinfo {pages} {1331} (\bibinfo {year} {1968})}\BibitemShut {NoStop}%
\bibitem [{\citenamefont {Ruffini}\ and\ \citenamefont
  {Bonazzola}(1969)}]{Ruffini:1969qy}%
  \BibitemOpen
  \bibfield  {author} {\bibinfo {author} {\bibfnamefont {R.}~\bibnamefont
  {Ruffini}}\ and\ \bibinfo {author} {\bibfnamefont {S.}~\bibnamefont
  {Bonazzola}},\ }\bibfield  {title} {\bibinfo {title} {{Systems of
  selfgravitating particles in general relativity and the concept of an
  equation of state}},\ }\href {https://doi.org/10.1103/PhysRev.187.1767}
  {\bibfield  {journal} {\bibinfo  {journal} {Phys. Rev.}\ }\textbf {\bibinfo
  {volume} {187}},\ \bibinfo {pages} {1767} (\bibinfo {year}
  {1969})}\BibitemShut {NoStop}%
\bibitem [{\citenamefont {Gleiser}(1988)}]{Gleiser:1988rq}%
  \BibitemOpen
  \bibfield  {author} {\bibinfo {author} {\bibfnamefont {M.}~\bibnamefont
  {Gleiser}},\ }\bibfield  {title} {\bibinfo {title} {{Stability of Boson
  Stars}},\ }\href {https://doi.org/10.1103/PhysRevD.38.2376} {\bibfield
  {journal} {\bibinfo  {journal} {Phys. Rev. D}\ }\textbf {\bibinfo {volume}
  {38}},\ \bibinfo {pages} {2376} (\bibinfo {year} {1988})},\ \bibinfo {note}
  {[Erratum: Phys.Rev.D 39, 1257 (1989)]}\BibitemShut {NoStop}%
\bibitem [{\citenamefont {Seidel}\ and\ \citenamefont
  {Suen}(1990)}]{Seidel:1990jh}%
  \BibitemOpen
  \bibfield  {author} {\bibinfo {author} {\bibfnamefont {E.}~\bibnamefont
  {Seidel}}\ and\ \bibinfo {author} {\bibfnamefont {W.-M.}\ \bibnamefont
  {Suen}},\ }\bibfield  {title} {\bibinfo {title} {{Dynamical Evolution of
  Boson Stars. 1. Perturbing the Ground State}},\ }\href
  {https://doi.org/10.1103/PhysRevD.42.384} {\bibfield  {journal} {\bibinfo
  {journal} {Phys. Rev. D}\ }\textbf {\bibinfo {volume} {42}},\ \bibinfo
  {pages} {384} (\bibinfo {year} {1990})}\BibitemShut {NoStop}%
\bibitem [{\citenamefont {Chavanis}(2011)}]{Chavanis:2011zi}%
  \BibitemOpen
  \bibfield  {author} {\bibinfo {author} {\bibfnamefont {P.-H.}\ \bibnamefont
  {Chavanis}},\ }\bibfield  {title} {\bibinfo {title} {{Mass-radius relation of
  Newtonian self-gravitating Bose-Einstein condensates with short-range
  interactions: I. Analytical results}},\ }\href
  {https://doi.org/10.1103/PhysRevD.84.043531} {\bibfield  {journal} {\bibinfo
  {journal} {Phys. Rev. D}\ }\textbf {\bibinfo {volume} {84}},\ \bibinfo
  {pages} {043531} (\bibinfo {year} {2011})},\ \Eprint
  {https://arxiv.org/abs/1103.2050} {arXiv:1103.2050 [astro-ph.CO]}
  \BibitemShut {NoStop}%
\bibitem [{\citenamefont {Harrison}\ \emph {et~al.}(2003)\citenamefont
  {Harrison}, \citenamefont {Moroz},\ and\ \citenamefont {Tod}}]{Harrison2003}%
  \BibitemOpen
  \bibfield  {author} {\bibinfo {author} {\bibfnamefont {R.}~\bibnamefont
  {Harrison}}, \bibinfo {author} {\bibfnamefont {I.}~\bibnamefont {Moroz}},\
  and\ \bibinfo {author} {\bibfnamefont {K.~P.}\ \bibnamefont {Tod}},\
  }\bibfield  {title} {\bibinfo {title} {A numerical study of the
  schr{\"o}dinger-newton equations},\ }\href
  {https://api.semanticscholar.org/CorpusID:118164871} {\bibfield  {journal}
  {\bibinfo  {journal} {Nonlinearity}\ }\textbf {\bibinfo {volume} {16}},\
  \bibinfo {pages} {101} (\bibinfo {year} {2003})},\ \Eprint
  {https://arxiv.org/abs/math-ph/0208045v1} {arXiv:math-ph/0208045v1 [math-ph]}
  \BibitemShut {NoStop}%
\bibitem [{\citenamefont {Guzman}\ and\ \citenamefont
  {Urena-Lopez}(2004)}]{Guzman:2004wj}%
  \BibitemOpen
  \bibfield  {author} {\bibinfo {author} {\bibfnamefont {F.~S.}\ \bibnamefont
  {Guzman}}\ and\ \bibinfo {author} {\bibfnamefont {L.~A.}\ \bibnamefont
  {Urena-Lopez}},\ }\bibfield  {title} {\bibinfo {title} {{Evolution of the
  Schrodinger-Newton system for a selfgravitating scalar field}},\ }\href
  {https://doi.org/10.1103/PhysRevD.69.124033} {\bibfield  {journal} {\bibinfo
  {journal} {Phys. Rev. D}\ }\textbf {\bibinfo {volume} {69}},\ \bibinfo
  {pages} {124033} (\bibinfo {year} {2004})},\ \Eprint
  {https://arxiv.org/abs/gr-qc/0404014} {arXiv:gr-qc/0404014} \BibitemShut
  {NoStop}%
\bibitem [{\citenamefont {Bernal}\ and\ \citenamefont
  {Guzman}(2006)}]{Bernal:2006it}%
  \BibitemOpen
  \bibfield  {author} {\bibinfo {author} {\bibfnamefont {A.}~\bibnamefont
  {Bernal}}\ and\ \bibinfo {author} {\bibfnamefont {F.~S.}\ \bibnamefont
  {Guzman}},\ }\bibfield  {title} {\bibinfo {title} {{Scalar Field Dark Matter:
  non-spherical collapse and late time behavior}},\ }\href
  {https://doi.org/10.1103/PhysRevD.74.063504} {\bibfield  {journal} {\bibinfo
  {journal} {Phys. Rev. D}\ }\textbf {\bibinfo {volume} {74}},\ \bibinfo
  {pages} {063504} (\bibinfo {year} {2006})},\ \Eprint
  {https://arxiv.org/abs/astro-ph/0608523} {arXiv:astro-ph/0608523}
  \BibitemShut {NoStop}%
\bibitem [{\citenamefont {Nambo}\ \emph {et~al.}(2024)\citenamefont {Nambo},
  \citenamefont {Diez-Tejedor}, \citenamefont {Roque},\ and\ \citenamefont
  {Sarbach}}]{Nambo:2024gvs}%
  \BibitemOpen
  \bibfield  {author} {\bibinfo {author} {\bibfnamefont {E.~C.}\ \bibnamefont
  {Nambo}}, \bibinfo {author} {\bibfnamefont {A.}~\bibnamefont {Diez-Tejedor}},
  \bibinfo {author} {\bibfnamefont {A.~A.}\ \bibnamefont {Roque}},\ and\
  \bibinfo {author} {\bibfnamefont {O.}~\bibnamefont {Sarbach}},\ }\bibfield
  {title} {\bibinfo {title} {{Linear stability of nonrelativistic
  self-interacting boson stars}},\ }\href
  {https://doi.org/10.1103/PhysRevD.109.104011} {\bibfield  {journal} {\bibinfo
   {journal} {Phys. Rev. D}\ }\textbf {\bibinfo {volume} {109}},\ \bibinfo
  {pages} {104011} (\bibinfo {year} {2024})},\ \Eprint
  {https://arxiv.org/abs/2402.07998} {arXiv:2402.07998 [gr-qc]} \BibitemShut
  {NoStop}%
\bibitem [{\citenamefont {Braaten}\ \emph
  {et~al.}(2016{\natexlab{a}})\citenamefont {Braaten}, \citenamefont
  {Mohapatra},\ and\ \citenamefont {Zhang}}]{Braaten:2015eeu}%
  \BibitemOpen
  \bibfield  {author} {\bibinfo {author} {\bibfnamefont {E.}~\bibnamefont
  {Braaten}}, \bibinfo {author} {\bibfnamefont {A.}~\bibnamefont {Mohapatra}},\
  and\ \bibinfo {author} {\bibfnamefont {H.}~\bibnamefont {Zhang}},\ }\bibfield
   {title} {\bibinfo {title} {{Dense Axion Stars}},\ }\href
  {https://doi.org/10.1103/PhysRevLett.117.121801} {\bibfield  {journal}
  {\bibinfo  {journal} {Phys. Rev. Lett.}\ }\textbf {\bibinfo {volume} {117}},\
  \bibinfo {pages} {121801} (\bibinfo {year} {2016}{\natexlab{a}})},\ \Eprint
  {https://arxiv.org/abs/1512.00108} {arXiv:1512.00108 [hep-ph]} \BibitemShut
  {NoStop}%
\bibitem [{\citenamefont {Zhang}(2019)}]{Zhang:2018slz}%
  \BibitemOpen
  \bibfield  {author} {\bibinfo {author} {\bibfnamefont {H.}~\bibnamefont
  {Zhang}},\ }\bibfield  {title} {\bibinfo {title} {{Axion Stars}},\ }\href
  {https://doi.org/10.3390/sym12010025} {\bibfield  {journal} {\bibinfo
  {journal} {Symmetry}\ }\textbf {\bibinfo {volume} {12}},\ \bibinfo {pages}
  {25} (\bibinfo {year} {2019})},\ \Eprint {https://arxiv.org/abs/1810.11473}
  {arXiv:1810.11473 [hep-ph]} \BibitemShut {NoStop}%
\bibitem [{\citenamefont {Braaten}\ and\ \citenamefont
  {Zhang}(2019)}]{Braaten:2019knj}%
  \BibitemOpen
  \bibfield  {author} {\bibinfo {author} {\bibfnamefont {E.}~\bibnamefont
  {Braaten}}\ and\ \bibinfo {author} {\bibfnamefont {H.}~\bibnamefont
  {Zhang}},\ }\bibfield  {title} {\bibinfo {title} {{Colloquium : The physics
  of axion stars}},\ }\href {https://doi.org/10.1103/RevModPhys.91.041002}
  {\bibfield  {journal} {\bibinfo  {journal} {Rev. Mod. Phys.}\ }\textbf
  {\bibinfo {volume} {91}},\ \bibinfo {pages} {041002} (\bibinfo {year}
  {2019})}\BibitemShut {NoStop}%
\bibitem [{\citenamefont {Detweiler}(1980)}]{Detweiler:1980uk}%
  \BibitemOpen
  \bibfield  {author} {\bibinfo {author} {\bibfnamefont {S.~L.}\ \bibnamefont
  {Detweiler}},\ }\bibfield  {title} {\bibinfo {title} {{KLEIN-GORDON EQUATION
  AND ROTATING BLACK HOLES}},\ }\href
  {https://doi.org/10.1103/PhysRevD.22.2323} {\bibfield  {journal} {\bibinfo
  {journal} {Phys. Rev. D}\ }\textbf {\bibinfo {volume} {22}},\ \bibinfo
  {pages} {2323} (\bibinfo {year} {1980})}\BibitemShut {NoStop}%
\bibitem [{\citenamefont {Arvanitaki}\ and\ \citenamefont
  {Dubovsky}(2011)}]{Arvanitaki:2010sy}%
  \BibitemOpen
  \bibfield  {author} {\bibinfo {author} {\bibfnamefont {A.}~\bibnamefont
  {Arvanitaki}}\ and\ \bibinfo {author} {\bibfnamefont {S.}~\bibnamefont
  {Dubovsky}},\ }\bibfield  {title} {\bibinfo {title} {{Exploring the String
  Axiverse with Precision Black Hole Physics}},\ }\href
  {https://doi.org/10.1103/PhysRevD.83.044026} {\bibfield  {journal} {\bibinfo
  {journal} {Phys. Rev. D}\ }\textbf {\bibinfo {volume} {83}},\ \bibinfo
  {pages} {044026} (\bibinfo {year} {2011})},\ \Eprint
  {https://arxiv.org/abs/1004.3558} {arXiv:1004.3558 [hep-th]} \BibitemShut
  {NoStop}%
\bibitem [{\citenamefont {Brito}\ \emph {et~al.}(2015)\citenamefont {Brito},
  \citenamefont {Cardoso},\ and\ \citenamefont {Pani}}]{Brito:2015oca}%
  \BibitemOpen
  \bibfield  {author} {\bibinfo {author} {\bibfnamefont {R.}~\bibnamefont
  {Brito}}, \bibinfo {author} {\bibfnamefont {V.}~\bibnamefont {Cardoso}},\
  and\ \bibinfo {author} {\bibfnamefont {P.}~\bibnamefont {Pani}},\ }\bibfield
  {title} {\bibinfo {title} {{Superradiance}: {New Frontiers in Black Hole
  Physics}},\ }\href {https://doi.org/10.1007/978-3-319-19000-6} {\bibfield
  {journal} {\bibinfo  {journal} {Lect. Notes Phys.}\ }\textbf {\bibinfo
  {volume} {906}},\ \bibinfo {pages} {pp.1} (\bibinfo {year} {2015})},\ \Eprint
  {https://arxiv.org/abs/1501.06570} {arXiv:1501.06570 [gr-qc]} \BibitemShut
  {NoStop}%
\bibitem [{\citenamefont {Bao}\ \emph {et~al.}(2022)\citenamefont {Bao},
  \citenamefont {Xu},\ and\ \citenamefont {Zhang}}]{Bao:2022hew}%
  \BibitemOpen
  \bibfield  {author} {\bibinfo {author} {\bibfnamefont {S.}~\bibnamefont
  {Bao}}, \bibinfo {author} {\bibfnamefont {Q.}~\bibnamefont {Xu}},\ and\
  \bibinfo {author} {\bibfnamefont {H.}~\bibnamefont {Zhang}},\ }\bibfield
  {title} {\bibinfo {title} {{Improved analytic solution of black hole
  superradiance}},\ }\href {https://doi.org/10.1103/PhysRevD.106.064016}
  {\bibfield  {journal} {\bibinfo  {journal} {Phys. Rev. D}\ }\textbf {\bibinfo
  {volume} {106}},\ \bibinfo {pages} {064016} (\bibinfo {year} {2022})},\
  \Eprint {https://arxiv.org/abs/2201.10941} {arXiv:2201.10941 [gr-qc]}
  \BibitemShut {NoStop}%
\bibitem [{\citenamefont {Bao}\ \emph {et~al.}(2023)\citenamefont {Bao},
  \citenamefont {Xu},\ and\ \citenamefont {Zhang}}]{Bao:2023xna}%
  \BibitemOpen
  \bibfield  {author} {\bibinfo {author} {\bibfnamefont {S.-S.}\ \bibnamefont
  {Bao}}, \bibinfo {author} {\bibfnamefont {Q.-X.}\ \bibnamefont {Xu}},\ and\
  \bibinfo {author} {\bibfnamefont {H.}~\bibnamefont {Zhang}},\ }\bibfield
  {title} {\bibinfo {title} {{Next-to-leading-order solution to Kerr-Newman
  black hole superradiance}},\ }\href
  {https://doi.org/10.1103/PhysRevD.107.064037} {\bibfield  {journal} {\bibinfo
   {journal} {Phys. Rev. D}\ }\textbf {\bibinfo {volume} {107}},\ \bibinfo
  {pages} {064037} (\bibinfo {year} {2023})},\ \Eprint
  {https://arxiv.org/abs/2301.05317} {arXiv:2301.05317 [gr-qc]} \BibitemShut
  {NoStop}%
\bibitem [{\citenamefont {Cardoso}\ \emph {et~al.}(2018)\citenamefont
  {Cardoso}, \citenamefont {Dias}, \citenamefont {Hartnett}, \citenamefont
  {Middleton}, \citenamefont {Pani},\ and\ \citenamefont
  {Santos}}]{Cardoso:2018tly}%
  \BibitemOpen
  \bibfield  {author} {\bibinfo {author} {\bibfnamefont {V.}~\bibnamefont
  {Cardoso}}, \bibinfo {author} {\bibfnamefont {{\'O}.~J.~C.}\ \bibnamefont
  {Dias}}, \bibinfo {author} {\bibfnamefont {G.~S.}\ \bibnamefont {Hartnett}},
  \bibinfo {author} {\bibfnamefont {M.}~\bibnamefont {Middleton}}, \bibinfo
  {author} {\bibfnamefont {P.}~\bibnamefont {Pani}},\ and\ \bibinfo {author}
  {\bibfnamefont {J.~E.}\ \bibnamefont {Santos}},\ }\bibfield  {title}
  {\bibinfo {title} {{Constraining the mass of dark photons and axion-like
  particles through black-hole superradiance}},\ }\href
  {https://doi.org/10.1088/1475-7516/2018/03/043} {\bibfield  {journal}
  {\bibinfo  {journal} {JCAP}\ }\textbf {\bibinfo {volume} {03}},\ \bibinfo
  {pages} {043}},\ \Eprint {https://arxiv.org/abs/1801.01420} {arXiv:1801.01420
  [gr-qc]} \BibitemShut {NoStop}%
\bibitem [{\citenamefont {Fernandez}\ \emph {et~al.}(2019)\citenamefont
  {Fernandez}, \citenamefont {Ghalsasi},\ and\ \citenamefont
  {Profumo}}]{Fernandez:2019qbj}%
  \BibitemOpen
  \bibfield  {author} {\bibinfo {author} {\bibfnamefont {N.}~\bibnamefont
  {Fernandez}}, \bibinfo {author} {\bibfnamefont {A.}~\bibnamefont
  {Ghalsasi}},\ and\ \bibinfo {author} {\bibfnamefont {S.}~\bibnamefont
  {Profumo}},\ }\bibfield  {title} {\bibinfo {title} {{Superradiance and the
  Spins of Black Holes from LIGO and X-ray binaries}},\ }\href@noop {} {\
  (\bibinfo {year} {2019})},\ \Eprint {https://arxiv.org/abs/1911.07862}
  {arXiv:1911.07862 [hep-ph]} \BibitemShut {NoStop}%
\bibitem [{\citenamefont {Ng}\ \emph {et~al.}(2021{\natexlab{a}})\citenamefont
  {Ng}, \citenamefont {Hannuksela}, \citenamefont {Vitale},\ and\ \citenamefont
  {Li}}]{Ng:2019jsx}%
  \BibitemOpen
  \bibfield  {author} {\bibinfo {author} {\bibfnamefont {K.~K.~Y.}\
  \bibnamefont {Ng}}, \bibinfo {author} {\bibfnamefont {O.~A.}\ \bibnamefont
  {Hannuksela}}, \bibinfo {author} {\bibfnamefont {S.}~\bibnamefont {Vitale}},\
  and\ \bibinfo {author} {\bibfnamefont {T.~G.~F.}\ \bibnamefont {Li}},\
  }\bibfield  {title} {\bibinfo {title} {{Searching for ultralight bosons
  within spin measurements of a population of binary black hole mergers}},\
  }\href {https://doi.org/10.1103/PhysRevD.103.063010} {\bibfield  {journal}
  {\bibinfo  {journal} {Phys. Rev. D}\ }\textbf {\bibinfo {volume} {103}},\
  \bibinfo {pages} {063010} (\bibinfo {year} {2021}{\natexlab{a}})},\ \Eprint
  {https://arxiv.org/abs/1908.02312} {arXiv:1908.02312 [gr-qc]} \BibitemShut
  {NoStop}%
\bibitem [{\citenamefont {Ng}\ \emph {et~al.}(2021{\natexlab{b}})\citenamefont
  {Ng}, \citenamefont {Vitale}, \citenamefont {Hannuksela},\ and\ \citenamefont
  {Li}}]{Ng:2020ruv}%
  \BibitemOpen
  \bibfield  {author} {\bibinfo {author} {\bibfnamefont {K.~K.~Y.}\
  \bibnamefont {Ng}}, \bibinfo {author} {\bibfnamefont {S.}~\bibnamefont
  {Vitale}}, \bibinfo {author} {\bibfnamefont {O.~A.}\ \bibnamefont
  {Hannuksela}},\ and\ \bibinfo {author} {\bibfnamefont {T.~G.~F.}\
  \bibnamefont {Li}},\ }\bibfield  {title} {\bibinfo {title} {{Constraints on
  Ultralight Scalar Bosons within Black Hole Spin Measurements from the
  LIGO-Virgo GWTC-2}},\ }\href {https://doi.org/10.1103/PhysRevLett.126.151102}
  {\bibfield  {journal} {\bibinfo  {journal} {Phys. Rev. Lett.}\ }\textbf
  {\bibinfo {volume} {126}},\ \bibinfo {pages} {151102} (\bibinfo {year}
  {2021}{\natexlab{b}})},\ \Eprint {https://arxiv.org/abs/2011.06010}
  {arXiv:2011.06010 [gr-qc]} \BibitemShut {NoStop}%
\bibitem [{\citenamefont {Cheng}\ \emph {et~al.}(2023)\citenamefont {Cheng},
  \citenamefont {Zhang},\ and\ \citenamefont {Bao}}]{Cheng:2022jsw}%
  \BibitemOpen
  \bibfield  {author} {\bibinfo {author} {\bibfnamefont {L.-d.}\ \bibnamefont
  {Cheng}}, \bibinfo {author} {\bibfnamefont {H.}~\bibnamefont {Zhang}},\ and\
  \bibinfo {author} {\bibfnamefont {S.-s.}\ \bibnamefont {Bao}},\ }\bibfield
  {title} {\bibinfo {title} {{Constraints on an axionlike particle from black
  hole spin superradiance}},\ }\href
  {https://doi.org/10.1103/PhysRevD.107.063021} {\bibfield  {journal} {\bibinfo
   {journal} {Phys. Rev. D}\ }\textbf {\bibinfo {volume} {107}},\ \bibinfo
  {pages} {063021} (\bibinfo {year} {2023})},\ \Eprint
  {https://arxiv.org/abs/2201.11338} {arXiv:2201.11338 [gr-qc]} \BibitemShut
  {NoStop}%
\bibitem [{\citenamefont {Arvanitaki}\ \emph {et~al.}(2015)\citenamefont
  {Arvanitaki}, \citenamefont {Baryakhtar},\ and\ \citenamefont
  {Huang}}]{Arvanitaki:2014wva}%
  \BibitemOpen
  \bibfield  {author} {\bibinfo {author} {\bibfnamefont {A.}~\bibnamefont
  {Arvanitaki}}, \bibinfo {author} {\bibfnamefont {M.}~\bibnamefont
  {Baryakhtar}},\ and\ \bibinfo {author} {\bibfnamefont {X.}~\bibnamefont
  {Huang}},\ }\bibfield  {title} {\bibinfo {title} {{Discovering the QCD Axion
  with Black Holes and Gravitational Waves}},\ }\href
  {https://doi.org/10.1103/PhysRevD.91.084011} {\bibfield  {journal} {\bibinfo
  {journal} {Phys. Rev. D}\ }\textbf {\bibinfo {volume} {91}},\ \bibinfo
  {pages} {084011} (\bibinfo {year} {2015})},\ \Eprint
  {https://arxiv.org/abs/1411.2263} {arXiv:1411.2263 [hep-ph]} \BibitemShut
  {NoStop}%
\bibitem [{\citenamefont {Yoshino}\ and\ \citenamefont
  {Kodama}(2014)}]{Yoshino:2013ofa}%
  \BibitemOpen
  \bibfield  {author} {\bibinfo {author} {\bibfnamefont {H.}~\bibnamefont
  {Yoshino}}\ and\ \bibinfo {author} {\bibfnamefont {H.}~\bibnamefont
  {Kodama}},\ }\bibfield  {title} {\bibinfo {title} {{Gravitational radiation
  from an axion cloud around a black hole: Superradiant phase}},\ }\href
  {https://doi.org/10.1093/ptep/ptu029} {\bibfield  {journal} {\bibinfo
  {journal} {PTEP}\ }\textbf {\bibinfo {volume} {2014}},\ \bibinfo {pages}
  {043E02} (\bibinfo {year} {2014})},\ \Eprint
  {https://arxiv.org/abs/1312.2326} {arXiv:1312.2326 [gr-qc]} \BibitemShut
  {NoStop}%
\bibitem [{\citenamefont {Yoshino}\ and\ \citenamefont
  {Kodama}(2015)}]{Yoshino:2014wwa}%
  \BibitemOpen
  \bibfield  {author} {\bibinfo {author} {\bibfnamefont {H.}~\bibnamefont
  {Yoshino}}\ and\ \bibinfo {author} {\bibfnamefont {H.}~\bibnamefont
  {Kodama}},\ }\bibfield  {title} {\bibinfo {title} {{Probing the string
  axiverse by gravitational waves from Cygnus X-1}},\ }\href
  {https://doi.org/10.1093/ptep/ptv067} {\bibfield  {journal} {\bibinfo
  {journal} {PTEP}\ }\textbf {\bibinfo {volume} {2015}},\ \bibinfo {pages}
  {061E01} (\bibinfo {year} {2015})},\ \Eprint
  {https://arxiv.org/abs/1407.2030} {arXiv:1407.2030 [gr-qc]} \BibitemShut
  {NoStop}%
\bibitem [{\citenamefont {Brito}\ \emph
  {et~al.}(2017{\natexlab{a}})\citenamefont {Brito}, \citenamefont {Ghosh},
  \citenamefont {Barausse}, \citenamefont {Berti}, \citenamefont {Cardoso},
  \citenamefont {Dvorkin}, \citenamefont {Klein},\ and\ \citenamefont
  {Pani}}]{Brito:2017wnc}%
  \BibitemOpen
  \bibfield  {author} {\bibinfo {author} {\bibfnamefont {R.}~\bibnamefont
  {Brito}}, \bibinfo {author} {\bibfnamefont {S.}~\bibnamefont {Ghosh}},
  \bibinfo {author} {\bibfnamefont {E.}~\bibnamefont {Barausse}}, \bibinfo
  {author} {\bibfnamefont {E.}~\bibnamefont {Berti}}, \bibinfo {author}
  {\bibfnamefont {V.}~\bibnamefont {Cardoso}}, \bibinfo {author} {\bibfnamefont
  {I.}~\bibnamefont {Dvorkin}}, \bibinfo {author} {\bibfnamefont
  {A.}~\bibnamefont {Klein}},\ and\ \bibinfo {author} {\bibfnamefont
  {P.}~\bibnamefont {Pani}},\ }\bibfield  {title} {\bibinfo {title}
  {{Stochastic and resolvable gravitational waves from ultralight bosons}},\
  }\href {https://doi.org/10.1103/PhysRevLett.119.131101} {\bibfield  {journal}
  {\bibinfo  {journal} {Phys. Rev. Lett.}\ }\textbf {\bibinfo {volume} {119}},\
  \bibinfo {pages} {131101} (\bibinfo {year} {2017}{\natexlab{a}})},\ \Eprint
  {https://arxiv.org/abs/1706.05097} {arXiv:1706.05097 [gr-qc]} \BibitemShut
  {NoStop}%
\bibitem [{\citenamefont {Brito}\ \emph
  {et~al.}(2017{\natexlab{b}})\citenamefont {Brito}, \citenamefont {Ghosh},
  \citenamefont {Barausse}, \citenamefont {Berti}, \citenamefont {Cardoso},
  \citenamefont {Dvorkin}, \citenamefont {Klein},\ and\ \citenamefont
  {Pani}}]{Brito:2017zvb}%
  \BibitemOpen
  \bibfield  {author} {\bibinfo {author} {\bibfnamefont {R.}~\bibnamefont
  {Brito}}, \bibinfo {author} {\bibfnamefont {S.}~\bibnamefont {Ghosh}},
  \bibinfo {author} {\bibfnamefont {E.}~\bibnamefont {Barausse}}, \bibinfo
  {author} {\bibfnamefont {E.}~\bibnamefont {Berti}}, \bibinfo {author}
  {\bibfnamefont {V.}~\bibnamefont {Cardoso}}, \bibinfo {author} {\bibfnamefont
  {I.}~\bibnamefont {Dvorkin}}, \bibinfo {author} {\bibfnamefont
  {A.}~\bibnamefont {Klein}},\ and\ \bibinfo {author} {\bibfnamefont
  {P.}~\bibnamefont {Pani}},\ }\bibfield  {title} {\bibinfo {title}
  {{Gravitational wave searches for ultralight bosons with LIGO and LISA}},\
  }\href {https://doi.org/10.1103/PhysRevD.96.064050} {\bibfield  {journal}
  {\bibinfo  {journal} {Phys. Rev. D}\ }\textbf {\bibinfo {volume} {96}},\
  \bibinfo {pages} {064050} (\bibinfo {year} {2017}{\natexlab{b}})},\ \Eprint
  {https://arxiv.org/abs/1706.06311} {arXiv:1706.06311 [gr-qc]} \BibitemShut
  {NoStop}%
\bibitem [{\citenamefont {Guo}\ \emph {et~al.}(2023)\citenamefont {Guo},
  \citenamefont {Bao},\ and\ \citenamefont {Zhang}}]{Guo:2022mpr}%
  \BibitemOpen
  \bibfield  {author} {\bibinfo {author} {\bibfnamefont {Y.-d.}\ \bibnamefont
  {Guo}}, \bibinfo {author} {\bibfnamefont {S.-s.}\ \bibnamefont {Bao}},\ and\
  \bibinfo {author} {\bibfnamefont {H.}~\bibnamefont {Zhang}},\ }\bibfield
  {title} {\bibinfo {title} {{Subdominant modes of the scalar superradiant
  instability and gravitational wave beats}},\ }\href
  {https://doi.org/10.1103/PhysRevD.107.075009} {\bibfield  {journal} {\bibinfo
   {journal} {Phys. Rev. D}\ }\textbf {\bibinfo {volume} {107}},\ \bibinfo
  {pages} {075009} (\bibinfo {year} {2023})},\ \Eprint
  {https://arxiv.org/abs/2212.07186} {arXiv:2212.07186 [gr-qc]} \BibitemShut
  {NoStop}%
\bibitem [{\citenamefont {Yang}\ and\ \citenamefont
  {Huang}(2023)}]{Yang:2023vwm}%
  \BibitemOpen
  \bibfield  {author} {\bibinfo {author} {\bibfnamefont {J.}~\bibnamefont
  {Yang}}\ and\ \bibinfo {author} {\bibfnamefont {F.~P.}\ \bibnamefont
  {Huang}},\ }\bibfield  {title} {\bibinfo {title} {{Gravitational waves from
  axions annihilation through quantum field theory}},\ }\href
  {https://doi.org/10.1103/PhysRevD.108.103002} {\bibfield  {journal} {\bibinfo
   {journal} {Phys. Rev. D}\ }\textbf {\bibinfo {volume} {108}},\ \bibinfo
  {pages} {103002} (\bibinfo {year} {2023})},\ \Eprint
  {https://arxiv.org/abs/2306.12375} {arXiv:2306.12375 [hep-ph]} \BibitemShut
  {NoStop}%
\bibitem [{\citenamefont {Guo}\ \emph {et~al.}(2024)\citenamefont {Guo},
  \citenamefont {Jia}, \citenamefont {Bao}, \citenamefont {Zhang},\ and\
  \citenamefont {Zhang}}]{Guo:2024dqd}%
  \BibitemOpen
  \bibfield  {author} {\bibinfo {author} {\bibfnamefont {Y.-D.}\ \bibnamefont
  {Guo}}, \bibinfo {author} {\bibfnamefont {N.}~\bibnamefont {Jia}}, \bibinfo
  {author} {\bibfnamefont {S.-S.}\ \bibnamefont {Bao}}, \bibinfo {author}
  {\bibfnamefont {H.}~\bibnamefont {Zhang}},\ and\ \bibinfo {author}
  {\bibfnamefont {X.}~\bibnamefont {Zhang}},\ }\bibfield  {title} {\bibinfo
  {title} {{Evolution and detection of vector superradiant instabilities}},\
  }\href {https://doi.org/10.1103/PhysRevD.110.083029} {\bibfield  {journal}
  {\bibinfo  {journal} {Phys. Rev. D}\ }\textbf {\bibinfo {volume} {110}},\
  \bibinfo {pages} {083029} (\bibinfo {year} {2024})},\ \Eprint
  {https://arxiv.org/abs/2407.00767} {arXiv:2407.00767 [gr-qc]} \BibitemShut
  {NoStop}%
\bibitem [{\citenamefont {Omiya}\ \emph {et~al.}(2024)\citenamefont {Omiya},
  \citenamefont {Takahashi}, \citenamefont {Tanaka},\ and\ \citenamefont
  {Yoshino}}]{Omiya:2024xlz}%
  \BibitemOpen
  \bibfield  {author} {\bibinfo {author} {\bibfnamefont {H.}~\bibnamefont
  {Omiya}}, \bibinfo {author} {\bibfnamefont {T.}~\bibnamefont {Takahashi}},
  \bibinfo {author} {\bibfnamefont {T.}~\bibnamefont {Tanaka}},\ and\ \bibinfo
  {author} {\bibfnamefont {H.}~\bibnamefont {Yoshino}},\ }\bibfield  {title}
  {\bibinfo {title} {{Deci-Hz gravitational waves from the self-interacting
  axion cloud around a rotating stellar mass black hole}},\ }\href
  {https://doi.org/10.1103/PhysRevD.110.044002} {\bibfield  {journal} {\bibinfo
   {journal} {Phys. Rev. D}\ }\textbf {\bibinfo {volume} {110}},\ \bibinfo
  {pages} {044002} (\bibinfo {year} {2024})},\ \Eprint
  {https://arxiv.org/abs/2404.16265} {arXiv:2404.16265 [gr-qc]} \BibitemShut
  {NoStop}%
\bibitem [{\citenamefont {Baumann}\ \emph {et~al.}(2019)\citenamefont
  {Baumann}, \citenamefont {Chia},\ and\ \citenamefont
  {Porto}}]{Baumann:2018vus}%
  \BibitemOpen
  \bibfield  {author} {\bibinfo {author} {\bibfnamefont {D.}~\bibnamefont
  {Baumann}}, \bibinfo {author} {\bibfnamefont {H.~S.}\ \bibnamefont {Chia}},\
  and\ \bibinfo {author} {\bibfnamefont {R.~A.}\ \bibnamefont {Porto}},\
  }\bibfield  {title} {\bibinfo {title} {{Probing Ultralight Bosons with Binary
  Black Holes}},\ }\href {https://doi.org/10.1103/PhysRevD.99.044001}
  {\bibfield  {journal} {\bibinfo  {journal} {Phys. Rev. D}\ }\textbf {\bibinfo
  {volume} {99}},\ \bibinfo {pages} {044001} (\bibinfo {year} {2019})},\
  \Eprint {https://arxiv.org/abs/1804.03208} {arXiv:1804.03208 [gr-qc]}
  \BibitemShut {NoStop}%
\bibitem [{\citenamefont {Berti}\ \emph {et~al.}(2019)\citenamefont {Berti},
  \citenamefont {Brito}, \citenamefont {Macedo}, \citenamefont {Raposo},\ and\
  \citenamefont {Rosa}}]{Berti:2019wnn}%
  \BibitemOpen
  \bibfield  {author} {\bibinfo {author} {\bibfnamefont {E.}~\bibnamefont
  {Berti}}, \bibinfo {author} {\bibfnamefont {R.}~\bibnamefont {Brito}},
  \bibinfo {author} {\bibfnamefont {C.~F.~B.}\ \bibnamefont {Macedo}}, \bibinfo
  {author} {\bibfnamefont {G.}~\bibnamefont {Raposo}},\ and\ \bibinfo {author}
  {\bibfnamefont {J.~L.}\ \bibnamefont {Rosa}},\ }\bibfield  {title} {\bibinfo
  {title} {{Ultralight boson cloud depletion in binary systems}},\ }\href
  {https://doi.org/10.1103/PhysRevD.99.104039} {\bibfield  {journal} {\bibinfo
  {journal} {Phys. Rev. D}\ }\textbf {\bibinfo {volume} {99}},\ \bibinfo
  {pages} {104039} (\bibinfo {year} {2019})},\ \Eprint
  {https://arxiv.org/abs/1904.03131} {arXiv:1904.03131 [gr-qc]} \BibitemShut
  {NoStop}%
\bibitem [{\citenamefont {Baumann}\ \emph {et~al.}(2020)\citenamefont
  {Baumann}, \citenamefont {Chia}, \citenamefont {Porto},\ and\ \citenamefont
  {Stout}}]{Baumann:2019ztm}%
  \BibitemOpen
  \bibfield  {author} {\bibinfo {author} {\bibfnamefont {D.}~\bibnamefont
  {Baumann}}, \bibinfo {author} {\bibfnamefont {H.~S.}\ \bibnamefont {Chia}},
  \bibinfo {author} {\bibfnamefont {R.~A.}\ \bibnamefont {Porto}},\ and\
  \bibinfo {author} {\bibfnamefont {J.}~\bibnamefont {Stout}},\ }\bibfield
  {title} {\bibinfo {title} {{Gravitational Collider Physics}},\ }\href
  {https://doi.org/10.1103/PhysRevD.101.083019} {\bibfield  {journal} {\bibinfo
   {journal} {Phys. Rev. D}\ }\textbf {\bibinfo {volume} {101}},\ \bibinfo
  {pages} {083019} (\bibinfo {year} {2020})},\ \Eprint
  {https://arxiv.org/abs/1912.04932} {arXiv:1912.04932 [gr-qc]} \BibitemShut
  {NoStop}%
\bibitem [{\citenamefont {Cardoso}\ \emph {et~al.}(2020)\citenamefont
  {Cardoso}, \citenamefont {Duque},\ and\ \citenamefont
  {Ikeda}}]{Cardoso:2020hca}%
  \BibitemOpen
  \bibfield  {author} {\bibinfo {author} {\bibfnamefont {V.}~\bibnamefont
  {Cardoso}}, \bibinfo {author} {\bibfnamefont {F.}~\bibnamefont {Duque}},\
  and\ \bibinfo {author} {\bibfnamefont {T.}~\bibnamefont {Ikeda}},\ }\bibfield
   {title} {\bibinfo {title} {{Tidal effects and disruption in superradiant
  clouds: a numerical investigation}},\ }\href
  {https://doi.org/10.1103/PhysRevD.101.064054} {\bibfield  {journal} {\bibinfo
   {journal} {Phys. Rev. D}\ }\textbf {\bibinfo {volume} {101}},\ \bibinfo
  {pages} {064054} (\bibinfo {year} {2020})},\ \Eprint
  {https://arxiv.org/abs/2001.01729} {arXiv:2001.01729 [gr-qc]} \BibitemShut
  {NoStop}%
\bibitem [{\citenamefont {Takahashi}\ \emph {et~al.}(2022)\citenamefont
  {Takahashi}, \citenamefont {Omiya},\ and\ \citenamefont
  {Tanaka}}]{Takahashi:2021yhy}%
  \BibitemOpen
  \bibfield  {author} {\bibinfo {author} {\bibfnamefont {T.}~\bibnamefont
  {Takahashi}}, \bibinfo {author} {\bibfnamefont {H.}~\bibnamefont {Omiya}},\
  and\ \bibinfo {author} {\bibfnamefont {T.}~\bibnamefont {Tanaka}},\
  }\bibfield  {title} {\bibinfo {title} {{Axion cloud evaporation during
  inspiral of black hole binaries: The effects of backreaction and
  radiation}},\ }\href {https://doi.org/10.1093/ptep/ptac044} {\bibfield
  {journal} {\bibinfo  {journal} {PTEP}\ }\textbf {\bibinfo {volume} {2022}},\
  \bibinfo {pages} {043E01} (\bibinfo {year} {2022})},\ \Eprint
  {https://arxiv.org/abs/2112.05774} {arXiv:2112.05774 [gr-qc]} \BibitemShut
  {NoStop}%
\bibitem [{\citenamefont {Baumann}\ \emph {et~al.}(2022)\citenamefont
  {Baumann}, \citenamefont {Bertone}, \citenamefont {Stout},\ and\
  \citenamefont {Tomaselli}}]{Baumann:2022pkl}%
  \BibitemOpen
  \bibfield  {author} {\bibinfo {author} {\bibfnamefont {D.}~\bibnamefont
  {Baumann}}, \bibinfo {author} {\bibfnamefont {G.}~\bibnamefont {Bertone}},
  \bibinfo {author} {\bibfnamefont {J.}~\bibnamefont {Stout}},\ and\ \bibinfo
  {author} {\bibfnamefont {G.~M.}\ \bibnamefont {Tomaselli}},\ }\bibfield
  {title} {\bibinfo {title} {{Sharp Signals of Boson Clouds in Black Hole
  Binary Inspirals}},\ }\href {https://doi.org/10.1103/PhysRevLett.128.221102}
  {\bibfield  {journal} {\bibinfo  {journal} {Phys. Rev. Lett.}\ }\textbf
  {\bibinfo {volume} {128}},\ \bibinfo {pages} {221102} (\bibinfo {year}
  {2022})},\ \Eprint {https://arxiv.org/abs/2206.01212} {arXiv:2206.01212
  [gr-qc]} \BibitemShut {NoStop}%
\bibitem [{\citenamefont {Tong}\ \emph {et~al.}(2022)\citenamefont {Tong},
  \citenamefont {Wang},\ and\ \citenamefont {Zhu}}]{Tong:2022bbl}%
  \BibitemOpen
  \bibfield  {author} {\bibinfo {author} {\bibfnamefont {X.}~\bibnamefont
  {Tong}}, \bibinfo {author} {\bibfnamefont {Y.}~\bibnamefont {Wang}},\ and\
  \bibinfo {author} {\bibfnamefont {H.-Y.}\ \bibnamefont {Zhu}},\ }\bibfield
  {title} {\bibinfo {title} {{Termination of superradiance from a binary
  companion}},\ }\href {https://doi.org/10.1103/PhysRevD.106.043002} {\bibfield
   {journal} {\bibinfo  {journal} {Phys. Rev. D}\ }\textbf {\bibinfo {volume}
  {106}},\ \bibinfo {pages} {043002} (\bibinfo {year} {2022})},\ \Eprint
  {https://arxiv.org/abs/2205.10527} {arXiv:2205.10527 [gr-qc]} \BibitemShut
  {NoStop}%
\bibitem [{\citenamefont {Takahashi}\ \emph {et~al.}(2023)\citenamefont
  {Takahashi}, \citenamefont {Omiya},\ and\ \citenamefont
  {Tanaka}}]{Takahashi:2023flk}%
  \BibitemOpen
  \bibfield  {author} {\bibinfo {author} {\bibfnamefont {T.}~\bibnamefont
  {Takahashi}}, \bibinfo {author} {\bibfnamefont {H.}~\bibnamefont {Omiya}},\
  and\ \bibinfo {author} {\bibfnamefont {T.}~\bibnamefont {Tanaka}},\
  }\bibfield  {title} {\bibinfo {title} {{Evolution of binary systems
  accompanying axion clouds in extreme mass ratio inspirals}},\ }\href
  {https://doi.org/10.1103/PhysRevD.107.103020} {\bibfield  {journal} {\bibinfo
   {journal} {Phys. Rev. D}\ }\textbf {\bibinfo {volume} {107}},\ \bibinfo
  {pages} {103020} (\bibinfo {year} {2023})},\ \Eprint
  {https://arxiv.org/abs/2301.13213} {arXiv:2301.13213 [gr-qc]} \BibitemShut
  {NoStop}%
\bibitem [{\citenamefont {Xu}\ \emph {et~al.}(2026{\natexlab{a}})\citenamefont
  {Xu}, \citenamefont {Brito}, \citenamefont {Della~Monica}, \citenamefont
  {Vicente},\ and\ \citenamefont {Yuan}}]{Xu:2026cky}%
  \BibitemOpen
  \bibfield  {author} {\bibinfo {author} {\bibfnamefont {Q.-X.}\ \bibnamefont
  {Xu}}, \bibinfo {author} {\bibfnamefont {R.}~\bibnamefont {Brito}}, \bibinfo
  {author} {\bibfnamefont {R.}~\bibnamefont {Della~Monica}}, \bibinfo {author}
  {\bibfnamefont {R.}~\bibnamefont {Vicente}},\ and\ \bibinfo {author}
  {\bibfnamefont {C.}~\bibnamefont {Yuan}},\ }\bibfield  {title} {\bibinfo
  {title} {{Resonances as signatures of scalar clouds in eccentric
  extreme-mass-ratio inspirals}},\ }\href@noop {} {\  (\bibinfo {year}
  {2026}{\natexlab{a}})},\ \Eprint {https://arxiv.org/abs/2605.03756}
  {arXiv:2605.03756 [gr-qc]} \BibitemShut {NoStop}%
\bibitem [{\citenamefont {Xu}\ \emph {et~al.}(2026{\natexlab{b}})\citenamefont
  {Xu}, \citenamefont {Brito}, \citenamefont {Della~Monica}, \citenamefont
  {Vicente},\ and\ \citenamefont {Yuan}}]{Xu:2026aic}%
  \BibitemOpen
  \bibfield  {author} {\bibinfo {author} {\bibfnamefont {Q.-X.}\ \bibnamefont
  {Xu}}, \bibinfo {author} {\bibfnamefont {R.}~\bibnamefont {Brito}}, \bibinfo
  {author} {\bibfnamefont {R.}~\bibnamefont {Della~Monica}}, \bibinfo {author}
  {\bibfnamefont {R.}~\bibnamefont {Vicente}},\ and\ \bibinfo {author}
  {\bibfnamefont {C.}~\bibnamefont {Yuan}},\ }\bibfield  {title} {\bibinfo
  {title} {{Relativistic effects in extreme-mass-ratio inspirals within scalar
  clouds: Eccentric and inclined orbits}},\ }\href@noop {} {\  (\bibinfo {year}
  {2026}{\natexlab{b}})},\ \Eprint {https://arxiv.org/abs/2606.21439}
  {arXiv:2606.21439 [gr-qc]} \BibitemShut {NoStop}%
\bibitem [{\citenamefont {Yoshino}\ and\ \citenamefont
  {Kodama}(2012)}]{Yoshino:2012kn}%
  \BibitemOpen
  \bibfield  {author} {\bibinfo {author} {\bibfnamefont {H.}~\bibnamefont
  {Yoshino}}\ and\ \bibinfo {author} {\bibfnamefont {H.}~\bibnamefont
  {Kodama}},\ }\bibfield  {title} {\bibinfo {title} {{Bosenova collapse of
  axion cloud around a rotating black hole}},\ }\href
  {https://doi.org/10.1143/PTP.128.153} {\bibfield  {journal} {\bibinfo
  {journal} {Prog. Theor. Phys.}\ }\textbf {\bibinfo {volume} {128}},\ \bibinfo
  {pages} {153} (\bibinfo {year} {2012})},\ \Eprint
  {https://arxiv.org/abs/1203.5070} {arXiv:1203.5070 [gr-qc]} \BibitemShut
  {NoStop}%
\bibitem [{\citenamefont {Baryakhtar}\ \emph {et~al.}(2021)\citenamefont
  {Baryakhtar}, \citenamefont {Galanis}, \citenamefont {Lasenby},\ and\
  \citenamefont {Simon}}]{Baryakhtar:2020gao}%
  \BibitemOpen
  \bibfield  {author} {\bibinfo {author} {\bibfnamefont {M.}~\bibnamefont
  {Baryakhtar}}, \bibinfo {author} {\bibfnamefont {M.}~\bibnamefont {Galanis}},
  \bibinfo {author} {\bibfnamefont {R.}~\bibnamefont {Lasenby}},\ and\ \bibinfo
  {author} {\bibfnamefont {O.}~\bibnamefont {Simon}},\ }\bibfield  {title}
  {\bibinfo {title} {{Black hole superradiance of self-interacting scalar
  fields}},\ }\href {https://doi.org/10.1103/PhysRevD.103.095019} {\bibfield
  {journal} {\bibinfo  {journal} {Phys. Rev. D}\ }\textbf {\bibinfo {volume}
  {103}},\ \bibinfo {pages} {095019} (\bibinfo {year} {2021})},\ \Eprint
  {https://arxiv.org/abs/2011.11646} {arXiv:2011.11646 [hep-ph]} \BibitemShut
  {NoStop}%
\bibitem [{\citenamefont {Omiya}\ \emph {et~al.}(2022)\citenamefont {Omiya},
  \citenamefont {Takahashi},\ and\ \citenamefont {Tanaka}}]{Omiya:2022mwv}%
  \BibitemOpen
  \bibfield  {author} {\bibinfo {author} {\bibfnamefont {H.}~\bibnamefont
  {Omiya}}, \bibinfo {author} {\bibfnamefont {T.}~\bibnamefont {Takahashi}},\
  and\ \bibinfo {author} {\bibfnamefont {T.}~\bibnamefont {Tanaka}},\
  }\bibfield  {title} {\bibinfo {title} {{Adiabatic evolution of the
  self-interacting axion field around rotating black holes}},\ }\href
  {https://doi.org/10.1093/ptep/ptac058} {\bibfield  {journal} {\bibinfo
  {journal} {PTEP}\ }\textbf {\bibinfo {volume} {2022}},\ \bibinfo {pages}
  {043E03} (\bibinfo {year} {2022})},\ \Eprint
  {https://arxiv.org/abs/2201.04382} {arXiv:2201.04382 [gr-qc]} \BibitemShut
  {NoStop}%
\bibitem [{\citenamefont {Omiya}\ \emph {et~al.}(2023)\citenamefont {Omiya},
  \citenamefont {Takahashi}, \citenamefont {Tanaka},\ and\ \citenamefont
  {Yoshino}}]{Omiya:2022gwu}%
  \BibitemOpen
  \bibfield  {author} {\bibinfo {author} {\bibfnamefont {H.}~\bibnamefont
  {Omiya}}, \bibinfo {author} {\bibfnamefont {T.}~\bibnamefont {Takahashi}},
  \bibinfo {author} {\bibfnamefont {T.}~\bibnamefont {Tanaka}},\ and\ \bibinfo
  {author} {\bibfnamefont {H.}~\bibnamefont {Yoshino}},\ }\bibfield  {title}
  {\bibinfo {title} {{Impact of multiple modes on the evolution of
  self-interacting axion condensate around rotating black holes}},\ }\href
  {https://doi.org/10.1088/1475-7516/2023/06/016} {\bibfield  {journal}
  {\bibinfo  {journal} {JCAP}\ }\textbf {\bibinfo {volume} {06}},\ \bibinfo
  {pages} {016}},\ \Eprint {https://arxiv.org/abs/2211.01949} {arXiv:2211.01949
  [gr-qc]} \BibitemShut {NoStop}%
\bibitem [{\citenamefont {Silveira}\ and\ \citenamefont
  {de~Sousa}(1995)}]{Silveira:1995dh}%
  \BibitemOpen
  \bibfield  {author} {\bibinfo {author} {\bibfnamefont {V.}~\bibnamefont
  {Silveira}}\ and\ \bibinfo {author} {\bibfnamefont {C.~M.~G.}\ \bibnamefont
  {de~Sousa}},\ }\bibfield  {title} {\bibinfo {title} {{Boson star rotation: A
  Newtonian approximation}},\ }\href {https://doi.org/10.1103/PhysRevD.52.5724}
  {\bibfield  {journal} {\bibinfo  {journal} {Phys. Rev. D}\ }\textbf {\bibinfo
  {volume} {52}},\ \bibinfo {pages} {5724} (\bibinfo {year} {1995})},\ \Eprint
  {https://arxiv.org/abs/astro-ph/9508034} {arXiv:astro-ph/9508034}
  \BibitemShut {NoStop}%
\bibitem [{\citenamefont {Schupp}\ and\ \citenamefont {van~der
  Bij}(1996)}]{Schupp:1995dy}%
  \BibitemOpen
  \bibfield  {author} {\bibinfo {author} {\bibfnamefont {B.}~\bibnamefont
  {Schupp}}\ and\ \bibinfo {author} {\bibfnamefont {J.~J.}\ \bibnamefont
  {van~der Bij}},\ }\bibfield  {title} {\bibinfo {title} {{An axially symmetric
  Newtonian boson star}},\ }\href
  {https://doi.org/10.1016/0370-2693(95)01327-X} {\bibfield  {journal}
  {\bibinfo  {journal} {Phys. Lett. B}\ }\textbf {\bibinfo {volume} {366}},\
  \bibinfo {pages} {85} (\bibinfo {year} {1996})},\ \Eprint
  {https://arxiv.org/abs/astro-ph/9508017} {arXiv:astro-ph/9508017}
  \BibitemShut {NoStop}%
\bibitem [{\citenamefont {Flores}\ \emph {et~al.}(2025)\citenamefont {Flores},
  \citenamefont {Stegner}, \citenamefont {Chabysheva},\ and\ \citenamefont
  {Hiller}}]{Flores:2024tnu}%
  \BibitemOpen
  \bibfield  {author} {\bibinfo {author} {\bibfnamefont {A.}~\bibnamefont
  {Flores}}, \bibinfo {author} {\bibfnamefont {C.}~\bibnamefont {Stegner}},
  \bibinfo {author} {\bibfnamefont {S.~S.}\ \bibnamefont {Chabysheva}},\ and\
  \bibinfo {author} {\bibfnamefont {J.~R.}\ \bibnamefont {Hiller}},\ }\bibfield
   {title} {\bibinfo {title} {{Schr{\"o}dinger-Newton solitons with axial
  symmetry}},\ }\href {https://doi.org/10.1103/hjtl-ypxt} {\bibfield  {journal}
  {\bibinfo  {journal} {Phys. Rev. D}\ }\textbf {\bibinfo {volume} {112}},\
  \bibinfo {pages} {044055} (\bibinfo {year} {2025})},\ \Eprint
  {https://arxiv.org/abs/2412.18769} {arXiv:2412.18769 [hep-th]} \BibitemShut
  {NoStop}%
\bibitem [{\citenamefont {Liebling}\ and\ \citenamefont
  {Palenzuela}(2023)}]{Liebling:2012fv}%
  \BibitemOpen
  \bibfield  {author} {\bibinfo {author} {\bibfnamefont {S.~L.}\ \bibnamefont
  {Liebling}}\ and\ \bibinfo {author} {\bibfnamefont {C.}~\bibnamefont
  {Palenzuela}},\ }\bibfield  {title} {\bibinfo {title} {{Dynamical boson
  stars}},\ }\href {https://doi.org/10.1007/s41114-023-00043-4} {\bibfield
  {journal} {\bibinfo  {journal} {Living Rev. Rel.}\ }\textbf {\bibinfo
  {volume} {26}},\ \bibinfo {pages} {1} (\bibinfo {year} {2023})},\ \Eprint
  {https://arxiv.org/abs/1202.5809} {arXiv:1202.5809 [gr-qc]} \BibitemShut
  {NoStop}%
\bibitem [{\citenamefont {Rindler-Daller}\ and\ \citenamefont
  {Shapiro}(2012)}]{Rindler-Daller:2011afd}%
  \BibitemOpen
  \bibfield  {author} {\bibinfo {author} {\bibfnamefont {T.}~\bibnamefont
  {Rindler-Daller}}\ and\ \bibinfo {author} {\bibfnamefont {P.~R.}\
  \bibnamefont {Shapiro}},\ }\bibfield  {title} {\bibinfo {title} {{Angular
  Momentum and Vortex Formation in Bose-Einstein-Condensed Cold Dark Matter
  Haloes}},\ }\href {https://doi.org/10.1111/j.1365-2966.2012.20588.x}
  {\bibfield  {journal} {\bibinfo  {journal} {Mon. Not. Roy. Astron. Soc.}\
  }\textbf {\bibinfo {volume} {422}},\ \bibinfo {pages} {135} (\bibinfo {year}
  {2012})},\ \Eprint {https://arxiv.org/abs/1106.1256} {arXiv:1106.1256
  [astro-ph.CO]} \BibitemShut {NoStop}%
\bibitem [{\citenamefont {Schobesberger}\ \emph {et~al.}(2021)\citenamefont
  {Schobesberger}, \citenamefont {Rindler-Daller},\ and\ \citenamefont
  {Shapiro}}]{Schobesberger:2021ghi}%
  \BibitemOpen
  \bibfield  {author} {\bibinfo {author} {\bibfnamefont {S.~O.}\ \bibnamefont
  {Schobesberger}}, \bibinfo {author} {\bibfnamefont {T.}~\bibnamefont
  {Rindler-Daller}},\ and\ \bibinfo {author} {\bibfnamefont {P.~R.}\
  \bibnamefont {Shapiro}},\ }\bibfield  {title} {\bibinfo {title} {{Angular
  momentum and the absence of vortices in the cores of fuzzy dark matter
  haloes}},\ }\href {https://doi.org/10.1093/mnras/stab1153} {\bibfield
  {journal} {\bibinfo  {journal} {Mon. Not. Roy. Astron. Soc.}\ }\textbf
  {\bibinfo {volume} {505}},\ \bibinfo {pages} {802} (\bibinfo {year}
  {2021})},\ \Eprint {https://arxiv.org/abs/2101.04958} {arXiv:2101.04958
  [astro-ph.GA]} \BibitemShut {NoStop}%
\bibitem [{\citenamefont {Purohit}\ \emph {et~al.}(2023)\citenamefont
  {Purohit}, \citenamefont {Natwariya}, \citenamefont {Bhatt},\ and\
  \citenamefont {Mehta}}]{Purohit:2023izj}%
  \BibitemOpen
  \bibfield  {author} {\bibinfo {author} {\bibfnamefont {K.~J.}\ \bibnamefont
  {Purohit}}, \bibinfo {author} {\bibfnamefont {P.~K.}\ \bibnamefont
  {Natwariya}}, \bibinfo {author} {\bibfnamefont {J.~R.}\ \bibnamefont
  {Bhatt}},\ and\ \bibinfo {author} {\bibfnamefont {P.~K.}\ \bibnamefont
  {Mehta}},\ }\bibfield  {title} {\bibinfo {title} {{Formation of a Bose star
  in a rotating cloud}},\ }\href {https://doi.org/10.1007/s10509-023-04253-8}
  {\bibfield  {journal} {\bibinfo  {journal} {Astrophys. Space Sci.}\ }\textbf
  {\bibinfo {volume} {368}},\ \bibinfo {pages} {97} (\bibinfo {year} {2023})},\
  \Eprint {https://arxiv.org/abs/2311.12789} {arXiv:2311.12789 [astro-ph.CO]}
  \BibitemShut {NoStop}%
\bibitem [{\citenamefont {Di~Giovanni}\ \emph {et~al.}(2020)\citenamefont
  {Di~Giovanni}, \citenamefont {Sanchis-Gual}, \citenamefont
  {Cerd{\'a}-Dur{\'a}n}, \citenamefont {Zilh{\~a}o}, \citenamefont {Herdeiro},
  \citenamefont {Font},\ and\ \citenamefont {Radu}}]{DiGiovanni:2020ror}%
  \BibitemOpen
  \bibfield  {author} {\bibinfo {author} {\bibfnamefont {F.}~\bibnamefont
  {Di~Giovanni}}, \bibinfo {author} {\bibfnamefont {N.}~\bibnamefont
  {Sanchis-Gual}}, \bibinfo {author} {\bibfnamefont {P.}~\bibnamefont
  {Cerd{\'a}-Dur{\'a}n}}, \bibinfo {author} {\bibfnamefont {M.}~\bibnamefont
  {Zilh{\~a}o}}, \bibinfo {author} {\bibfnamefont {C.}~\bibnamefont
  {Herdeiro}}, \bibinfo {author} {\bibfnamefont {J.~A.}\ \bibnamefont {Font}},\
  and\ \bibinfo {author} {\bibfnamefont {E.}~\bibnamefont {Radu}},\ }\bibfield
  {title} {\bibinfo {title} {{Dynamical bar-mode instability in spinning
  bosonic stars}},\ }\href {https://doi.org/10.1103/PhysRevD.102.124009}
  {\bibfield  {journal} {\bibinfo  {journal} {Phys. Rev. D}\ }\textbf {\bibinfo
  {volume} {102}},\ \bibinfo {pages} {124009} (\bibinfo {year} {2020})},\
  \Eprint {https://arxiv.org/abs/2010.05845} {arXiv:2010.05845 [gr-qc]}
  \BibitemShut {NoStop}%
\bibitem [{\citenamefont {Siemonsen}\ and\ \citenamefont
  {East}(2021)}]{Siemonsen:2020hcg}%
  \BibitemOpen
  \bibfield  {author} {\bibinfo {author} {\bibfnamefont {N.}~\bibnamefont
  {Siemonsen}}\ and\ \bibinfo {author} {\bibfnamefont {W.~E.}\ \bibnamefont
  {East}},\ }\bibfield  {title} {\bibinfo {title} {{Stability of rotating
  scalar boson stars with nonlinear interactions}},\ }\href
  {https://doi.org/10.1103/PhysRevD.103.044022} {\bibfield  {journal} {\bibinfo
   {journal} {Phys. Rev. D}\ }\textbf {\bibinfo {volume} {103}},\ \bibinfo
  {pages} {044022} (\bibinfo {year} {2021})},\ \Eprint
  {https://arxiv.org/abs/2011.08247} {arXiv:2011.08247 [gr-qc]} \BibitemShut
  {NoStop}%
\bibitem [{\citenamefont {Dmitriev}\ \emph {et~al.}(2021)\citenamefont
  {Dmitriev}, \citenamefont {Levkov}, \citenamefont {Panin}, \citenamefont
  {Pushnaya},\ and\ \citenamefont {Tkachev}}]{Dmitriev:2021utv}%
  \BibitemOpen
  \bibfield  {author} {\bibinfo {author} {\bibfnamefont {A.~S.}\ \bibnamefont
  {Dmitriev}}, \bibinfo {author} {\bibfnamefont {D.~G.}\ \bibnamefont
  {Levkov}}, \bibinfo {author} {\bibfnamefont {A.~G.}\ \bibnamefont {Panin}},
  \bibinfo {author} {\bibfnamefont {E.~K.}\ \bibnamefont {Pushnaya}},\ and\
  \bibinfo {author} {\bibfnamefont {I.~I.}\ \bibnamefont {Tkachev}},\
  }\bibfield  {title} {\bibinfo {title} {{Instability of rotating Bose
  stars}},\ }\href {https://doi.org/10.1103/PhysRevD.104.023504} {\bibfield
  {journal} {\bibinfo  {journal} {Phys. Rev. D}\ }\textbf {\bibinfo {volume}
  {104}},\ \bibinfo {pages} {023504} (\bibinfo {year} {2021})},\ \Eprint
  {https://arxiv.org/abs/2104.00962} {arXiv:2104.00962 [gr-qc]} \BibitemShut
  {NoStop}%
\bibitem [{\citenamefont {Braaten}\ \emph
  {et~al.}(2016{\natexlab{b}})\citenamefont {Braaten}, \citenamefont
  {Mohapatra},\ and\ \citenamefont {Zhang}}]{Braaten:2016kzc}%
  \BibitemOpen
  \bibfield  {author} {\bibinfo {author} {\bibfnamefont {E.}~\bibnamefont
  {Braaten}}, \bibinfo {author} {\bibfnamefont {A.}~\bibnamefont {Mohapatra}},\
  and\ \bibinfo {author} {\bibfnamefont {H.}~\bibnamefont {Zhang}},\ }\bibfield
   {title} {\bibinfo {title} {{Nonrelativistic Effective Field Theory for
  Axions}},\ }\href {https://doi.org/10.1103/PhysRevD.94.076004} {\bibfield
  {journal} {\bibinfo  {journal} {Phys. Rev. D}\ }\textbf {\bibinfo {volume}
  {94}},\ \bibinfo {pages} {076004} (\bibinfo {year} {2016}{\natexlab{b}})},\
  \Eprint {https://arxiv.org/abs/1604.00669} {arXiv:1604.00669 [hep-ph]}
  \BibitemShut {NoStop}%
\bibitem [{\citenamefont {Braaten}\ \emph {et~al.}(2018)\citenamefont
  {Braaten}, \citenamefont {Mohapatra},\ and\ \citenamefont
  {Zhang}}]{Braaten:2018lmj}%
  \BibitemOpen
  \bibfield  {author} {\bibinfo {author} {\bibfnamefont {E.}~\bibnamefont
  {Braaten}}, \bibinfo {author} {\bibfnamefont {A.}~\bibnamefont {Mohapatra}},\
  and\ \bibinfo {author} {\bibfnamefont {H.}~\bibnamefont {Zhang}},\ }\bibfield
   {title} {\bibinfo {title} {{Classical Nonrelativistic Effective Field
  Theories for a Real Scalar Field}},\ }\href
  {https://doi.org/10.1103/PhysRevD.98.096012} {\bibfield  {journal} {\bibinfo
  {journal} {Phys. Rev. D}\ }\textbf {\bibinfo {volume} {98}},\ \bibinfo
  {pages} {096012} (\bibinfo {year} {2018})},\ \Eprint
  {https://arxiv.org/abs/1806.01898} {arXiv:1806.01898 [hep-ph]} \BibitemShut
  {NoStop}%
\bibitem [{\citenamefont {Namjoo}\ \emph {et~al.}(2018)\citenamefont {Namjoo},
  \citenamefont {Guth},\ and\ \citenamefont {Kaiser}}]{Namjoo:2017nia}%
  \BibitemOpen
  \bibfield  {author} {\bibinfo {author} {\bibfnamefont {M.~H.}\ \bibnamefont
  {Namjoo}}, \bibinfo {author} {\bibfnamefont {A.~H.}\ \bibnamefont {Guth}},\
  and\ \bibinfo {author} {\bibfnamefont {D.~I.}\ \bibnamefont {Kaiser}},\
  }\bibfield  {title} {\bibinfo {title} {{Relativistic Corrections to
  Nonrelativistic Effective Field Theories}},\ }\href
  {https://doi.org/10.1103/PhysRevD.98.016011} {\bibfield  {journal} {\bibinfo
  {journal} {Phys. Rev. D}\ }\textbf {\bibinfo {volume} {98}},\ \bibinfo
  {pages} {016011} (\bibinfo {year} {2018})},\ \Eprint
  {https://arxiv.org/abs/1712.00445} {arXiv:1712.00445 [hep-ph]} \BibitemShut
  {NoStop}%
\bibitem [{\citenamefont {Wang}(2001)}]{Wang:2001wq}%
  \BibitemOpen
  \bibfield  {author} {\bibinfo {author} {\bibfnamefont {X.~Z.}\ \bibnamefont
  {Wang}},\ }\bibfield  {title} {\bibinfo {title} {{Cold bose stars:
  Selfgravitating Bose-Einstein condensates}},\ }\href
  {https://doi.org/10.1103/PhysRevD.64.124009} {\bibfield  {journal} {\bibinfo
  {journal} {Phys. Rev. D}\ }\textbf {\bibinfo {volume} {64}},\ \bibinfo
  {pages} {124009} (\bibinfo {year} {2001})}\BibitemShut {NoStop}%
\bibitem [{\citenamefont {Baratta}\ \emph {et~al.}(2023)\citenamefont
  {Baratta}, \citenamefont {Dean}, \citenamefont {Dokken}, \citenamefont
  {Habera}, \citenamefont {Hale}, \citenamefont {Richardson}, \citenamefont
  {Rognes}, \citenamefont {Scroggs}, \citenamefont {Sime},\ and\ \citenamefont
  {Wells}}]{BarattaEtal2023}%
  \BibitemOpen
  \bibfield  {author} {\bibinfo {author} {\bibfnamefont {I.~A.}\ \bibnamefont
  {Baratta}}, \bibinfo {author} {\bibfnamefont {J.~P.}\ \bibnamefont {Dean}},
  \bibinfo {author} {\bibfnamefont {J.~S.}\ \bibnamefont {Dokken}}, \bibinfo
  {author} {\bibfnamefont {M.}~\bibnamefont {Habera}}, \bibinfo {author}
  {\bibfnamefont {J.~S.}\ \bibnamefont {Hale}}, \bibinfo {author}
  {\bibfnamefont {C.~N.}\ \bibnamefont {Richardson}}, \bibinfo {author}
  {\bibfnamefont {M.~E.}\ \bibnamefont {Rognes}}, \bibinfo {author}
  {\bibfnamefont {M.~W.}\ \bibnamefont {Scroggs}}, \bibinfo {author}
  {\bibfnamefont {N.}~\bibnamefont {Sime}},\ and\ \bibinfo {author}
  {\bibfnamefont {G.~N.}\ \bibnamefont {Wells}},\ }\href
  {https://doi.org/10.5281/zenodo.10447666} {\bibinfo {title} {{DOLFINx}: the
  next generation {FEniCS} problem solving environment}},\ \bibinfo
  {howpublished} {preprint} (\bibinfo {year} {2023})\BibitemShut {NoStop}%
\bibitem [{\citenamefont {Logg}\ \emph {et~al.}(2012)\citenamefont {Logg},
  \citenamefont {Mardal}, \citenamefont {Wells} \emph {et~al.}}]{LoggEtal2012}%
  \BibitemOpen
  \bibfield  {author} {\bibinfo {author} {\bibfnamefont {A.}~\bibnamefont
  {Logg}}, \bibinfo {author} {\bibfnamefont {K.-A.}\ \bibnamefont {Mardal}},
  \bibinfo {author} {\bibfnamefont {G.~N.}\ \bibnamefont {Wells}}, \emph
  {et~al.},\ }\href {https://doi.org/10.1007/978-3-642-23099-8} {\emph
  {\bibinfo {title} {Automated Solution of Differential Equations by the Finite
  Element Method}}}\ (\bibinfo  {publisher} {Springer},\ \bibinfo {year}
  {2012})\BibitemShut {NoStop}%
\bibitem [{\citenamefont {Roman}\ \emph {et~al.}(2026)\citenamefont {Roman},
  \citenamefont {Campos}, \citenamefont {Dalcin}, \citenamefont {Romero},\ and\
  \citenamefont {Tomas}}]{slepc-users-manual}%
  \BibitemOpen
  \bibfield  {author} {\bibinfo {author} {\bibfnamefont {J.~E.}\ \bibnamefont
  {Roman}}, \bibinfo {author} {\bibfnamefont {C.}~\bibnamefont {Campos}},
  \bibinfo {author} {\bibfnamefont {L.}~\bibnamefont {Dalcin}}, \bibinfo
  {author} {\bibfnamefont {E.}~\bibnamefont {Romero}},\ and\ \bibinfo {author}
  {\bibfnamefont {A.}~\bibnamefont {Tomas}},\ }\href@noop {} {\emph {\bibinfo
  {title} {{SLEPc} Users Manual, Chap.~3}}},\ \bibinfo {type} {Tech. Rep.}\
  \bibinfo {number} {DSIC-II/24/02 - Revision 3.25}\ (\bibinfo  {institution}
  {D. Sistemes Inform\`atics i Computaci\'o, Universitat Polit\`ecnica de
  Val\`encia},\ \bibinfo {year} {2026})\BibitemShut {NoStop}%
\bibitem [{\citenamefont {Hughes}(2000)}]{hughes2000finite}%
  \BibitemOpen
  \bibfield  {author} {\bibinfo {author} {\bibfnamefont {T.}~\bibnamefont
  {Hughes}},\ }\href {https://books.google.com.sg/books?id=yarmSc7ULRsC} {\emph
  {\bibinfo {title} {The Finite Element Method: Linear Static and Dynamic
  Finite Element Analysis}}}\ (\bibinfo  {publisher} {Dover Publications},\
  \bibinfo {year} {2000})\BibitemShut {NoStop}%
\bibitem [{\citenamefont {Yoshida}(1990)}]{Yoshida:1990zz}%
  \BibitemOpen
  \bibfield  {author} {\bibinfo {author} {\bibfnamefont {H.}~\bibnamefont
  {Yoshida}},\ }\bibfield  {title} {\bibinfo {title} {{Construction of higher
  order symplectic integrators}},\ }\href
  {https://doi.org/10.1016/0375-9601(90)90092-3} {\bibfield  {journal}
  {\bibinfo  {journal} {Phys. Lett. A}\ }\textbf {\bibinfo {volume} {150}},\
  \bibinfo {pages} {262} (\bibinfo {year} {1990})}\BibitemShut {NoStop}%
\bibitem [{Note1()}]{Note1}%
  \BibitemOpen
  \bibinfo {note} {The interpolation onto the cubic Cartesian grid preserves
  only the discrete fourfold rotational symmetry of the grid, leading to
  numerical leakage from \(s=1\) into \(s=1\pm 4j\), with \(j=1,2,\protect
  \ldots \). At a resolution of $256^3$, these spurious normalized populations
  remain at approximately $10^{-10}$ during the linear stage and decrease with
  grid refinement. They are negligible compared with the physically growing
  sidebands and do not affect the extracted instability rates.}\BibitemShut
  {Stop}%
\bibitem [{\citenamefont {Pitaevskii}\ and\ \citenamefont
  {Stringari}(2016)}]{PitaevskiiStringari2016}%
  \BibitemOpen
  \bibfield  {author} {\bibinfo {author} {\bibfnamefont {L.}~\bibnamefont
  {Pitaevskii}}\ and\ \bibinfo {author} {\bibfnamefont {S.}~\bibnamefont
  {Stringari}},\ }\href
  {https://doi.org/10.1093/acprof:oso/9780198758884.001.0001} {\emph {\bibinfo
  {title} {Bose-Einstein Condensation and Superfluidity}}}\ (\bibinfo
  {publisher} {Oxford University Press},\ \bibinfo {year} {2016})\BibitemShut
  {NoStop}%
\bibitem [{\citenamefont {Castin}(2001)}]{Castin2001}%
  \BibitemOpen
  \bibfield  {author} {\bibinfo {author} {\bibfnamefont {Y.}~\bibnamefont
  {Castin}},\ }\bibinfo {title} {Bose-einstein condensates in atomic gases:
  Simple theoretical results},\ in\ \href
  {https://doi.org/10.1007/3-540-45338-5_1} {\emph {\bibinfo {booktitle}
  {Coherent atomic matter waves}}}\ (\bibinfo  {publisher} {Springer Berlin
  Heidelberg},\ \bibinfo {year} {2001})\ p.\ \bibinfo {pages} {1–136},\
  \Eprint {https://arxiv.org/abs/cond-mat/0105058} {arXiv:cond-mat/0105058
  [cond-mat]} \BibitemShut {NoStop}%
\bibitem [{\citenamefont {Sallatti}\ \emph {et~al.}(2026)\citenamefont
  {Sallatti}, \citenamefont {Tomio}, \citenamefont {Pelinovsky},\ and\
  \citenamefont {Gammal}}]{darksoliton}%
  \BibitemOpen
  \bibfield  {author} {\bibinfo {author} {\bibfnamefont {R.~W.}\ \bibnamefont
  {Sallatti}}, \bibinfo {author} {\bibfnamefont {L.}~\bibnamefont {Tomio}},
  \bibinfo {author} {\bibfnamefont {D.~E.}\ \bibnamefont {Pelinovsky}},\ and\
  \bibinfo {author} {\bibfnamefont {A.}~\bibnamefont {Gammal}},\ }\bibfield
  {title} {\bibinfo {title} {Stability of dark solitons in a bubble
  bose-einstein condensate},\ }\href {https://doi.org/10.1103/1j8m-96pz}
  {\bibfield  {journal} {\bibinfo  {journal} {Phys. Rev. A}\ }\textbf {\bibinfo
  {volume} {113}},\ \bibinfo {pages} {L041303} (\bibinfo {year} {2026})},\
  \Eprint {https://arxiv.org/abs/2511.04385v2} {arXiv:2511.04385v2 [cond-mat]}
  \BibitemShut {NoStop}%
\bibitem [{\citenamefont {Gao}\ and\ \citenamefont
  {Cai}(2020)}]{GAO2020109058}%
  \BibitemOpen
  \bibfield  {author} {\bibinfo {author} {\bibfnamefont {Y.}~\bibnamefont
  {Gao}}\ and\ \bibinfo {author} {\bibfnamefont {Y.}~\bibnamefont {Cai}},\
  }\bibfield  {title} {\bibinfo {title} {Numerical methods for bogoliubov-de
  gennes excitations of bose-einstein condensates},\ }\href
  {https://doi.org/https://doi.org/10.1016/j.jcp.2019.109058} {\bibfield
  {journal} {\bibinfo  {journal} {Journal of Computational Physics}\ }\textbf
  {\bibinfo {volume} {403}},\ \bibinfo {pages} {109058} (\bibinfo {year}
  {2020})}\BibitemShut {NoStop}%
\bibitem [{\citenamefont {Xie}\ and\ \citenamefont
  {Zhang}(2026)}]{XIE2026114813}%
  \BibitemOpen
  \bibfield  {author} {\bibinfo {author} {\bibfnamefont {M.}~\bibnamefont
  {Xie}}\ and\ \bibinfo {author} {\bibfnamefont {Y.}~\bibnamefont {Zhang}},\
  }\bibfield  {title} {\bibinfo {title} {Computing the bogoliubov-de gennes
  excitations of two-component bose-einstein condensates},\ }\href
  {https://doi.org/https://doi.org/10.1016/j.jcp.2026.114813} {\bibfield
  {journal} {\bibinfo  {journal} {Journal of Computational Physics}\ }\textbf
  {\bibinfo {volume} {556}},\ \bibinfo {pages} {114813} (\bibinfo {year}
  {2026})},\ \Eprint {https://arxiv.org/abs/2506.12688} {arXiv:2506.12688
  [math]} \BibitemShut {NoStop}%
\bibitem [{\citenamefont {Sadaka}\ \emph {et~al.}(2024)\citenamefont {Sadaka},
  \citenamefont {Kalt}, \citenamefont {Danaila},\ and\ \citenamefont
  {Hecht}}]{SADAKA2024108948}%
  \BibitemOpen
  \bibfield  {author} {\bibinfo {author} {\bibfnamefont {G.}~\bibnamefont
  {Sadaka}}, \bibinfo {author} {\bibfnamefont {V.}~\bibnamefont {Kalt}},
  \bibinfo {author} {\bibfnamefont {I.}~\bibnamefont {Danaila}},\ and\ \bibinfo
  {author} {\bibfnamefont {F.}~\bibnamefont {Hecht}},\ }\bibfield  {title}
  {\bibinfo {title} {A finite element toolbox for the bogoliubov-de gennes
  stability analysis of bose-einstein condensates},\ }\href
  {https://doi.org/https://doi.org/10.1016/j.cpc.2023.108948} {\bibfield
  {journal} {\bibinfo  {journal} {Computer Physics Communications}\ }\textbf
  {\bibinfo {volume} {294}},\ \bibinfo {pages} {108948} (\bibinfo {year}
  {2024})},\ \Eprint {https://arxiv.org/abs/2303.05350} {arXiv:2303.05350
  [cond-mat]} \BibitemShut {NoStop}%
\bibitem [{\citenamefont {Smerzi}\ \emph {et~al.}(1997)\citenamefont {Smerzi},
  \citenamefont {Fantoni}, \citenamefont {Giovanazzi},\ and\ \citenamefont
  {Shenoy}}]{Smerzi1997}%
  \BibitemOpen
  \bibfield  {author} {\bibinfo {author} {\bibfnamefont {A.}~\bibnamefont
  {Smerzi}}, \bibinfo {author} {\bibfnamefont {S.}~\bibnamefont {Fantoni}},
  \bibinfo {author} {\bibfnamefont {S.}~\bibnamefont {Giovanazzi}},\ and\
  \bibinfo {author} {\bibfnamefont {S.~R.}\ \bibnamefont {Shenoy}},\ }\bibfield
   {title} {\bibinfo {title} {Quantum coherent atomic tunneling between two
  trapped bose-einstein condensates},\ }\href
  {https://doi.org/10.1103/PhysRevLett.79.4950} {\bibfield  {journal} {\bibinfo
   {journal} {Phys. Rev. Lett.}\ }\textbf {\bibinfo {volume} {79}},\ \bibinfo
  {pages} {4950} (\bibinfo {year} {1997})},\ \Eprint
  {https://arxiv.org/abs/cond-mat/9706221v1} {arXiv:cond-mat/9706221v1
  [cond-mat]} \BibitemShut {NoStop}%
\bibitem [{\citenamefont {Ostrovskaya}\ \emph {et~al.}(2000)\citenamefont
  {Ostrovskaya}, \citenamefont {Kivshar}, \citenamefont {Lisak}, \citenamefont
  {Hall}, \citenamefont {Cattani},\ and\ \citenamefont
  {Anderson}}]{Ostrovskaya2000}%
  \BibitemOpen
  \bibfield  {author} {\bibinfo {author} {\bibfnamefont {E.~A.}\ \bibnamefont
  {Ostrovskaya}}, \bibinfo {author} {\bibfnamefont {Y.~S.}\ \bibnamefont
  {Kivshar}}, \bibinfo {author} {\bibfnamefont {M.}~\bibnamefont {Lisak}},
  \bibinfo {author} {\bibfnamefont {B.}~\bibnamefont {Hall}}, \bibinfo {author}
  {\bibfnamefont {F.}~\bibnamefont {Cattani}},\ and\ \bibinfo {author}
  {\bibfnamefont {D.}~\bibnamefont {Anderson}},\ }\bibfield  {title} {\bibinfo
  {title} {Coupled-mode theory for bose-einstein condensates},\ }\href
  {https://doi.org/10.1103/PhysRevA.61.031601} {\bibfield  {journal} {\bibinfo
  {journal} {Phys. Rev. A}\ }\textbf {\bibinfo {volume} {61}},\ \bibinfo
  {pages} {031601(R)} (\bibinfo {year} {2000})},\ \Eprint
  {https://arxiv.org/abs/cond-mat/9910185v2} {arXiv:cond-mat/9910185v2
  [cond-mat]} \BibitemShut {NoStop}%
\bibitem [{\citenamefont {Castin}\ and\ \citenamefont
  {Dum}(1996)}]{Castin:1996zz}%
  \BibitemOpen
  \bibfield  {author} {\bibinfo {author} {\bibfnamefont {Y.}~\bibnamefont
  {Castin}}\ and\ \bibinfo {author} {\bibfnamefont {R.}~\bibnamefont {Dum}},\
  }\bibfield  {title} {\bibinfo {title} {{Bose-Einstein Condensates in Time
  Dependent Traps}},\ }\href {https://doi.org/10.1103/PhysRevLett.77.5315}
  {\bibfield  {journal} {\bibinfo  {journal} {Phys. Rev. Lett.}\ }\textbf
  {\bibinfo {volume} {77}},\ \bibinfo {pages} {5315} (\bibinfo {year}
  {1996})},\ \Eprint {https://arxiv.org/abs/2607.18944} {arXiv:2607.18944
  [cond-mat.quant-gas]} \BibitemShut {NoStop}%
\bibitem [{\citenamefont {Asakawa}\ \emph {et~al.}(2024)\citenamefont
  {Asakawa}, \citenamefont {Ishihara},\ and\ \citenamefont
  {Tsubota}}]{Asakawa:2023tkm}%
  \BibitemOpen
  \bibfield  {author} {\bibinfo {author} {\bibfnamefont {K.}~\bibnamefont
  {Asakawa}}, \bibinfo {author} {\bibfnamefont {H.}~\bibnamefont {Ishihara}},\
  and\ \bibinfo {author} {\bibfnamefont {M.}~\bibnamefont {Tsubota}},\
  }\bibfield  {title} {\bibinfo {title} {{Collective Excitations of
  Self-Gravitating Bose{\textendash}Einstein Condensates: Breathing Mode and
  Appearance of Anisotropy under Self-Gravity}},\ }\href
  {https://doi.org/10.1093/ptep/ptae078} {\bibfield  {journal} {\bibinfo
  {journal} {PTEP}\ }\textbf {\bibinfo {volume} {2024}},\ \bibinfo {pages}
  {063J01} (\bibinfo {year} {2024})},\ \Eprint
  {https://arxiv.org/abs/2307.14018} {arXiv:2307.14018 [cond-mat.quant-gas]}
  \BibitemShut {NoStop}%
\end{thebibliography}%

\end{document}